\documentclass[11pt]{article}

\usepackage{geometry}
\usepackage{graphicx}
\usepackage{amssymb}
\usepackage{amsfonts}
\usepackage{dcolumn}
\usepackage{bm}
\usepackage{float}
\usepackage{footnote}
\usepackage{dcolumn}
\usepackage{amsmath}
\usepackage{amsfonts}
\usepackage{bm}
\usepackage{cite}
\usepackage{epsfig}
\usepackage{amssymb}
\usepackage{latexsym}
\usepackage[normalem]{ulem}
\usepackage{authblk}

\renewcommand{\vec}[1]{{\bm #1}}

\newcommand{\ket}[1]{\left| #1 \right>} % for Dirac bras
\numberwithin{equation}{section}
\numberwithin{figure}{section}

\pdfoutput=1
\usepackage[utf8]{inputenc}
\usepackage{amsfonts}
\usepackage{amssymb}
\usepackage{braket}
\usepackage{tikz}
\usepackage{caption}
\usepackage{subcaption}
\usepackage{color}
\usepackage{graphicx}
\usepackage{indentfirst}
\numberwithin{equation}{section}
\numberwithin{figure}{section}
\numberwithin{table}{section}
\newcommand{\st}{\stackrel}

\usepackage{diagbox}

\usepackage{mathtools}

\usepackage[normalem]{ulem}

\usepackage{hyperref} % adds hyper links inside the generated pdf file
\hypersetup{
 colorlinks=true,
 linkcolor=blue,
 anchorcolor = blue,
 citecolor = blue,
 filecolor = blue,
 urlcolor = blue
}

\makeatletter
\newcommand\Row[1]{%
  \par\nobreak\nointerlineskip\vskip-\fboxrule%
  \@tfor\@tempa:=#1 \do {\csname ChessBox\@tempa\endcsname\kern-\fboxrule}}
\define@key{chessB}{side}{\def\ChessSide{#1}}
\define@key{chessB}{colori}{\def\ChessColori{#1}}
\define@key{chessB}{colorii}{\def\ChessColorii{#1}}
\setkeys{chessB}{
  side=1.5em,
  colori=black!70,
  colorii=white}
\makeatother

\newenvironment{Chessboard}[1][]
  {\setkeys{chessB}{#1}%
  \par\medskip\setlength\parindent{0pt}}
  {\par\medskip}

\newcommand\beq {\begin{eqnarray}}
\newcommand\eeq {\end{eqnarray}}
\newcommand\be            {\begin{equation}}
\newcommand\bea           {\begin{equation}\begin{array}l\displaystyle}
\newcommand\ee            {\end{equation}}
\newcommand\bes           {\begin{subequations}}
\newcommand\esu           {\end{subequations}}

\renewcommand{\(}{\left(}
\renewcommand{\)}{\right)}
\renewcommand{\[}{\left[}
\renewcommand{\]}{\right]}

\newcommand{\bigx}[1]{\bBigg@{#1}}

\newcommand\lab           {\label }

\renewcommand\th         {\theta}

\newcommand\red[1]{{\color{red}{#1}} }

\newcommand\goto{\rightarrow} 
\newcommand\vev[1]{{\langle#1\rangle}}

\def\3pt#1#2#3{{\langle{#1}\vert{#2}\vert{#3}\rangle}}

\newcommand{\EQ}{\begin{equation}}
\newcommand{\EN}{\end{equation}}

\def\tilde{\widetilde}
\def\bar{\overline}
\def\hat{\widehat}
\def\*{\star}
\def\[{\left[}
\def\]{\right]}
\def\({\left(}      

\def\){\right)}

\def\frac#1#2{\dfrac{#1}{#2}}

\def\vev#1{\langle #1 \rangle}
\def\ket#1{ | #1 \rangle}

\def\2pi{\hbox{$2\pi i$}}

\def\dsl{\raise.15ex\hbox{/}\kern-.57em\partial}
\def\Dsl{\,\raise.15ex\hbox{/}\mkern-.13.5mu D}
\def\be{\beta}

\def\2pi{\hbox{$2\pi i$}}

\def\dsl{\raise.15ex\hbox{/}\kern-.57em\partial}
\def\Dsl{\,\raise.15ex\hbox{/}\mkern-.13.5mu D}

\def\barray{\begin{eqnarray}}
\def\earray{\end{eqnarray}}
\def\beq{\begin{equation}}
\def\eeq{\end{equation}}

\def\AA{\leavevmode\setbox0=\hbox{h}
\dimen0=\ht0 \advance\dimen0 by-1ex\rlap{\raise.67\dimen0\hbox{\char'27}}A}

\def\red#1{{\color{red}{#1}}}

\def\red#1{{\color{red}{#1}}}

\def\be{\begin{equation}}
\def\ee{\end{equation}}
\def\bea{\begin{eqnarray}}
\def\eea{\end{eqnarray}}

\def\ba{\begin{aligned}}
\def\ea{\end{aligned}}

\renewcommand\th          {\theta}

\newcommand{\vt}{\vartheta}

\begin{document}
\title{
Quantum Integrability vs Experiments: \\
Correlation Functions and Dynamical Structure Factors
}
\author[1]{M. Lencs{\'e}s}
\author[2]{G. Mussardo}
\author[3,4,5]{G. Tak{\'a}cs}

\affil[1]{ Wigner Research Centre for Physics,\\
Konkoly-Thege Miklós u. 29-33, 1121 Budapest, Hungary}
\affil[2]{SISSA and INFN, Sezione di Trieste,\\ via Bonomea 265, I-34136, Trieste, Italy}
\affil[3]{Department of Theoretical Physics,\\ Institute of Physics, Budapest University of Technology and Economics,\\ 1111 Budapest, M{\H u}egyetem rkp. 3, Hungary}
\affil[4]{MTA-BME Quantum Dynamics and Correlations Research Group,\\ Budapest University of Technology and Economics, 1111 Budapest,\\ M{\H u}egyetem rkp. 3, Hungary}
\affil[5]{BME-MTA Statistical Field Theory ‘Lend{\"u}let’ Research Group,\\ Budapest University of Technology and Economics,\\ 1111 Budapest, M{\H u}egyetem rkp. 3, Hungary}

\date{March 26, 2023}
\maketitle

\begin{abstract}
Integrable Quantum Field Theories can be solved exactly using bootstrap techniques based on their elastic and factorisable $S$-matrix. While knowledge of the scattering amplitudes reveals the exact spectrum of particles and their on-shell dynamics, the expression of the matrix elements of the various operators allows the reconstruction of off-shell quantities such as two-point correlation functions with a high level of precision. In this review, we summarise results relevant to the contact point between theory and experiment providing a number of quantities that can be computed theoretically with great accuracy. We concentrate on universal amplitude ratios which can be determined from the measurement of generalised susceptibilities, and dynamical structure factors, which can be accessed experimentally e.g. via inelastic neutron scattering or nuclear magnetic resonance. Besides an overview of the subject and a summary of recent advances, we also present new results regarding generalised susceptibilities in the tricritical Ising universality class.
\end{abstract}

\newpage
\tableofcontents
\clearpage

\section{Introduction}
Recent years have brought outstanding progress on the subject of low-dimensional physics, in particular concerning the dynamics of models with local interactions satisfying general conditions of translation and rotational invariance. At their critical point, these models also display conformal invariance which allows us to determine the spectrum of the anomalous dimensions of their operators, identify the relevant operators, and compute multi-point correlation functions and partition functions relative to various boundary conditions \cite{BPZ,FQS1,NPB240:312,cardy861,cardy862}. 

Off-critical deformations can be achieved using one or several of the relevant operators, permitting to analyse the scaling region associated with their class of universality. For deformations inducing a mass gap, it is useful to regard the deformed theory as a quantum field theory with a spectrum of massive excitations interacting by quantum scattering.  While the resulting quantum field theory is generally  non-integrable -- showing all the familiar and general phenomena of particle productions or decays, also accompanied  by the  presence of resonances -- as shown by Zamolodchikov's seminal paper \cite{1989IJMPA...4.4235Z}, certain deformations give rise to integrable quantum field theories. The on-shell dynamics of these theories is fully encoded into the elastic and factorisable $S$-matrix. Nowadays, the exact scattering amplitudes of several important models are known  (c.f. \cite{Mussardo:2020rxh}  and references therein). Knowledge of the amplitudes allows us to determine also the number of different particles and their mass spectrum. 

However, in order to find an experimental confirmation of the theoretical description of the integrable directions in the scaling region it is necessary to go off the mass shell and determine the correlation functions. Most of the experimentally relevant information is encoded into the two-point correlations functions, therefore here we focus our attention on these quantities. On the one hand, knowledge of the two-point correlation functions enables access to the various susceptibilities of the model via the fluctuation-dissipation theorem and therefore to various important  universal ratios of Renormalisation Group (see, for instance, \cite{cardy_1996}). On the other hand, two-point correlation functions can be directly probed at various energy and momentum scales using, e.g., neutron or light scattering on a physical material sample associated with the universality class. 

Therefore it is crucial to determine these two-point correlation functions as accurately as possible. This can be achieved using the spectral representation \cite{Peskin:1995ev}, due to the generically fast convergence of the corresponding series \cite{CMpol}, which can be constructed from the matrix elements of the various operators, the so-called Form Factors. In integrable quantum field theories, the exact Form Factors can be constructed by exploiting their analytic properties \cite{KW,Smirnov}, which allows the derivation of accurate experimental predictions.

In this review, we discuss the various experimentally relevant quantities which describe the off-critical dynamics of an integrable deformation of the critical point, taking the magnetic deformation of the Ising model and the thermal deformation of the Tricritical Ising model with a temperature as our main examples. Besides the intrinsic interest of these two models, they also display a particle spectrum and dynamics ruled by the exceptional Lie algebras $E_8$ and $E_7$, which makes the calculation of their Form Factors particularly rich and interesting.

\section{Experimentally relevant quantities from statistical field theory}
\subsection{Field theory description of the scaling region}
Consider a $D$-dimensional statistical model near its critical point. Under a general set of conditions, such as the locality of the interactions, together with the translation and rotation invariance of the system, the scaling region nearby the critical point can be described by the Euclidean action
\begin{equation}
\mathcal{A}^{(E)}=\mathcal{A}^{(E)}_\text{CFT}+\sum\limits_{i=1}^n g_i\int d^Dx\, \Phi_i(x)
\label{eq:PCFT_Euclidean}\end{equation}
where $\mathcal{A}^{(E)}_\text{CFT}$ is the fixed-point action corresponding to a Conformal Field Theory (CFT), while the $\Phi_i(x)$ are relevant fields of conformal dimension $\Delta_i$ normalised by the condition that their two-point functions have the short-distance behaviour
\begin{equation}
    \langle \Phi_i(x)\Phi_j(0)\rangle\simeq \frac{\delta_{ij}}{|x|^{4\Delta_i}}\,,
    \quad |x| \rightarrow 0 \,.
    \label{eq:cft_normalisation}
\end{equation}
Simple dimensional analysis shows that the coupling constants behave as $g_i\sim \mu^{D-2\Delta_i}$ in terms of an arbitrary mass scale $\mu$.  Consider first the case with a single perturbing field $\Phi_i(x)$ giving rise to a finite correlation length $\xi$. This quantity can be expressed as
\begin{equation}
    \xi= a (K_i g_i)^{-\tfrac{1}{D-2\Delta_i}}=a \xi_i^0 g_i^{-\tfrac{1}{D-2\Delta_i}}\,,
\end{equation}
where
\begin{equation}
    \xi_i^0= K_i^{-\tfrac{1}{D-2\Delta_i}}
\end{equation}
and $a=1/\mu$ is a length scale and the dimensionless quantities $K_i$ are non-universal metric factors which depend on the specific realisation of the universality class and the choice of units for the couplings $g_i$. In the presence of several perturbations, the correlation length can be written as
\begin{equation}
    \xi= a (K_i g_i)^{-\tfrac{1}{D-2\Delta_i}}\mathcal{L}_i\left( \frac{K_j g_j}{(K_ig_i)^{\phi_{ji}}}\right) \,.
\end{equation}
Such a representation exists for any choice of the selected perturbing field $i$, where 
\begin{equation}
    \phi_{ji}=\frac{D-2\Delta_j}{D-2\Delta_i}
\end{equation}
are called crossover exponents, while the functions $\mathcal{L}_i(x)$ are universal homogeneous scaling functions of their arguments. Consider now the free energy $f(g_1,\dots g_n)$ defined as 
\begin{equation}
  e^{-f(g_1,\dots g_n)}=\int D\phi \exp -\left\{ \mathcal{A}_\text{CFT}+\sum\limits_{i=1}^n g_i\int \Phi_i(x)d^Dx \right\}\,.
\end{equation}
In the thermodynamic limit, the assumption of hyperscaling implies that the singular part of the free energy density is proportional to the $D$th power of the correlation length, which leads to several equivalent parameterisations of the form
\begin{equation}
     f_i(g_1,\dots g_n)= (K_i g_i)^{-\tfrac{D}{D-2\Delta_i}}\mathcal{F}_i\left( \frac{K_j g_j}{(K_ig_i)^{\phi_{ji}}}\right)
\label{eq:free_energy_scaling}\end{equation}
where, similarly to the $\mathcal{L}_i(x)$, the functions $\mathcal{F}_i(x)$ are universal homogeneous scaling functions of their arguments.

\subsection{Critical exponents and universal ratios}\label{sec:uniratios}

The scaling form \eqref{eq:free_energy_scaling} of the free energy allows extraction of the scaling behaviour of several important quantities, such as Vacuum Expectation Values (VEV) and also generalised susceptibilities. Considering for simplicity a single perturbation $g_i$ with all other couplings vanishing, i.e. $g_j=0$ for $j\neq i$, and denote the expectation values under such perturbation by $\langle\dots\rangle_i$. Then the one-point functions of the fields $\Phi_j$ can be expressed as 
\begin{equation}
    \vev{\Phi_j}_i=-\left.\frac{\partial f_i}{\partial g_j}\right|_{g_{l\neq i}=0}=B_{ji}g_i^{\tfrac{2\Delta_j}{D-2\Delta_i}}\,,
\end{equation}
with 
\begin{equation}
    B_{ji}\sim K_j K_i^{\tfrac{2\Delta_j}{D-2\Delta_i}}\,,
\end{equation}
which can also be inverted in the form
\begin{equation}
    g_i=D_{ij}\vev{\Phi_j}_i^{\tfrac{D-2\Delta_i}{2\Delta_j}} \,,
    \quad \, 
    D_{ij}\sim K_i^{-1}K_j^{\tfrac{2\Delta_i-D}{2\Delta_j}}\,.
\end{equation}
Similarly, the generalised susceptibilities can be computed as 
\begin{equation}
    \hat{\Gamma}^i_{jk}=-\left.\frac{\partial^2 f_i}{\partial g_j\partial g_k}\right|_{g_{l\neq i}=0}
    ={\Gamma}^i_{jk} g_i^{\tfrac{2\Delta_j+2\Delta_k-D}{D-2\Delta_i}}\,,
\end{equation}
with 
\begin{equation}
    {\Gamma}^i_{jk} \sim K_j K_k K_i^{\tfrac{2\Delta_j+2\Delta_k-D}{D-2\Delta_i}}\,.
\end{equation}
Notice that the generalised susceptibilities can also be computed by 
means of the fluctuation-dissipation theorem in terms of the connected two-point functions as follows: 
\EQ
\frac{\partial}{\partial g_i} \langle \varphi_j\rangle_i = 
- \int d^2 x \langle \varphi_i(x) \varphi_j(0) \rangle_c^i \,.
\label{derflu}
\EN 
The one-point functions and susceptibilities are obviously non-universal quantities since they contain the metric factors $K_i$. However, their behaviour as functions of the coupling $g_i$ is given by power laws dictated by the critical exponents associated with the universality class. Furthermore, it is possible to define the so-called universal amplitude ratios demanding that the metric factors cancel, such as  \cite{prlFMS,PRE63:016103,Delfino:2002lkc,Zamcth2,DelfinoCardy1,DelfinoIsing,CMpol,SFW1}:
\begin{align}
(R_c)^i_{jk}&=\frac{\Gamma^i_{ii}\Gamma^i_{jk}}{B_{ji}B_{ki}}
\qquad
(R_\chi)^i_{j}=\Gamma^i_{jj}D_{jj}B_{ji}^{\tfrac{D-4\Delta_j}{2\Delta_j}}
\qquad
(R_\xi)^i=\left(\Gamma^i_{ii}\right)^{1/D}\xi^0_i
\nonumber\\
(Q_2)^i_{jk}&=\frac{\Gamma^i_{jj}}{\Gamma^k_{jj}} \left( \frac{\xi^0_k}{\xi^0_j} \right)^{D-4\Delta_j}
\qquad
(R_A)^i_{j}=\Gamma^i_{jj}D_{ii}^{\tfrac{4\Delta_j+2\Delta_i-2D}{D-2\Delta_i}}B_{ij}^{\tfrac{2\Delta_j-D}{\Delta_i}}\,.
\end{align}
In contrast to the critical exponents which are characteristic of the critical point itself, these quantities carry information about the scaling region. Moreover, they typically vary significantly between different universality classes, in contrast to the critical exponents which usually assume small values that only differ by a small amount between different universality classes. In addition, the universal ratios are much more accessible to experiments as they require only measurements at some fixed value of the coupling driving the system off criticality, while critical exponents can only be determined from data spanning several decades in the values of the couplings $g_i$. 
Furthermore, the number of independent universal ratios is generally much larger than the number of independent critical exponents. As a result of these properties, universal ratios allow for a more convenient and precise determination of the universality class.

\subsection{Dynamical Structure Factors}
While the field theory \eqref{eq:PCFT_Euclidean} can be considered as a description of a $D$-dimensional (classical) statistical model in equilibrium, it can also be continued to real (Minkowski) time to describe the dynamics of a quantum statistical system in $d=D-1$ spatial dimensions with the real-time action
\begin{equation}
    \mathcal{A}=\mathcal{A}_\text{CFT}-\sum\limits_{i=1}^n g_i\int dt \,d^dx\,\Phi_i(t,\vec{x})\,.
\label{eq:PCFT_realtime}\end{equation}
This allows access to other experimentally relevant quantities called dynamical structure factors (DSF) which describe the response of the system under probes such as inelastic neutron or Raman scattering experiments. The DSFs are given as Fourier transform of real-time two-point functions of appropriate operator fields $\mathcal{O}_i$ and $\mathcal{O}_j$: 
\begin{equation}
    {\cal S}^{\mathcal{O}_i\mathcal{O}_j}(\omega,\vec{q})=\int dt \,d^dx\, e^{i\omega t - i \vec{q}\cdot\vec{x}}
    \langle\mathcal{O}_i(\vec{x},t)\mathcal{O}_j(\vec{0},0)\rangle
\label{eq:dsf}
\end{equation}
As we discuss in the next sections, these quantities can be computed efficiently by employing the spectral representation of the correlation functions built upon the matrix elements of the operator in the basis of asymptotic multi-particle states, the so-called Form Factors. 

\section{S-matrix bootstrap}\label{sec:Smatrix}

In this section, we briefly recall the $S$-matrix theory of two-dimensional integrable models which leads, in particular, to the exact spectrum of the massive excitations away from the critical point.  The key point of this formalism is the self-consistent bootstrap method for computing the exact expressions of all scattering amplitudes and the mass of the particles, which was pioneered by A.B. Zamolodchikov \cite{1989IJMPA...4.4235Z} (for a review and an extended list of references c.f. \cite{Mussardo:2020rxh}). In this paper, our attention is focused on two main examples, the Ising Model  in an external magnetic field at $T=T_c$ (related to the exceptional Lie algebra $E_8$), and the Tricritical Ising Model away from its critical temperature (related to the exceptional Lie algebra $E_7$).  

\subsection{Asymptotic states and rapidity} 
Integrable models are characterised by an infinite number of local conserved quantities $Q_s$, expressed as 
\EQ
Q_s = \int  \left[ T_{s+1} \, dz + \Theta_{s-1} \,d\overline z\right] \,
\EN
where $T_{s+1}$ and $\Theta_{s-1}$ are certain local fields satisfying the continuity equations 
\EQ
\partial_{\overline z} T_{s+1} =\partial_z \Theta_{s-1}\,,
\label{conservationlawspin}
\EN
where the positive integer $s$ denotes the spin of the fields. In integrable models the scattering processes which involve $n$ incoming particles are 
elastic and factorizable in terms of the $n (n-1)/2$ two-body scattering amplitudes. The momenta of the particles involved in scattering processes are {\em on-shell} and in $(1+1)$ dimensions there exists an efficient parameterization of the dispersion relation $E^2 - p^2 = m^2$ in terms of the rapidity variable  $\theta$, which for a particle of mass $m_i$ is given by 
\EQ
p_i^{(0)} =m_i \cosh\theta_i \,,  
\quad
p_i^{(1)} =m_i \sinh\theta_i
\,.
\label{rapdisper}
\EN 
In terms of the rapidity parameter, Lorentz transformations are rotations with a hyperbolic angle $\alpha$ acting as $\theta \rightarrow \theta + \alpha$. The asymptotic $n$-particle states can be written as    
\EQ
|A_{a_1}(\theta_1)  A_{a_2}(\theta_2) \ldots A_{a_n}(\theta_n) \rangle 
\,,
\label{asimstate}
\EN 
where the symbol $A_{a_i}(\theta_i)$ denotes a particle of type $a_i$ moving with rapidity $\theta_i$. An initial asymptotic state is given by a set of free particles at $t \rightarrow -\infty$. Since in the $(1 + 1)$ dimensional theories, the actual motion takes place along a line, the fastest particle has to be on the farthest left-hand side of all the others, while the slowest must be on the right-hand side of all the others, with the remaining particles ordered according to the value of their rapidities between those two. To express such a situation in a more formal way, it is appropriate to consider the symbols  $A_{a_i}(\theta_i)$ as non-commuting symbols, whose order is associated with the spatial ordering of the particles that they represent. In this way, an initial asymptotic state can be written as  
\EQ
| A_{a_1}(\theta_1)  A_{a_2}(\theta_2) \ldots A_{a_n}(\theta_n) \rangle 
\,,
\label{asimstatein}
\EN 
where the rapidities are listed in a {\em decreasing} order  
\EQ
\theta_1 \geq \theta_2 \geq \theta_3 \cdots \geq \theta_n \,.
\label{statiin}
\EN 
Similarly, a final asymptotic state is made of free particles at $t \rightarrow +\infty$. Hence each particle must be on the left-hand side of all the others that move faster than it. The final asymptotic states can be then represented by 
\EQ
| A_{a_1}(\theta_1)  A_{a_2}(\theta_2) \ldots A_{a_n}(\theta_n) \rangle 
\,,
\label{asimstateout}
\EN 
but this time with an {\em increasing} order of the rapidities, i.e.  
\EQ
\theta_1 \leq \theta_2 \leq \theta_3 \cdots \leq \theta_n \,.
\label{statiout}
\EN 
Obviously, each state (\ref{asimstate}) can always be re-ordered by means of a certain number of commutations of the symbols $A_i(\theta_i)$ between neighbour particles where each commutation can be interpreted as a scattering process of two particles, as discussed below. 
It is customary to normalise these states as   
\EQ
\langle A_i(\theta_1) |A_j(\theta_2) \rangle =2\pi \delta_{ij} 
\delta(\theta_1 - \theta_2) \,.
\label{normonepartheta}
\EN 
Consequently, the density of states with rapidities  $(\theta, \theta + d\theta)$  is given by $d\theta/2\pi$. 

\subsection{Analytic properties of the $2$-particle $S$-matrix}
Without losing in generality, in integrable QFT it is enough to analyse the $2$-particle $S$-matrix. Concerning the analytic behaviour of this amplitude, in addition to the delta function $\delta^{(2)}(p_1 + p_2 - p_3 -p_4)$ expressing the conservation of the total energy and momentum, Lorentz invariance imposes that the scattering amplitude depends on the particle momenta only on their invariant combinations, given by the Mandelstam variables $s$ and $t$  which for the process 
$$
A_i \, A_j \,\rightarrow \,
A_k \, A_l \,,
$$ 
are given by 
\EQ
\begin{array}{c}
s(\theta_{ij}) = (p_1 + p_2)^2 =
m^2_i + m^2_j + 2 \, m_i \, m_j \,\cosh\theta_{ij}
\,, \\
\theta_{ij} = \theta_i - \theta_j
\,.
\end{array}
\label{smantheta}
\EN 
For physical processes, $\theta_{ij}$ assumes real values, and consequently, $s$ is also real with values $s \geq (m_i + m_j)^2$. The 
Mandelstam variable $t$ is instead given by 
\EQ
\begin{array}{c}
t(\theta_{ij}) = (p_1 - p_2)^2 =
m^2_i + m^2_j - 2 \, m_i \, m_j \,\cosh\theta_{ij}
\,. 
\end{array}
\label{tmantheta}
\EN 
Consequently, we can switch between the $s$ to the $t$-channels by the analytic continuation related to crossing symmetry
\EQ
t(\theta) =s(i \pi - \theta) \,,
\label{contanalist}
\EN 
In $(1+1)$-dimensional systems, the two-particle $S$-matrix elements are defined by
\EQ
|A_i(\theta_1) A_j(\theta_2) \rangle =
S_{ij}^{kl}(\theta) 
|A_k(\theta_2)  A_l(\theta_1) \rangle \,, 
\label{Sijkl}
\EN 
with $\theta = \theta_{12}$ and $\theta_1 > \theta_2$, consistently with the definition of the initial and final asymptotic states previously discussed. In this equation, a summation is understood for indices $k$ and $l$ for particles that are not distinguished from $i$ and $j$ by any eigenvalues of the conserved charges. Note that the dependence of the $S$-matrix on the difference of the rapidities is a consequence of Lorentz invariance. There relation between the $S$-matrix defined above and the one written in terms of the original Mandelstam variable ${\mathcal S}(s)$ is given by the Jacobian of the transformation $s(\theta)$
\EQ
{\mathcal S}_{ij}^{kl}(s) =  4 m_i m_j \,\sinh\theta\, 
S_{ij}^{kl}(\theta) \,.
\label{calSjacobian}
\EN 
The scattering amplitudes satisfy the physical conditions of unitarity  
\EQ
\sum_{n,m} S_{ij}^{nm}(\theta) \, S_{nm}^{kl}(-\theta) =
\delta_i^k \, \delta_j^l \,, 
\label{unitarietaijkl}
\EN 
and crossing symmetry  
\EQ
S_{i\,j}^{k\,l}(\theta) =
S_{i \,\bar l}^{k \,\bar j}(i \pi - \theta) \,,
\label{crossinijkl}
\EN 
where the bar upon the indices denotes the anti-particles. Notice that the unitarity and crossing symmetry equations can be analytically continued for arbitrary values of $\theta$ and therefore they hold in the complex $\theta$ plane. Moreover, the definition of the $S$-matrix can be reinterpreted as an algebra (known as Faddev-Zamolodchikov algebra) for the symbols $A_a(\theta)$
 \EQ
A_i(\theta_1) \, A_j(\theta_2) =
S_{ij}^{kl}(\theta) \,
A_k(\theta_2) \, A_l(\theta_1) \,. 
\label{FadZam}
\EN 
In other words,  the scattering processes can be equivalently interpreted as commutation relations among the operators that create the particles. The well-known Yang-Baxter equations which relate three amplitudes are nothing else than the associativity condition of this algebra. 

The elastic $S$ matrices are meromorphic functions in the complex plane of $\theta$. The bound states correspond to simple poles with positive residue\footnote{It is worth mentioning that this concept can be generalised both to the cases of poles with negative residues and higher order poles.} along the imaginary segment $(0,i \pi)$ of the $\theta$ variable. Consider a $S$-matrix with incoming particles $A_i$ and $A_j$ that has a simple pole in the $s$-channel at $\theta = i \, u_{ij}^n$. In correspondence with this pole, the amplitude can be expressed as   
\EQ
S_{ij}^{kl} \,\simeq \, i \,\frac{R^{(n)}}{\theta - i u_{ij}^n} 
\,, 
\EN 
with the residue $R^{(n)}$ related to the {\em on-shell} vertex functions of the incoming particles and the bound state $A_n$
\EQ
R^{(n)} = \Gamma_{ij}^n \, \Gamma_{kl}^n \,.
\EN 
A non-zero value of $\Gamma_{ij}^n$ obviously implies a pole singularity in the other two amplitudes $S_{in}$ and $S_{jn}$ as well, where the poles are now due to the bound states $A_j$ and $A_i$. Since in the bootstrap approach, the bound states are on the same footing as the asymptotic states, there is an important relation among the masses of the system: if $\theta = i u_{ij}^n$ is the position of the pole in the scattering of the particles $A_i$ and $A_j$, the mass of the bound state is given by 
\EQ
m^2_n =m_i^2 + m_j^2 + 2 m_i m_j \, \cos u_{ij}^n \,.
\label{triangolodellemasse}
\EN 
This relation is simply obtained by substituting in the Mandelstam variable $s$ given in eqn. (\ref{smantheta}) the resonance condition $\theta = i u_{ij}^n$. Notice that this formula expressed a well-known geometrical relation, known as Carnot theorem, among the sides of a triangle (here equal to the values of the masses), where $u_{ij}^n$ is one of the external angles of such a triangle.
From this geometrical interpretation, it is easy to show that the positions of the poles in the three channels satisfy  
\EQ
u_{ij}^n + u_{in}^j + u_{jn}^i = 2 \pi \,. 
\EN 
As it is the case for the models discussed later, the elastic $S$-matrix of $(1+1)$-dimensional systems may also have higher order poles, whose interpretation stays in the singularities coming from multiple scattering processes. 

\subsection{Diagonal $S$-matrices and bootstrap equations}
For the models we consider in the following, the discussion of the $S$-matrix can be simplified. The reason is that they have a non-degenerate mass spectrum and moreover all particles are neutral and uniquely identified in terms of their different eigenvalues with respect to the conserved charges. Under these conditions, the elasticity of the scattering processes enforces the vanishing of the reflection amplitude. As a result, the $S$-matrix turns out to be completely diagonal and the Yang--Baxter equations are then identically satisfied. The unitarity and crossing symmetry conditions simplify as follows 
\EQ
S_{ab}(\theta) \, S_{ab}(-\theta) =1 
\,
,
\,
S_{a b}(\theta) =
S_{a b }(i\pi- \theta) \,. 
\label{unicrossempl}
\EN 
These two equations imply that the amplitudes $S_{ab}(\theta)$ are periodic functions of $\theta$ with period $2 \pi i$: in this case, the Riemann surface of the $S$-matrix consists of a double covering of the complex plane $s$. There general solution of Eqs. (\ref{unicrossempl}) can be expressed in terms of products of the meromorphic functions 
\EQ
f_x(\theta)  =
\frac
{\tanh\frac{1}{2}(\theta + i \pi x)}
{\tanh\frac{1}{2}(\theta - i \pi x)} =\frac{\sinh\theta + i \sin\pi x}{\sinh\theta - i \sin\pi x}\,.
\label{ffunctionS}
\EN 
The simple poles of these functions are at $\theta = i \pi x$ and $\theta = i \pi (1 -x)$, and are related by the crossing transformation. Due to their periodicity the parameter $x$ can always be chosen as $-1 \leq x \leq 1$. They have also simple zeros at  $-i \pi x$ and $-i \pi (1-x)$. Hence, as a consequence of the unitarity and crossing symmetry equations, any amplitude $S_{ab}(\theta)$ of a diagonal $S$-matrix can be expressed as   
\EQ
S_{ab}(\theta) =\prod_{x \in {\mathcal A}_{ab}} f_x(\theta) \,,
\label{prodottof}
\EN 
with ${\mathcal A}_{ab}$ a discrete subset of the interval $(-1,1]$. 

Unitarity and crossing symmetry equations, however, do not fix the position of the poles i.e. the sets  ${\mathcal A}_{ab}$. This can be achieved by an additional dynamical condition, provided by the bootstrap principle that posits that the bound states are on the same footing as the asymptotic states. As a consequence, the amplitudes that involve the bound states can be obtained in terms of the amplitudes of the external particles and vice versa. This translates into an additional equation satisfied by the scattering amplitudes    
\EQ
S_{i \bar l}(\theta) =
S_{ij}(\theta + i {\bar u}_{jl}^k) \, 
S_{ik}(\theta - i {\bar u}_{lk}^j) \,,
\label{bootstrapboundstate}
\EN 
where  
\EQ
{\bar u}_{ab}^c \,\equiv \,\pi - u_{ab}^c \,.
\EN 
Therefore, a consistent $S$-matrix must have a set of poles for all amplitudes $S_{ab}$ compatible with the bootstrap equation  (\ref{bootstrapboundstate}), that can be interpreted in terms of bound states or multi-particle scattering processes of the asymptotic particles themselves. The masses of the particles are determined by the relation (\ref{triangolodellemasse}). In practice, this means starting from the amplitude that involves the lightest particle, therefore with the simplest pole structure, and then iteratively applying the bootstrap equations   (\ref{bootstrapboundstate}) in order to get the scattering amplitudes involving the bound states of higher mass. This is how the scattering theories of the Ising model in a magnetic field and the Tricritical Ising Model away from the critical temperature can be constructed. 

Since the $\theta$ dependence of the elementary building block $f_{\alpha}(\theta)$ is fixed, later to denote the building blocks $f_x(\theta)$ of the $S$-matrix it is convenient to use the notation 
\EQ
(f_{\alpha}(\theta))^{p_{\alpha}}\,\equiv \,\st{\bf a}{(\alpha)}^{p_\alpha}
\label{convention}
\EN
where $\alpha$ denotes the location of the simple pole, $p_{\alpha}$ its multiplicity $p_\alpha$ and the top index ${\bf a}$ indicates the particle species corresponding to the bound state $A_a$. 

\section{Correlation functions and Dynamical Structure Factors}

\subsection{Form Factor Equations}\label{subsec:FFeqs}
Here we briefly summarise the basic functional equations satisfied by matrix elements of local fields on particle asymptotic states in an integrable quantum field theory \cite{KW,Smirnov,DMIMMF,Koubek:1993ke,AMVcluster,DSC}. 

The form factors (FF) of a local field $\Phi(x)$ are defined as
\EQ
F^{\Phi}_{a_1,\ldots ,a_n}(\th_1,\ldots,\th_n) = \langle 0 |
\Phi(0) |A_{a_1}(\th_1),\ldots,A_{a_n}(\th_n) \rangle\,.
\label{form}
\EN
General matrix elements of local fields can be expressed in terms of form factors using crossing symmetry. For a scalar operator $\Phi(x)$, relativistic invariance requires that its FF depend only on the rapidity differences $\th_i-\th_j$. Form Factors satisfy several equations which are listed below for the case of scalar operators. 
\paragraph{Monodromy properties.}  
Interchanging two particles is equivalent to their scattering, while an analytic continuation of rapidities by $i\pi$ corresponds to crossing symmetry. As a result, the FF satisfy the monodromy equations 
\EQ
\begin{array}{lll}
F^{\Phi}_{a_1,..,a_i,a_{i+1},.. a_n}(\th_1,..,\th_i, \th_{i+1}, ..
, \th_{n}) &=& \,S_{a_i a_{i+1}}(\th_i-\th_{i+1}) \,
F^{\Phi}_{a_1,..a_{i+1},a_i,..a_n}(\th_1,..,\th_{i+1},\th_i ,.., \th_{n})
\, ,\\
F^{\Phi}_{a_1,a_2,\ldots a_n}(\th_1+2 \pi i, \dots, \th_{n-1}, \th_{n} ) &=&
F^{\Phi}_{a_2,a_3,\ldots,a_n,a_1}(\th_2 ,\ldots,\th_{n}, \th_1)
\, .
\end{array}
\label{permu1}\\
\EN
\paragraph{Singularity structures.}
The FF must have pole singularities related to those of the $S$-matrix. The most common ones are first-order poles which fall into three different classes:
\begin{enumerate}
    \item {\bf Kinematical poles.} These are related to the annihilation processes of a pair of particle  and anti-particle states, located at $\th_{a}-\th_{\overline a} = i\pi$ with their residue given by a recursive equation between the $(n+2)$-particle FF and the $n$-particle FF
\begin{align}
 -i\,\lim_{\tilde\th \rightarrow \th}
(\tilde\th - \th)
&F^{\Phi}_{a,\overline a,a_1,\ldots,a_n}(\tilde\th + i\pi,\th,\th_1,\ldots,\th_n)
\nonumber\\
&=
\left(1 - e^{2\pi i \omega_a}\,\prod_{j=1}^{n} S_{a,a_j}(\th -\th_j)\right)\,
F^{\Phi}_{a,\ldots,a_n}(\th_1,\ldots,\th_{n})  \,, \label{recursive}
\end{align}
where $\omega_a$ is the index of mutual semi-locality of the operator $\Phi$ with respect to the particle $A_a$. 
\item {\bf Bound state poles.} The second type of simple poles is related to the presence of bound states appearing as simple poles in the $S$-matrix. If $\th = i u_{ab}^c$ and
$\Gamma_{ab}^c$ are the resonance angle and the three-particle coupling of the fusion $A_a \times A_b \rightarrow A_c$ respectively, then
FF involving the particles $A_a$ and $A_b$ also has a pole at $\th = i u_{ab}^c$ and its residue gives rise to a recursive equation between 
the $(n+2)$-particle FF and the $(n+1)$-particle FF
\EQ
-i \,\lim_{\th_{ab} \rightarrow i u_{ab}^c}
(\th_{ab} - i u_{ab}^c) \,
F^{\Phi}_{a,b,a_i,\ldots,a_n}\left(\th_a, \th_b ,\th_1,\ldots,\th_n\right) =
\Gamma_{ab}^c \,
F^{\Phi}_{c,a_i,\ldots,a_n}\left(\th_c,\th_1,\ldots,\th_n\right)\,,
\label{recurb}
\EN
where
$\th_c = (\th_a \overline u_{bc}^a + \th_b \overline u_{ac}^b)/u_{ab}^c$. In general, the FF may also have higher-order poles but, in order to address 
them, let's first discuss the general way of parameterising the FF and the special role played by the $2$-particle FF.
\item {\bf Multi-particle poles.} A third kind of simple poles in the FF are related to double poles in the S-matrix which reflect multi-particle scattering processes. For a double pole in the S-matrix $S_{ab}$ located at an angle $\phi\in(0,\pi)$ which can be written $\phi=u_{ad}^c+u_{bd}^e-\pi$ for some $c$, $d$ and $e$, the FF has a simple pole at $\theta=i\phi$ \cite{DMIMMF}. For a two-particle FF, the corresponding residue is given by 
\beq
-i\lim_{\theta_{ab}\to i\phi}(\theta_{ab}-i\phi)F^{\varphi}_{ab}(\theta_{ab})=\Gamma_{ad}^c\Gamma_{bd}^eF^\varphi_{ce}(i\gamma),\label{multiparticlepole}
\eeq
where $\gamma-\pi-u_{cd}^a-u_{de}^b$.
\end{enumerate}

In general, form factors can also have higher order poles, corresponding to third and higher order poles in the S-matrix; however, we omit these details as they are not needed in the construction of the form factors used here, referring the reader to \cite{DMIMMF} for details.

\paragraph{Bound on the asymptotic growth.} If $\Delta_{\Phi}$  is the conformal dimension of the scalar operator ${\Phi}(x)$ then its FF satisfies  
\EQ
\lim_{|\th_i|\goto\infty}
F^{\Phi}_{a_1,\ldots,a_n}(\th_1,\ldots,\th_n) \sim \,
e^{y_{\Phi}|\th_i|}\,.
\lab{bound}
\EN
with 
\EQ
y_{\Phi}\,\leq\,\Delta_\Phi\,.
\lab{bbb}
\EN

\paragraph{Cluster property.} In a massive field theory, the form factors of relevant operators $\Phi_i$ are expected to satisfy\footnote{Up to phases related to the normalisation of the states.}  the asymptotic factorisation \cite{Koubek:1993ke,AMVcluster,DSC}
\begin{eqnarray}
&&\lim_{\alpha\rightarrow\infty} F_{r+l}^{\Phi_a}(\theta_1 +\alpha,\ldots,\theta_r+\alpha,\theta_{r+1},\ldots,\theta_{r+l}) \nonumber\\
&&=
F^{\Phi_b}_r(\theta_1,\ldots,\theta_r) \, F_l^{\Phi_c}(\theta_{r+1},\ldots,\theta_{r+l})\,,
\label{aympfact}
\end{eqnarray}
where $\Phi_a, \Phi_b, \Phi_c$ label fields of the same conformal dimension. The origin of this identity simply comes from the fact that both functions on the right-hand side satisfy the same form factor axioms and therefore are defining matrix elements of some operators $\Phi_b$ and $\Phi_c$ with an appropriate normalisation. 

\paragraph{Operator content.} As noticed in \cite{Cardy:1990pc}, 
the equations satisfied by the Form Factors admit several solutions, each of them which can be put in correspondence with an operator of the theory. In other words, similarly to what happens in Conformal Field Theory \cite{BPZ1984} where the operator content comes from the irreducible representations of the Virasoro algebra, given that an operator is identified by its matrix elements on a complete set of states, for integrable deformations of the fixed point action the operator content can be extracted by looking at the different solutions of the Form Factors. This property will be exploited in the following in order to pin down the various operators of the Ising and Tricritical Ising Models. The validity of the solutions can be cross-checked by computing the scaling dimension of the operator using a sum rule following from the $\Delta$-theorem \cite{DSC}.

\subsection{Solving the Form Factor Equations}\label{subsec:solvingff}

The two-particle Form Factors are the basic building blocks from which the general solutions of the Form Factor Equations are obtained. The Form Factor Equations imply that the $2$-particle FF $F^{\Phi}_{ab}(\th)$ are meromorphic functions of the rapidity difference defined in the strip ${\it Im}\,\th\in[0,\pi)$, with their monodromy dictated by the equations (\ref{permu1})
\EQ
F^{\Phi}_{ab}(\th)=S_{ab}(\th)F^{\Phi}_{ab}(-\th)\,,
\lab{w1}
\EN
\EQ
F^{\Phi}_{ab}(i\pi+\th)=F^{\Phi}_{ab}(i\pi-\th)\,.
\lab{w2}
\EN
If $F^{\it min}_{ab}(\th)$ denotes a solution of these equations, free of poles and zeros in the strip ${\cal S}$, the general solution $F^{\Phi}_{ab}(\th)$ can be written as 
\EQ
F^{\Phi}_{ab}(\th)=\frac{Q^{\Phi}_{ab}(\th)}{D_{ab}(\th)}
F^{min}_{ab}(\th)\,,
\lab{param}
\EN
where $D_{ab}(\th)$ and $Q^{\Phi}_{ab}(\th)$ are polynomials in $\cosh\th$: the former is fixed by the singularity structure of $S_{ab}(\th)$ while the latter carries the whole information about the operator ${\Phi}(x)$. For excitations which are non-local with respect to the operator $\Phi$, the form factor above will include an extra term $\cosh\theta/2$ (either in the numerator or in the denominator, according to the asymptotic behaviour in $\theta$ of the form factor) which is even under $\theta \rightarrow -\theta$ but changes a sign under the transformation $\theta \rightarrow \theta + 2 \pi i$ which probes the mutual non-locality of $\Phi$ with respect to the excitations. 

For arbitrary number $n$ ($n \geq 3$) of particles, the form factor can be parameterised as
\be
\ba
F^{\Phi}_{a_{1},...,a_{n}}(\theta_{1},...,\theta_{n})=Q^{\Phi}_{a_{1},...,a_{n}}(\theta_{1},...,\theta_{n}) \prod_{i \leq j}\dfrac{F_{a_{i}a_{j}}^{min}(\theta_{i}-\theta_{j})}{(e^{\theta_{i}}+e^{\theta_{j}})^{\delta_{a_{i}a_{j}}}D_{a_{i}a_{j}}(\theta_{i}-\theta_{j})}\,,
\label{FF_Ansatz}
\ea
\ee
where $Q_{a_{1},...,a_{n}}^{\Phi}(\theta)$ are symmetric polynomials in $\cosh\theta$. It must be noticed, however, that since the various particles of the theory appear as bound states of some scattering channel which can all be traced back to the amplitude $S_{11}(\theta)$ involving the fundamental particle $A_1$, all the Form Factors of the theory can be then obtained once all the $n$-particle Form Factors involving just the lightest particle $A_1$ are known.

The parameterisation \eqref{FF_Ansatz} solves all the monodromy properties, while the singularity equations \eqref{recursive}, \eqref{recurb} and \eqref{multiparticlepole} result in recursive equations between the polynomials, which allow for their explicit construction.

\subsection{Spectral expansion for correlation functions and Dynamical Structure Factors}\label{Structurefactor}

The importance of form factors consists in the fact that for a set of local operators $\mathcal{O}_i$, their two-point functions $\langle \mathcal{O}_i(x,\tau)\mathcal{O}_j(0,0)\rangle$ can be evaluated in terms of the so-called spectral representation. Such a representation is obtained by inserting a resolution of the identity in terms of asymptotic multi-particle states between the two operators
\begin{eqnarray}
	{\mathbb{I}} = 
	\sum_{n=1}^\infty \sum_{\{a_i\}}
	\prod_{k=1}^{\mathcal{N}} \left(\frac{1}{N_k^{\{a_i\}}!}\right)
	\int \frac{d\theta_1}{2\pi}\cdots \frac{d\theta_n}{2\pi} |a_1,\theta_1;\dots;a_n,\theta_n\rangle\langle a_n,\theta_n;\dots;a_1,\theta_1|\,.
	\label{eq:identity_resolution}
\end{eqnarray}
The sum is over states involving an arbitrary number $n$ of particles with arbitrary rapidities $\theta_i,~i=1,\dots,n$ and particle species labelled by the integers $1\leq a_i\leq \mathcal{N}$ such that $a_1\leq a_2 \leq \dots \leq a_n$, while $N_k^{\{a_i\}}$ is the number of particles of type $k$ in the set $\{a_i\}$. 

Using the completeness of states expressed by Eq.\, \eqref{eq:identity_resolution}, the Euclidean correlation functions can be expressed as
\begin{align}
    \langle \mathcal{O}_i(x)\mathcal{O}_j(0)\rangle=&\sum_{n=1}^{\infty}\sum_{\{a_i\}}
	\prod_{k=1}^{\mathcal{N}} \left(\frac{1}{N_k^{\{a_i\}}!}\right)
	\int \frac{d\theta_1}{2\pi}\cdots \frac{d\theta_n}{2\pi}
	F^{\mathcal{O}_i}_{a_1,a_2,\dots,a_n}(\theta_1,\theta_2,\dots,\theta_n)F^{\mathcal{O}_j*}_{a_1,a_2,\dots,a_n}(\theta_1,\theta_2,\dots,\theta_n)\nonumber\\
      &\times\exp\left\{-|x|\sum\limits_{i=1}^n m_{a_i}\cosh\theta_i\right\}
\end{align}
In the case of real-time dynamics, the dynamical structure factors (DSF) are defined as the Fourier transforms of the corresponding correlators
\beq
	{\cal S}^{ij}(\omega,q) = \int dt dx e^{i\omega t - i q x}
    \langle \mathcal{O}_i(t,x)\mathcal{O}_j(0,0)\rangle \,.
\eeq 
which, using \eqref{eq:identity_resolution}, can be written as a sum over $n$-particle contributions 
\begin{eqnarray}
    {\cal S}^{ij}(\omega,q) &=& \sum_{n=1}^{\infty} {\cal D}^{ij}_n(\omega, q)\\
	{\cal S}^{ij}_n(\omega, q) &=& 
	\sum_{\{a_i\}}
	\prod_{k=1}^{\mathcal{N}} \left(\frac{1}{N_k^{\{a_i\}}!}\right)
	\int \frac{d\theta_1}{2\pi}\cdots \frac{d\theta_n}{2\pi}
	F^{\mathcal{O}_i}_{a_1,a_2,\dots,a_n}(\theta_1,\theta_2,\dots,\theta_n)F^{\mathcal{O}_j*}_{a_1,a_2,\dots,a_n}(\theta_1,\theta_2,\dots,\theta_n)
	\nonumber\\ && \times(2\pi)^2\delta\left(\omega - \sum_{i=1}^n E_i\right)\,\delta\left(q-\sum_{i=1}^n P_i\right)\,.
 \label{dsf_spectral_exp}
\end{eqnarray}
For dynamical structure factors note that we can restrict the evaluation of ${\cal S}^{ij}(\omega,q)$ to $q=0$ since the general expression can be obtained by Lorentz invariance. 
Although it is generally not possible to evaluate all terms in the above infinite sums, it can be nevertheless argued that the spectral series typically converge quite rapidly \cite{CMpol}, i.e. most of the spectral weight comes from the first few contributions with small number $n$ of particles. 
Furthermore, in a gapped theory, truncating the sum \eqref{dsf_spectral_exp} to those states with energies $E_n < \omega$ leads to an exact result up to $\omega$ due to the presence of energy thresholds.  

Let us consider the contributions of the $n=1,2$ and $3$-particle FF. Assuming that the masses of the particles are ordered as $m_1\leq m_2\leq\dots$, this gives the exact spectral function for energies $\omega < 4m_1$. The single particle contributions to ${\cal S}^{ij}(\omega,q=0)$ takes the simple form
\begin{eqnarray}
	{\cal S}^{ij}_1(\omega,q=0)= \sum_{a}\frac{2\pi}{m_a}\delta(\omega-m_a)F^i_aF^{j*}_a
	\label{eq:S1}
\end{eqnarray}
where $a$ here indexes all the different particles that couple to $\mathcal{O}_i$ and $\mathcal{O}_j$, i.e. they give a coherent contribution of isolated delta-function peaks.  

The two-particle contributions take the form
\begin{eqnarray}
	{\cal S}^{ij}_2(\omega,q=0) &=& \sum_{a_1 \leq a_2}\left(\frac{1}{2}\right)^{\delta_{a_1,a_2}}\frac{\Theta(\omega-(m_{a_1}+m_{a_2}))}{m_{a_1}m_{a_2}|\sinh(\theta_1-\theta_2)|}F^i_{a_1,a_2}(\theta_1-\theta_2)F^{j*}_{a_2,a_1}(\theta_1-\theta_2)
	\label{eq:S2}
\end{eqnarray}
where 
\EQ
\theta_1-\theta_2 = 
{\rm arccosh} \,\left(
\frac{\omega^2 - m^2_{a_i} - m^2_{a_j}}{2 m_{a_i} m_{a_j}}\right)
\EN
from energy and momentum conservation. The two-particle contribution is an incoherent continuum, which for a given two-particle pair $(a_1,a_2)$ opens at the threshold for $\omega = m_{a_1}+m_{a_2}$.  As $\omega$ approaches this threshold from above, the kinematical prefactor  generally introduces a square-root van Hove singularity in the spectral function; however, this singularity can be washed out in particular cases when the two-particle form factor  vanishes as $\theta_1-\theta_2\rightarrow 0$.

The three-particle contributions can be written as \cite{2021PhRvB.103w5117W}
\begin{eqnarray}
	{\cal S}^{ij}_3(\omega,q=0)=\sum_{a_1\leq a_2 \leq a_3} \left(\prod_a \frac{1}{N_a^{\{a_1,a_2,a_3\}} !} \right)
	\int\frac{d\theta_3}{2\pi} \frac{F^{\mathcal{O}_i}_{a_1,a_2,a_3}(\theta_1,\theta_2,\theta_3)F^{\mathcal{O}_j*}_{a_1,a_2,a_3}(\theta_1,\theta_2,\theta_3)}{m_{a_1}m_{a_2}|\sinh(\theta_1-\theta_2)|}
	\label{eq:S3}
\end{eqnarray}
where $N_a^{\{a_1,a_2,a_3\}}$ is the number of times the species label $a$ occurs in the set $\{a_1,a_2,a_3\}$, and the rapidities $\theta_1,\theta_2,\theta_3$ satisfy the kinematic constraints
\begin{eqnarray}
	\omega &=& m_{a_1} \cosh\theta_1+m_{a_2} \cosh\theta_2+m_{a_3} \cosh\theta_3\nonumber\\
	0 &=& m_{a_1} \sinh\theta_1+m_{a_2} \sinh\theta_2+m_{a_3} \sinh\theta_3\,.
	\label{eq:3ptkinematics}
\end{eqnarray}
For three particles of equal mass $m_{a_1}=m_{a_2}=m_{a_3}=m$, the solution of the kinematic constraints can be written explicitly. The integration range of $\theta_3$ is restricted to 
\begin{equation}
    \cosh\theta_3 \leq \frac{\omega^2-3 m^2}{2m\omega}\,,
    \label{eq:S3kinematicrestriction}
\end{equation}
where \ref{eq:3ptkinematics} have two solutions related by swapping the sign of $\theta_{12}$, given by 
\begin{eqnarray}
 \cosh\theta_{12}&=&\frac{\omega^{2}-2m\omega\cosh\theta_{3}-m^2}{2m^2}\text{ with the choice }\theta_{12}\geq0\nonumber\\
	\cosh\theta_{1}&=&\frac{2\sqrt{\left(3+4\cosh\theta_{12}+\cosh2\theta_{3}\right)\cosh^{4}\frac{\theta_{12}}{2}}-\sinh\theta_{3}\sinh\theta_{12}}{2\left(1+\cosh\theta_{12}\right)}
\end{eqnarray}
with the sign of $\theta_{1}$ chosen so that $\sinh\theta_{1}+\sinh\left(\theta_{1}-\theta_{12}\right)=-\sinh\theta_{3}$  is satisfied. 

Finally we report here the case for the four-particle contributions using a notation similar to \eqref{eq:S3} \cite{2021PhRvB.103w5117W}
\begin{align}
	{\cal S}^{ij}_4(\omega,q=0)=\sum_{a_1\leq a_2 \leq a_3 \leq a_4} &\left(\prod_a \frac{1}{N_a^{\{a_1,a_2,a_3,a_4\}} !} \right)\nonumber\\
	&\times\int\frac{d\theta_3}{2\pi} \int\frac{d\theta_4}{2\pi}\frac{F^{\mathcal{O}_i}_{a_1,a_2,a_3,a_4}(\theta_1,\theta_2,\theta_3,\theta_4)F^{\mathcal{O}_j*}_{a_1,a_2,a_3,a_4}(\theta_1,\theta_2,\theta_3,\theta_3)}{m_{a_1}m_{a_2}|\sinh(\theta_1-\theta_2)|}
	\label{eq:S4}
\end{align}
where the rapidities $\theta_1,\theta_2,\theta_3,\theta_4$ satisfy the kinematic constraints
\begin{eqnarray}
	\omega &=& m_{a_1} \cosh\theta_1+m_{a_2} \cosh\theta_2+m_{a_3} \cosh\theta_3+m_{a_4} \cosh\theta_4\nonumber\\
	0 &=& m_{a_1} \sinh\theta_1+m_{a_2} \sinh\theta_2+m_{a_3} \sinh\theta_3+m_{a_4} \sinh\theta_4\,.
	\label{eq:4ptkinematics}
\end{eqnarray}
With all the information gathered here and in the previous sections, let's now address the two classes of universality of our interest.

\section{Ising Model}\label{sec:Ising}
In this section, we are going to discuss the universality class of the Ising Model. The Euclidean field theory describes the vicinity of the critical point of the two-dimensional classical Ising model which is defined by the partition function
\begin{align}
    Z_\text{CIM}&=\sum_{s_i=\pm 1} \exp\left\{-\frac{1}{T}\mathcal{H}(\{s_i\})\right\}\nonumber\\
    & \mathcal{H}_\text{CIM}(\{s_i\})=-J\sum_{\langle i,j\rangle}s_is_j-H\sum_i s_i
    \label{eq:2DIM}
\end{align}
where the indices $i,j$ run over the two-dimensional lattice and $\langle i,j\rangle$ denotes pairs of nearest neighbour sites. For zero magnetic field, $H$ the model has a critical point at a critical temperature $T_c$; at high temperatures $T>T_c$ the system is in a paramagnetic phase, while at low temperatures $T<T_c$ it is in a ferromagnetic phase.

When continued to real-time, the corresponding Minkowski field theory describes the dynamics of the one-dimensional quantum Ising spin chain governed by the Hamiltonian
\begin{equation}
    \hat{H}_\text{IC}=-\mathcal{J}\sum_{i} \left\{\sigma_i^z \sigma_{i+1}^z+ h_T \sigma_i^x+ h_L \sigma_i^z
    \right\}
    \label{eq:TFIM}
\end{equation}
(where $\sigma_i^{x,y,z}$ are the Pauli matrices) which has a quantum phase transition for zero longitudinal field $h_L$ at the quantum critical point (QCP) $h_T=1$, with $h_T<1$ corresponding to the ferromagnetic phase, while $h_T>1$ to the paramagnetic phase.

The quantum chain Hamiltonian \eqref{eq:TFIM} can be obtained from the transfer matrix representation of the partition function of the two-dimensional classical Ising model \eqref{eq:2DIM}, with the transverse field $h_T$ corresponding to the temperature $T$, and the longitudinal field $h_L$ corresponding to the magnetic field $H$.

\subsection{Majorana fermions and order/disorder fields}\label{subsec:IM_CFT}
At the critical point, the Ising model is described by the first minimal unitary conformal model \cite{BPZ} with central charge $c=1/2$ and the Kac table of the conformal dimensions reported in Table~\ref{KACISING}.  
\begin{table}
\begin{center}
\bgroup
\setlength{\tabcolsep}{0.5em}
\begin{tabular}{|c||c|c|c|}
\hline
& & & \\[-1.3em]
2 & $\frac{1}{2}$ & $\frac{1}{16}$ & $0$   \\
& & & \\[-1.3em]
\hline
& & & \\[-1.3em]
1 &$0$ & $\frac{1}{16}$  & $\frac{1}{2}$  \\
& & & \\[-1.3em]
\hline
\hline
\diagbox[dir=SW,width=2em,height=2.5em]{$r$}{$s$}& 1 & 2 & 3  \\
\hline
\end{tabular}
\egroup
\end{center}
\caption{Kac table of the first minimal unitary model of CFT, corresponding to the Ising model. Irreducible representations of the Virasoro algebra are labelled by indices $r=1,2$ and $s=1,2,3$.}
\label{KACISING}
\end{table}
In order to discuss the operator content of this theory, let us introduce the following notations for the fields
\begin{eqnarray}
&& {\bf 1}:  (0,0) \hspace{7mm} ,  \hspace{7mm} \,\psi  :   \left(\frac{1}{2},0\right) \hspace{7mm} , \hspace{5mm}  \bar\psi  :  \left(0,\frac{1}{2}\right) \\
&& \epsilon  :  \left(\frac{1}{2},\frac{1}{2}\right) \hspace{3mm} ,   \hspace{7mm}  \sigma  =  \left(\frac{1}{16},\frac{1}{16}\right)  \hspace{3mm} , \hspace{3mm}   \mu  :  \left(\frac{1}{16},\frac{1}{16}\right) \nonumber\,,
\label{eq:Ising_fields}
\end{eqnarray}
where $(\Delta,\bar\Delta)$ are the conformal weights provided by the Kac table, with each pair identifying a physical operator obtained by combining the analytic and anti-analytic parts.
For our purposes, it is important to notice the presence of the fermionic fields $\psi$ and $\bar\psi$, which are the analytic and the anti-analytic components of the two-dimensional Majorana fermion field $\Psi = (\psi,\bar\psi)$. As is well known,  the critical action of the Ising model can be written\footnote{In the following $z = x_1 +i x_2$ and $\overline z = x_1 - i x_2$.} 
\EQ
{\mathcal A}_\text{CI} =\frac{1}{2\pi} \int d^2 x \left[\psi \,\partial_{\bar z}\, \psi + 
\bar\psi \,\partial_z\,\bar\psi\right] \,.
\label{AZIONECIAK}
\EN 
This action is quadratic and therefore free. The primary field $\epsilon$ can be identified as the fermion bilinear $\bar{\psi}\psi$. The corresponding equations of motion
\EQ
\begin{array}{l}
\partial_z \,\bar \psi=0 \,,\\
\partial_{\bar z} \psi =0 \,,
\end{array}
\label{eqqqqmot}
\EN 
show that the two spinor components are decoupled, and $\psi$ depends only on $z$ while $\bar\psi$ depends only on $\bar z$.  
The analytic and anti-analytic components of the stress-energy tensor which accompany the action (\ref{AZIONECIAK}) are  
\EQ
T=-\frac{1}{2} :\psi \partial_{z} \psi: 
\; \,, \,\;
\bar T =-\frac{1}{2} :\bar\psi\partial_{\bar z}\bar\psi:\,.
\EN 
Focusing on the analytic sector alone, the OPE of $\psi$ with itself is given by
\EQ
\psi(z_1) \psi(z_2) =\frac{1}{z_1-z_2} + \cdots .
\label{sviloppsi}
\EN 
The mode expansion of the Taylor--Laurent series reads 
\EQ
\psi(z) =\sum_{n=-\infty}^{\infty} \frac{\psi_n}{z^{n+1/2}}\,,
\label{sviluppopsi}
\EN
where 
\EQ
\psi_n =\oint_{C}\,\frac{dz}{2\pi i} z^{n-1/2}\,\psi(z) \,,
\label{contpsi}
\EN
with a closed contour $C$ around the origin. The modes satisfy the anticommutation relations 
\EQ
\{\psi_n,\psi_m\} =\delta_{n+m,0} \nonumber \,.
\label{heisenbergalgebra} 
\EN
The fermion field $\psi(z)$ can satisfy two different monodromy properties since it is naturally defined on the double covering of the complex plane with a branch cut starting from a point, here assumed to be the origin:
\EQ
\psi(e^{2\pi i}\,z)=\pm \,\psi(z) \, .
\label{cPA}
\EN 
 The first case defines the  so-called Neveu-Schwarz (NS) sector, while the second defines the so-called Ramond (R) sector. In the Neveu-Schwarz sector, the mode expansion of the field is given in terms of half-integer indices, while in the Ramond sector the indices $n$ 
of the (\ref{sviluppopsi}) are instead integers 
\EQ
\begin{array}{lllll}
\psi(e^{2\pi i}\,z) =\psi(z) &,& n \in \mathbb{Z} + \tfrac{1}{2} &,& (NS) \\
\psi(e^{2\pi i}\,z) =-\psi(z) &,&  n \in \mathbb{Z} &.& (R) \\
\end{array}
\EN
The conformally invariant vacuum is the ground state $\ket{0}$ of the NS sector. It is also convenient to introduce the operator $(-1)^F$, where $F$ is the fermionic number, defined in terms of its anti-commutation with the field $\psi$
$$
(-1)^F \,\psi (z) =-\psi(z) \, (-1)^F \,.
$$
This operator satisfies $\left((-1)^F\right)^2=1$ and 
\EQ
\left\{ (-1)^F,\psi_n\right\}=0 \hspace{3mm},\, \forall n 
\EN
Let's focus the attention on the Ramond sector, i.e. when the field satisfies the anti-periodic boundary conditions. In such a case, it is necessary to take into account the presence of the zero modes of the field that satisfies 
\EQ
\{\psi_0,\psi_0 \}=1
\,
,
\,
\{(-1)^F,\psi_0\} =0 
\label{modozerooo}
\EN 
The ground state of the Ramond sector realises a representation of the two-dimensional algebra given by $\psi_0$ and $(-1)^F$. The smallest irreducible representation is spanned by two states created from $\ket{0}$ by a doublet of degenerate operators $\sigma$ and $\mu$, the so-called {\em order and disorder operators}, and the associated states
\EQ
\ket{\sigma}=\sigma(z=0)\ket{0}\quad\ket{\mu}=\mu(z=0)\ket{0}\,,
\EN
which have the same conformal weight as given in \eqref{eq:Ising_fields}. The conjugate states are given by
\begin{eqnarray}
    \langle \sigma |& = &\lim_{z\rightarrow\infty }\langle 0| \sigma(z)|z|^{1/4} \\
    \langle \mu | &=& \lim_{z\rightarrow\infty }\langle 0| \mu(z)|z|^{1/4}\,.
\end{eqnarray}
In this space, a $2 \times 2$ matrix representation of  $\psi_0$ and $(-1)^F$ is given by  
\EQ
\psi_0 =\frac{1}{\sqrt{2}}\left(\begin{array}{cc}
0 & 1 \\
1 & 0 
\end{array}
\right)
\,
,
\,
(-1)^F =\left(\begin{array}{cc}
1 & 0 \\
0 & -1
\end{array}
\right) \,.
\label{bidipsi0}
\EN
In this representation the fields $\sigma$ and $\mu$ are eigenvectors of $(-1)^F$ with eigenvalue $+1$ and $-1$ respectively. The OPE of the fermionic fields with the order/disorder fields is given by \cite{DiFMS}
\begin{align}
&\psi(z) \sigma(w,\bar{w}) = \frac{e^{i\pi/4}}{\sqrt{2}}\,(z-w)^{-1/2} \,\mu(w,\bar{w}) + \dots 
\,
;
\nonumber\\
&\psi(z) \mu(w,\bar{w}) \,= \,\frac{e^{-i\pi/4}}{\sqrt{2}}\,(z-w)^{-1/2} \,\sigma(w,\bar{w}) + \dots 
\,
;
\nonumber\\
&\bar{\psi}(\bar{z}) \sigma(w,\bar{w}) = \frac{e^{-i\pi/4}}{\sqrt{2}}\,(\bar{z}-\bar{w})^{-1/2} \,\mu(w,\bar{w}) + \dots 
\,
;
\nonumber\\
&\bar{\psi}(\bar{z}) \mu(w,\bar{w}) \,= \,\frac{e^{i\pi/4}}{\sqrt{2}}\,(\bar{z}-\bar{w})^{-1/2} \,\sigma(w,\bar{w}) + \dots\, .
\label{cutpsi}
\end{align}
Note that there is a square root branch cut that changes sign if the fermion field is transported around the location of $\sigma$ or $\mu$; the latter can be interpreted as defect-creating operators changing the fermion boundary condition. Similar branch cuts appear in the OPE involving the order and disorder field \cite{DiFMS}
\begin{align}
    \sigma(z,\bar{z})\mu(w,\bar{w})=
    \frac{e^{i\pi/4}(z-w)^{(1/2)}\psi(w)+
    e^{-i\pi/4}(\bar{z}-\bar{w})^{(1/2)}\bar{\psi}(\bar{w})}
    {\sqrt{2}|z-w|^{1/4}}+\dots\,.
\end{align}

The full OPE algebra of the mutually local scalar fields of order and energy operators (omitting the structure constants and the dependence on the coordinates) is then
\EQ
\begin{array}{lll}
\sigma \, \sigma &= &{\bf 1} + \epsilon \\
\epsilon \,\epsilon &= &{\bf 1} \\
\epsilon \,\sigma &= &\sigma 
\end{array}
\label{algebrascalare1}
\EN
An equivalently set of mutually local scalar fields is given by the disorder and the energy operators, with the OPE algebra
\EQ
\begin{array}{lll}
\mu \,\mu &= & {\bf 1} + \epsilon \\
\epsilon \,\mu &= &\mu \\
\epsilon \,\epsilon &= &{\bf 1}
\end{array}
\label{algebrascalare2}
\EN 
These algebras highlight that the Ising model has two independent $\mathbb{Z}_2$ spin symmetries:
\begin{itemize}
    \item one of them flips the sign of the order operator 
    \begin{equation}
        \sigma \rightarrow -\sigma\,,\,\mu \rightarrow \mu\,,
    \end{equation}
    \item while the other flips the sign of the disorder field
    \begin{equation}
        \sigma \rightarrow \sigma\,,\,\mu \rightarrow -\mu\,,
    \end{equation}
\end{itemize}
$\epsilon $ is even under the two spin $\mathbb{Z}_2$ symmetries stated above. 

Moreover, at its critical point the Ising model is also invariant under the Kramers--Wannier duality transformation, under which $\epsilon \leftrightarrow - \epsilon$ and $\sigma \leftrightarrow \mu$. The odd parity of $\epsilon$ under the duality transformation naturally explains the absence of $\epsilon$ in the operator product expansion of this field with itself. 

As a final remark, we observe that the OPE algebra (\ref{algebrascalare1}) of the scalar fields can be also interpreted as the algebra of the composite operators of a $\varphi^4$ Landau-Ginzburg theory -- a theory notoriously associated to the universality class of the Ising model \cite{Zamolodchikov:1986db}. Choosing the field identification $\sigma \equiv \varphi$, the operator product expansion yields $:\varphi^2: = \epsilon$ and $:\varphi^3: =\partial_z\,\partial_{\bar z} \varphi$. Hence, the conformal model can be seen as the exact fixed-point solution of the field theory associated to the Lagrangian 
\EQ
{\mathcal L} =\frac{1}{2} (\partial_{\mu} \varphi)^2 + g \varphi^4\,.
\EN

\subsection{Thermal deformation}
Adding a perturbation by the energy density operator to the critical Ising action (\ref{AZIONECIAK})
\EQ
\mathcal{A}_\tau=\mathcal{A}_\text{CI}+\tau \int d^2 x\:\epsilon(x)
\label{eq:ising_thermal}
\EN
corresponds to adding a mass term $m=2\pi\tau$ for the fermion due to the relation $\epsilon=\bar{\psi}\psi$, with $\tau$ parameterising the deviation from the critical point: $\tau > 0$ drives the system in the paramagnetic phase, while $\tau < 0$ in the ferromagnetic phase, with the two phases related by the Kramers-Wannier duality. The spontaneous magnetisation in the ferromagnetic phase, i.e., the expectation value of the order parameter is \cite{FATEEV1998652}
\EQ
\langle\sigma\rangle = 2^{1/12} e^{-1/8}\mathcal{A}^{3/2}\dots m^{1/8}\,, 
\label{eq:vev_ising_thermal}
\EN
where $\mathcal{A}=1.282427\dots$ is Glaisher's constant. Note that here and in all subsequent formulas the fields are normalised according to \eqref{eq:cft_normalisation} which uniquely specifies their expectation values, as well as the mass gap, in terms of the QFT coupling constants.

The Kramers-Wannier duality is also related to the aforementioned $\mathbb{Z}_2$ spin symmetries of the Ising model and their breaking in the following way: 
\begin{enumerate}
    \item[(i)] for the paramagnetic phase the $\mathbb{Z}_2$ spin symmetry $\sigma \rightarrow - \sigma$ on the order operator is exact while the $\mathbb{Z}_2$ spin symmetry $\mu \rightarrow - \mu$ on the disorder operator is spontaneously broken, therefore for we have a non-zero vacuum expectation value $\langle \mu \rangle$; 
    \item[(ii] for the ferromagnetic phase the $\mathbb{Z}_2$ symmetry of the order parameter $\sigma$ is spontaneously broken while the $Z_2$ symmetry of the disorder operator is exact, therefore for we have a non-zero vacuum expectation value $\langle \sigma\rangle$. 
\end{enumerate}
At the critical point, both vacuum expectation values vanish, $\langle \sigma\rangle = \langle \mu \rangle = 0$. 

\subsubsection{Form Factors of Relevant Local Operators}\label{sec:IsingFF}

Since the resulting quantum field theory is integrable, its corresponding form factors can be computed exactly \cite{PhysRevD.19.2477,Yurov:1990kv}. Due to the self-duality of the model, the two phases can be discussed on equal footing. We choose to focus our attention on the paramagnetic phase where there is no spontaneous symmetry breaking of the $\mathbb{Z}_2$ spin symmetry. In this case, there is a unique ground state and only one massive particle excitation $A$ in the spectrum, with a two-body $S$-matrix $S = -1$. The particle $A$ can be considered as created by the magnetisation operator $\sigma(x)$, therefore it is odd under the $\mathbb{Z}_2$ symmetry of the Ising model.

Let's consider first the energy operator $\epsilon$. Due to its quadratic relation with the massive free Majorana fermion, the FF of this operator are particularly simple 
\EQ
F_n(\theta_1,\dots,\theta_n)=
\begin{cases}
-i m \sinh\frac{\theta_1-\theta_2}{2}\,,&n=2\\
0\,,&\text{otherwise.}
\end{cases}
\EN
In the high-temperature phase, the order parameter $\sigma(x)$ is odd under the unbroken $\mathbb{Z}_2$ symmetry while the disorder operator $\mu(x)$ is even. Hence, $\sigma(x)$ has matrix elements on states with an odd number of particles, $F_{2n+1}^{\sigma}$, whereas $\mu(x)$ on those with an even number of particles, $F_{2n}^{\mu}$. Their form factors can be obtained by solving the form factor  equations and are given by \cite{Yurov:1990kv}
\EQ
F_n(\theta_1,\ldots,\theta_n) =H_n \,\prod_{i < j}^n \,\tanh\frac{\theta_i - \theta_j}{2} 
\,,
\label{FFmagnetizzazione}
\EN 
where the normalisation coefficients satisfy the recursive equation  
$$
H_{n+2} =i\, H_n \,.
$$
The solutions with $n$ even are therefore fixed by choosing $F_0 = H_0$, namely by the non-zero value of the vacuum expectation of the disorder operator 
\EQ
F_{0} =\langle 0 | \mu(0) | 0 \rangle =\langle \mu \rangle
\,,
\EN 
while those with $n$ odd are determined by the real constant $F_1$ 
relative to the one-particle matrix element of $\sigma(x)$ 
\EQ
F_1 =\langle 0 | \sigma(0) | A(0) \rangle \,.
\EN  
which are related by
\EQ
F_1^2 = i\, F_0^2 \,.
\EN
We note that the above form factors automatically satisfy the cluster property \eqref{aympfact}. The clustering of the form factors of the order/disorder operators is simply determined by the parity of the number of particles from the $\mathbb{Z}_2$ symmetry of these operators. In the case of the Ising model, considering the disorder operator $\mu$, and splitting the set of particles into two sets of odd numbers of particles we have\footnote{In the following we have decided to normalise the FF in terms of the VEV of the disorder operator and this is the reason of $\langle \mu \rangle$ in the formulas below.} 
\begin{eqnarray}
&&\lim_{\alpha\rightarrow\infty} F_{2r+2l+2}^{\mu}(\theta_1 +\alpha,\ldots,\theta_{2r+1}+\alpha,\theta_{2r+2},\ldots,\theta_{2r+2l+2}) \nonumber\\
&&=\frac{1}{\langle \mu\rangle}
F^{\sigma}_{2r+1}(\theta_1,\ldots,\theta_{2r+1}) \, F_{2l+1}^{\sigma}(\theta_{2r+2},\ldots,\theta_{2r+2l+2})\,,
\label{aympfact1}
\end{eqnarray}
while for splitting into two sets of even number of particles we have
\begin{eqnarray}
&&\lim_{\alpha\rightarrow\infty} F_{2r+2l}^{\mu}(\theta_1 +\alpha,\ldots,\theta_{2r}+\alpha,\theta_{2r+1},\ldots,\theta_{2r+2l}) \nonumber\\
&&=\frac{1}{\langle \mu\rangle}
F^{\mu}_{2r}(\theta_1,\ldots,\theta_{2r}) \, F_{2l}^{\mu}(\theta_{2r+1},\ldots,\theta_{2r+2l})\,.
\label{aympfact2}
\end{eqnarray}

Adopting the conformal normalisation of both operators
\EQ
\langle \sigma(x) \sigma(0) \rangle \, = \, 
\langle \mu(x) \mu(0) \rangle \simeq \frac{1}{|x|^{1/4}} 
\,
,
\,
|x| \rightarrow 0 
\EN 
the vacuum expectation value $F_0= \langle\sigma\rangle$ is known exactly and is given by \eqref{eq:vev_ising_thermal}.

We finally note that the form factors in the low-temperature phase can be obtained by a simple application of Kramers--Wannier duality which results in swapping the identification of the order and disorder fields.

\subsubsection{Dynamical Structure Factors of the thermal deformation of the Ising Model}\label{DSFIM}

Here we compute the dynamical structure factors corresponding to order/disorder correlation functions
\begin{eqnarray}
    {\cal S}^{\sigma\sigma}(\omega,q)&=&\int dx dt e^{i\omega t - i q x}
    \langle\sigma(x,t)\sigma(0,0)\rangle
    \nonumber\\
    {\cal S}^{\mu\mu}(\omega,q)&=&\int dx dt e^{i\omega t - i q x}
    \langle\mu(x,t)\mu(0,0)\rangle\,.
\label{eq:ising_dsf}
\end{eqnarray}
We consider the frequency dependence at zero-momentum transfer ($q=0$), which determines the full structure factor due to Lorentz invariance, and compute it using a spectral expansion in terms of form factors as described in Subsection \ref{Structurefactor}. 

Let us concentrate on the high-temperature phase (results for the other phase can be obtained by exploiting Kramers-Wannier duality). Here $\mu$ has only even-particle matrix elements and $\sigma$ has only odd-particle ones. Therefore truncating the expansion states with up to three particles gives 
\begin{eqnarray}
	{\cal S}^{\mu \mu}(\omega,q=0) &=& {\cal S}^{\mu \mu}_2(\omega,q=0)+\dots\\
	{\cal S}^{\sigma \sigma}(\omega,q=0) &=& {\cal S}^{\sigma \sigma}_1(\omega,q=0) + {\cal S}^{\sigma \sigma}_3(\omega,q=0)+\dots
\end{eqnarray}
For the disorder operator substituting the form factors from Subsection \ref{sec:IsingFF} into \eqref{eq:S2} gives the two-particle contribution
\beq
	{\cal S}^{\mu\mu}_2(\omega,q=0) = \frac{F_0^2}{\omega^3}\sqrt{\omega^2-4m^2}\,\Theta(\omega-2m)
\eeq
which has a square root behaviour $\propto\sqrt{\omega-2m}$ at threshold.

\begin{figure}
	\centering
	\begin{subfigure}{0.45\textwidth}
		\includegraphics{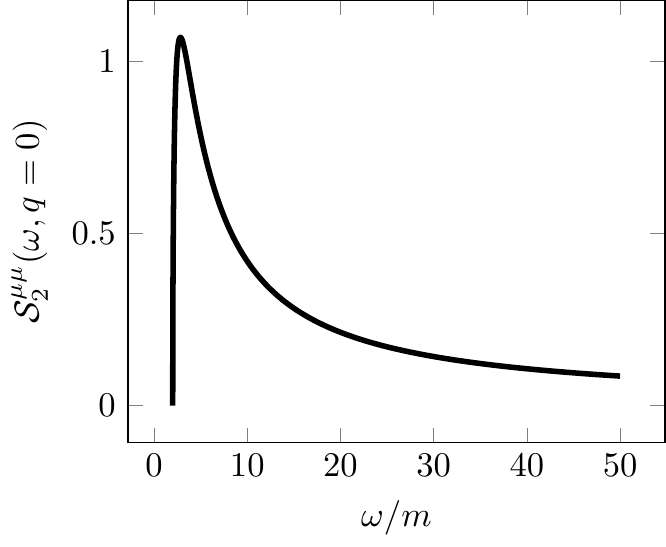}
		\caption{Two-particle contribution}
		\label{fig:isings2}
	\end{subfigure}
	\begin{subfigure}{0.45\textwidth}
		\includegraphics{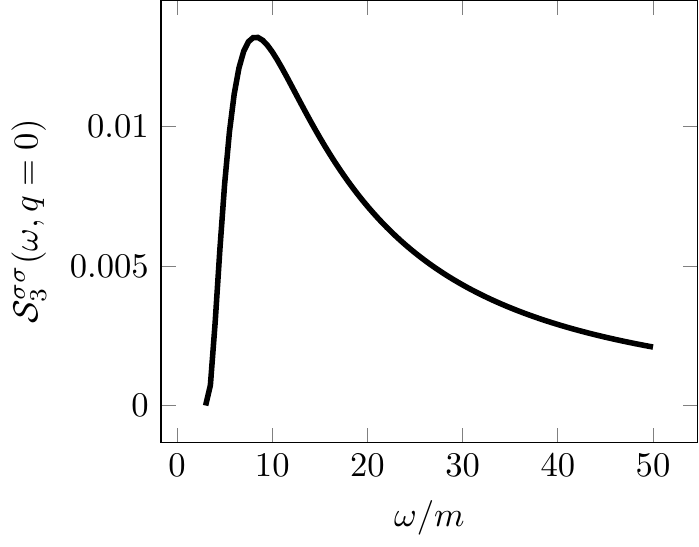}
		\caption{Three-particle contribution}
		\label{fig:isings3}
	\end{subfigure}
	\caption{Two- and three-particle contributions to the dynamical structure factor  of disorder (a) and order operators (b) in the thermal deformation of the 
	Ising model. The values of the dynamical structure factors are shown in units of the mass gap $m$ using the expression \eqref{eq:vev_ising_thermal} for $F_0$.}
\end{figure}

For the order operator \eqref{eq:S1} gives the following one-particle contribution:
\beq
	{\cal S}^{\sigma\sigma}_1(\omega,q=0)=\frac{2\pi F_0^2}{m}\delta(\omega-m)\,.
\eeq
The three-particle contribution \eqref{eq:S3} is a bit more complicated, but after some manipulations, it can be evaluated as 
\begin{equation}
	{\cal S}^{\sigma\sigma}_3(\omega,q=0) = \frac{F_0^2}{6\pi m^2}\int_{\theta_{12}>0}\frac{d\theta_{3}}{\left|\sinh\theta_{12}\right|}\prod_{i<j}^{3}\tanh^{2}\frac{\theta_{ij}}{2}\,,
	\label{eq:S3Isingresult}
\end{equation}
where integration is performed over the branch $\theta_{12}>0$ and the identical contribution from the other branch is taken into account by including a factor  of 2. The behaviour of the three-particle contribution at the threshold is given by 
\beq
	\mathcal{S}^{\sigma\sigma}_3(\omega,k=0)=\frac{F_0^2}{192\sqrt{3}m^5}(\omega-3m)^3+O((\omega-3m)^4)
\eeq
The functions $\mathcal{S}^{\mu\mu}_2$ and $\mathcal{S}^{\sigma\sigma}_3$ are shown in Figures~\ref{fig:isings2} and~\ref{fig:isings3}.

\subsection{Magnetic deformation of the Ising model} 

Possibly the best example of a very detailed and accurate experimental confirmation of a very rich theoretical prediction is given by the magnetic deformation of the Ising model. In this section, we discuss the most relevant features, both on the theoretical and experimental sides, of this deformation. We invite however the reader to 
consult the original literature for further details \cite{1989IJMPA...4.4235Z, DMIMMF, Delfino:1996jr,2010Sci...327..177C, 2020arXiv200513302Z,   2020PhRvB.101v0411Z,2021PhRvB.103w5117W}.

On the lattice, the Minkowski version of this field theory can be realised as the one-dimensional quantum Ising spin chain \eqref{eq:TFIM} at its quantum critical point (QCP) $h_T=1$, perturbed by a longitudinal field $h_L$. The scaling limit of the model is described by the Ising conformal field theory perturbed by its $\mathbb{Z}_2$ odd spin operator 
$\sigma$, namely
\EQ
\mathcal{A}_{E_{8}}=\mathcal{A}_{c=1/2} + h\int d^2 x\,\sigma(x)\,,
\label{eq:action_E8}
\EN
where the operator $\sigma(x)$ and the field $h$ are the rescaled field theory versions of the lattice magnetisation operator $\sigma_i^z$ and longitudinal magnetic field $h_L,$ respectively.  As shown by Zamolodchikov~\cite{1989IJMPA...4.4235Z}, this leads to a massive integrable quantum field theory, the so-called $E_{8}$ model. The exact mass gap can be expressed as \cite{FATEEV199445}
\EQ
m_1=\kappa |h|^{8/15}\,,\quad \kappa=\frac{4\sin(\pi/5)\Gamma(1/5)}{\Gamma(2/3)\Gamma(8/15)}
\left(
\frac{4\pi^2\Gamma(3/4)\Gamma(13/16)^2}{\Gamma(1/4)\Gamma(3/16)^2}
\right)^{4/15}=4.40490857\dots\,,
\label{eq:gap_E8}
\EN
while the exact expectation vacuum values of the relevant fields are given by \cite{russian2}
\EQ
\langle\sigma\rangle = -1.27758\dots |h|^{1/15}\,,\quad 
\langle\epsilon\rangle = 2.00314\dots |h|^{8/15}\,.
\label{eq:vev_E8}
\EN
Let's briefly discuss the $S$-matrix of this deformation of the Ising model. 

\subsubsection{The $E_8$ $S$-matrix}
The hallmark of this model
is the presence of eight stable particle excitations
with the mass $m_1$ of the lightest particle given in \eqref{eq:gap_E8}, while all the other masses can also be expressed in terms of $m_1$ exactly, as shown in Eq.\,(\ref{masssse}) below. 
It can be argued that, for the magnetic deformation of the Ising model, the spins $s$ of the conserved charges (see Eq.\,(\ref{conservationlawspin})) take the values 
\EQ
s = 1, 7, 11, 13, 17, 19, 23, 29 \,\,\,(\rm{mod}  \,30)
\EN
The absence  of the spin $s=3$ in this sequence permits to have the so-called $\Phi^3$ property, namely the possibility that the fundamental particle is a bound state of itself. 
This means that in the $S$-matrix $S_{11}(\theta)$ of the fundamental particle $A_1$ there could a the pole at $\theta = 2 \pi i/3 $. Moreover, the absence of the spin $s=5$ suggests the hypothesis that there is another particle $A_2$, a bound state of $A_1 \times A_1$, where $A_1$ can also be considered as a bound state of $A_2 \times A_2$! This bootstrap chain fixes uniquely the mass ratio of the two particles 
\EQ
\frac{m_2}{m_1} = 2 \cos\frac{\pi}{5} =\frac{\sqrt{5} +1}{2} \,,
\EN 
which is then equal to the golden ratio. This means that in the amplitude $S_{11}(\theta)$ there is also another pole at $\theta = 2 \pi i/5$. However, in order to satisfy the 
bootstrap equation 
\EQ
S_{11}(\theta) = S_{11}\left(\theta - i \frac{\pi}{3}\right) \, S_{11}\left(\theta + i \frac{\pi}{3}\right),
\EN 
coming from the channel $A_1 \times A_1 \rightarrow A_1 \rightarrow A_1 \times A_1$, it is necessary to add at least another pole, placed at $\theta = i \pi/15$, which implies there is at least a third particle $A_3$ in the spectrum. Hence the 
amplitude $S_{11}(\theta)$ of the fundamental particle can be taken as \cite{1989IJMPA...4.4235Z}
\EQ
S_{11}(\theta) =\st{\bf 1}{\left(\frac{2}{3}\right)} \, \st{\bf 2}{\left(\frac{2}{5}\right)} \, \st{\bf 3}{\left(\frac{1}{15}\right)}
\,,
\label{eq:s11e8}
\EN
where we have used the notation set in Eq.~(\ref{convention}). The remaining $2$-body $S$-amplitudes ($36$ in total) can be computed by a recursive application of the bootstrap equations (\ref{bootstrapboundstate}) following the poles corresponding to the bound states. This procedure requires certain care because an inevitable feature of the bootstrap equations is the presence of higher-order poles. While the single poles can be associated with one-particle intermediate states, the higher order poles are explained by intermediate multi-scattering processes via the Coleman-Thun mechanism \cite{Coleman:1978kk}. In order to close the bootstrap all of these singularities must be accounted for \cite{MC,Braden:1989bu}, resulting in a final theory which has $8$ particles, whose mass spectrum coincides with that of an affine Toda field theory based on the exceptional algebra $E_8$. The full set of the $S$-matrix amplitudes for this model is reported in Tables \ref{tab:e8smat1}-\ref{tab:e8smat2} which can be found in Appendix \ref{Appendix:FF2ptE8}. 

\begin{figure}[t]
\begin{center}
\begin{picture}(290,40)
\thicklines
%\put(0,30){\line(1,0){290}}
%\put(0,30){\line(0,1){100}}
%\put(290,30){\line(0,1){100}}
%\put(0,130){\line(1,0){290}}
\put(90,20){\line(1,0){120}}
\put(130,20){\line(0,1){20}}
\put(90,20){\circle*{3}} 
\put(110,20){\circle*{3}} 
\put(130,20){\circle*{3}} 
\put(150,20){\circle*{3}} 
\put(170,20){\circle*{3}} 
\put(190,20){\circle*{3}} 
\put(210,20){\circle*{3}} 
\put(130,40){\circle*{3}} 
\put(90,10){\makebox(0,0){$m_2$}}
\put(110,10){\makebox(0,0){$m_6$}}
\put(130,10){\makebox(0,0){$m_8$}}
\put(150,10){\makebox(0,0){$m_7$}}
\put(170,10){\makebox(0,0){$m_5$}}
\put(190,10){\makebox(0,0){$m_3$}}
\put(210,10){\makebox(0,0){$m_1$}}
\put(120,40){\makebox(0,0){$m_4$}}
\end{picture}
\end{center}
\caption{Dynkin diagram of $E_8$, showing the association of the masses to the vertices. }\label{PERRON-FROBENIUSE8}
\end{figure}
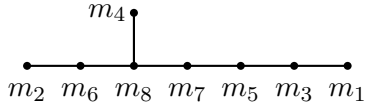

The exact masses of the eight particles are given by
\bea
m_1 &=& \kappa |h|^{8/15}\nonumber \\
m_2 &=& 2 m_1 \cos\frac{\pi}{5} = (1.6180339887..) \,m_1\nonumber\\
m_3 &=& 2 m_1 \cos\frac{\pi}{30} = (1.9890437907..) \,m_1\nonumber\\
m_4 &=& 2 m_2 \cos\frac{7\pi}{30} = (2.4048671724..) \,m_1\label{masssse}\\
m_5 &=& 2 m_2 \cos\frac{2\pi}{15} = (2.9562952015..) \,m_1\nonumber\\
m_6 &=& 2 m_2 \cos\frac{\pi}{30} = (3.2183404585..) \,m_1\nonumber\\
m_7 &=& 4 m_2 \cos\frac{\pi}{5}\cos\frac{7\pi}{30} = (3.8911568233..) \,m_1\
\nonumber\\
m_8 &=& 4 m_2 \cos\frac{\pi}{5}\cos\frac{2\pi}{15} = (4.7833861168..) \,m_1
\nonumber
\eea
It is worth noticing that these masses are in one-to-one correspondence with the entries of the Perron-Frobenius vector of the incidence matrix of the corresponding Dynkin diagram shown in Figure \ref{PERRON-FROBENIUSE8}. Notice that in this bootstrap system only the first three particles have a mass less than the lowest threshold  $2 m_1$. The stability of the particles with a mass higher than the threshold $2 m_1$ is entirely due to the integrability of the theory: indeed, moving away from the critical temperature $T_c$, i.e. coupling the model also to the energy density field $\varepsilon(x)$, all the particles with mass above the threshold $2m_1$ become unstable and decay \cite{Delfino:2005bh,2006NuPhB.748..485P}.  

\subsubsection{Form Factors in the $E_8$ model}\label{subsec:e8ff}
Following the procedure outlined in Subsection \ref{subsec:solvingff}, the two-particle FF can be written as
\be
F_{ab}^{\Phi}(\theta)=\dfrac{Q_{ab}^{\Phi}(\theta)}{D_{ab}(\theta)}F_{ab}^{min}(\theta)\,,
\label{s3}
\ee
where the minimal form factor is
 \be
F_{ab}^{min}(\theta)=\left(-i\sinh\left(\dfrac{\theta}{2}\right)\right)^{\delta_{ab}}\prod_{\alpha}(G_{\alpha}(\theta))^{p_{\alpha}},
\label{s4}
\ee
with
\be
G_{\alpha}(\theta)=\exp\left\{ 2\int_{0}^{\infty} \dfrac{dt}{t} \dfrac{\cosh(\alpha-t/2)}{\cosh(t/2)\sinh(t)}{\sin}^2\dfrac{(i\pi-\theta) t}{2\pi}\right\},
\label{s5}
\ee
while the pole factor $D_{ab}(\theta)$ can be written as
\be
D_{ab}(\theta)=\prod_{\alpha}(P_{\alpha}(\theta))^{i_{\alpha}}(P_{1-\alpha}(\theta))^{j_{\alpha}},
\label{s6}
\ee
with
\be
\ba
&i_{\alpha}=n+1\,, \, j_{\alpha}=n, \, \text{if }  p_{\alpha}=2n+1 \\
&i_{\alpha}=n\,,\, j_{\alpha}=n, \, \,\text{if } p_{\alpha}=2n
\label{s7}
\ea
\ee
and
\be
P_{\alpha}(\theta)=\dfrac{\cos\frac{\pi \alpha}{30}-\cos\theta}{2 \,{\cos}^{2}{\dfrac{\pi \alpha}{60 }}}.
\label{s8}
\ee
The parameters $\alpha$ and $p_{\alpha}$ are those listed in Tables \ref{tab:e8smat1} and \ref{tab:e8smat2}.

The single-particle FF can be obtained using the bound state singularities as
\be
F_{s}^{\Phi}=\frac{\text{Res}(F^{\Phi}_{ab}(\theta)\vert_{\theta=i u_{ab}^{c}})}{i\Gamma_{ab}^{c}}
\label{s9}
\ee
with $\Gamma_{ab}^{c}=\sqrt{i\text{Res}(S_{ab}(\theta)\vert_{\theta=i u_{ab}^{c}})}$.

To generate all form factors it is sufficient to construct multi-particle FF containing only the lightest particle $A_1$, which can be parameterised as
\be
\ba
F_{n}^{\Phi}(\theta_1,\theta_2,\dots \theta_n)\equiv F_{\underbrace{\scriptstyle 1\dots1}_n}^{\Phi}(\theta_1,\theta_2,\dots \theta_n)
=H_n\frac{\Lambda_n(x_1,\dots ,x_n)}{(\omega_n(x_1,\dots ,x_n))^n} \prod_{i<j}^{n}\frac{F_{11}^{\text{min}}(\theta_i-\theta_j)}{D_{11}(\theta_i-\theta_j)(x_i+x_j)}\,,
\label{FF1}
\ea
\ee
since all others can be obtained by application of the bound state singularity equation \eqref{recurb}. Here $x\equiv\exp(\vt)$, $\omega_n$ denotes the elementary symmetric polynomials generated by
\be
\prod_{k=1}^{n}(x+x_k)=\sum_{j=0}^{n}x^{n-j}\omega_j(x_1,\dots ,x_n)\,,
\label{FF2}
\ee
and $H_n$ is a coefficient chosen to simplify the recursion relations for the polynomials $\Lambda_n$, which are  $\Lambda_n(x_1,\dots ,x_n)$ that is an $n$-variable symmetric polynomial that can be expressed in terms of the elementary symmetric polynomials $\omega$ and carry all the information on the particular operator $\Phi$. From \eqref{s6}, $D_{11}$ can be expressed as
\be
D_{11}(\vt)=P_{2/3}(\vt)P_{2/5}(\vt)P_{1/15}(\vt)\,,
\label{FF3}
\ee
while from \eqref{s4} the minimal form factor can be written as
\be
F_{11}^\text{min}(\vt)=-i\sinh(\vt/2)G_{2/3}(\vt)G_{2/5}(\vt)G_{1/15}(\vt)\,.
\label{FF4}
\ee
Following \eqref{recurb}, the pole of the fundamental amplitude $S_{11}$ \eqref{eq:s11e8} corresponding to $A_1$ results in the recurrence relation
\be
\frac{\Lambda_{n+2}(x e^{i\pi/3},x e^{-i\pi/3},x_1,\dots ,x_n)}{x^4\prod_{i=1}^{n}(x-e^{-11 i\pi/15}x_j)(x-e^{11 i\pi/15}x_j)(x+x_j)}=(-1)^n\Lambda_{n+1}(x,x_1,\dots ,x_n)\,,
\label{FF5}
\ee
provided the $H_n$ are chosen to satisfy
\be
\ba
\frac{H_{n+2}}{H_{n+1}}=\frac{\Gamma_{11}^1 \sin \left(\frac{2 \pi }{15}\right) \sin \left(\frac{11 \pi }{30}\right) \sin \left(\frac{8 \pi }{15}\right) \sin \left(\frac{3 \pi }{10}\right)}{2\cos^{2}(\pi/3)\cos^{2}(\pi/5)\cos^{2}(\pi/30)G_{11}(2\pi i/3)}\times\\
\times\left[\frac{\sin^{2}(11\pi/30)\gamma}{4\cos^{2}(\pi/3)\cos^{2}(\pi/5)\cos^{2}(\pi/30)}\right]^n\,.
\label{FF6}
\ea
\ee
The kinematical residue equation \eqref{recursive} yields the recurrence relation
\be
(-1)^n \Lambda_{n+2}(-x,x,x_1,\dots ,x_n)=\mathcal{A}_n U(x,x_1,\dots ,x_n)\Lambda_n(x_1,\dots ,x_n)
\label{FF7}
\ee
with
\be
\ba
U(x,x_1,\dots ,x_n)=\frac{1}{2}x^5&\sum_{k_1,k_2,\dots ,k_6=0}^{n}(-1)^{k_1+k_3+k_5} x^{6n-(k_1+\dots +k_6)}\\
&\times\sin(\frac{\pi}{15}(10(k_1-k_2)+6(k_3-k_4)+(k_5-k_6)))\omega_{k_1}\dots \omega_{k_6}\,,
\label{FF8}
\ea
\ee
and
\be
\mathcal{A}_n=\frac{4\gamma\sin ^2\left(\frac{11 \pi }{30}\right) \left(\cos \left(\frac{\pi }{3}\right) \cos \left(\frac{\pi }{5}\right) \cos \left(\frac{\pi }{30}\right)\right)^2 \left(G_{11}\left(\frac{2 \pi  i}{3}\right)\right)^2}{\left(\Gamma_{11}^1 \sin \left(\frac{2 \pi }{15}\right) \sin \left(\frac{11 \pi }{30}\right) \sin \left(\frac{8 \pi }{15}\right) \sin \left(\frac{3 \pi }{10}\right)\right)^2} \left(\frac{\sin \left(\frac{2 \pi }{3}\right) \sin \left(\frac{2 \pi }{5}\right) \sin \left(\frac{\pi }{15}\right)}{8 \sin ^4\left(\frac{11 \pi }{30}\right) G_{11}(0)\gamma^2}\right)^n\,.
\label{FF9}
\ee
Once the two-particle form factor is specified, the explicit computation of FF with more particles of type $A_1$ proceeds by solving the recurrence relations \eqref{FF5} and \eqref{FF7}. 

To determine the initial condition for the recursion, recall that the universality class of the Ising model has two relevant scaling fields $\sigma(x)$ and $\varepsilon(x)$ with conformal weights $1/16$ and $1/2$ respectively, which imply that their FF satisfy the asymptotic condition \eqref{bound} with $y_\Phi=0$. Since both operators have FF satisfying the recurrence relations Eq.~(\ref{FF5}) and Eq.~(\ref{FF7}), these recursions must admit {\em two} different solutions with the appropriate asymptotic growth. Contrary to the thermal deformation, the magnetic deformation breaks the ${\mathbb Z}_2$ spin symmetry, it cannot be used to distinguish the two operators, and the general solution corresponds to a field $\Phi$ that is the linear combination of the two relevant scaling fields of the model
\be
\Phi = \alpha \sigma + \beta \varepsilon\,.
\label{FF10}
\ee
This means that the most general expression for $Q_{11}(\theta)$ satisfying all the conditions for the Form Factor of the field $\Phi$ is given by 
\cite{Delfino:1996jr}
\be
Q_{11}^{\Phi}(\theta) =c_{11}^1 \cosh\theta + c_{11}^0
,\, c_{11}^1 \neq 0
\ee
All this amounts to say that the solutions of the FF bootstrap for the relevant scalar operators of the IMMF form a two–dimensional linear space, which is expected from the fact that such operators can only correspond to linear combinations of $\sigma(x)$ and $\varepsilon(x)$. Solutions corresponding to different operators up to normalisation can be labelled by the ratio
\be
z=\frac{c_{11}^0}{c_{11}^1}
\ee
with particular value of $z$ corresponding to the magnetic field $\sigma(x)$ and another one corresponding to the thermal field $\varepsilon(x)$. These two different values were first obtained in \cite{Delfino:1996jr} employing the clustering property (\ref{aympfact}) of the Form Factors.

In fact, for $\sigma(x)$ the above problem is even simpler since this field is proportional to the trace of the stress-energy tensor:
\EQ
\Theta(x)=2\pi h\left( 2-2\Delta_\sigma\right)\sigma(x)\,,
\EN
and energy-momentum conservation implies that the corresponding form factor must contain a factor $P^+ P^-$ with $P^\pm = \displaystyle\sum_{i=1}^{n}p_i^\pm$ and $p^\pm = p^0\pm p^1 = m e^{\pm \vt}$ \cite{1991NuPhB.348..619Z}. Consequently, for $\sigma(x)$ the Ansatz can be reduced by solving only for symmetric polynomials containing the factor $\omega_1(x_1,...,x_n)\omega_{n-1}(x_1,...,x_n)$, due to $P^+P^-=\omega_1 \omega_{n-1}/\omega_n$. This observation is enough to solve for the polynomials of the $\sigma$ operator without employing the clustering property, resulting in the following value of $z$ corresponding to the magnetic field \cite{Delfino:1996jr}
\be
z_{\sigma} = \frac{2 m_1^2 + m_3 m_7}{2 m_1^2} =4.869840...
\label{zsigma}
\ee
However, to determine the value of $z$ corresponding to $\varepsilon$ it is necessary to invoke the cluster property. Here we simply quote the numerical value 
determined in \cite{Delfino:1996jr}
\be
z_{\varepsilon} = 1.255585...
\label{zepsilon}
\ee
The remaining Form Factors of the field $\sigma(x)$ can be obtained by solving the recursive equations starting from $Q_{11}^{\phi}(\theta)$ with the value $z_\sigma$ given in Eq. (\ref{zsigma}), while for the remaining Form Factors of the field $\varepsilon(x)$ the starting point must be set using $z_{\varepsilon}$ given in Eq. (\ref{zepsilon}). For more details, the interested reader is invited to consult the work \cite{2021PhRvB.103w5117W}.

\subsubsection{DSF of the magnetic deformation of the Ising Model and experiments}
\begin{figure}[t]
	\centering
	\includegraphics[width=11cm]{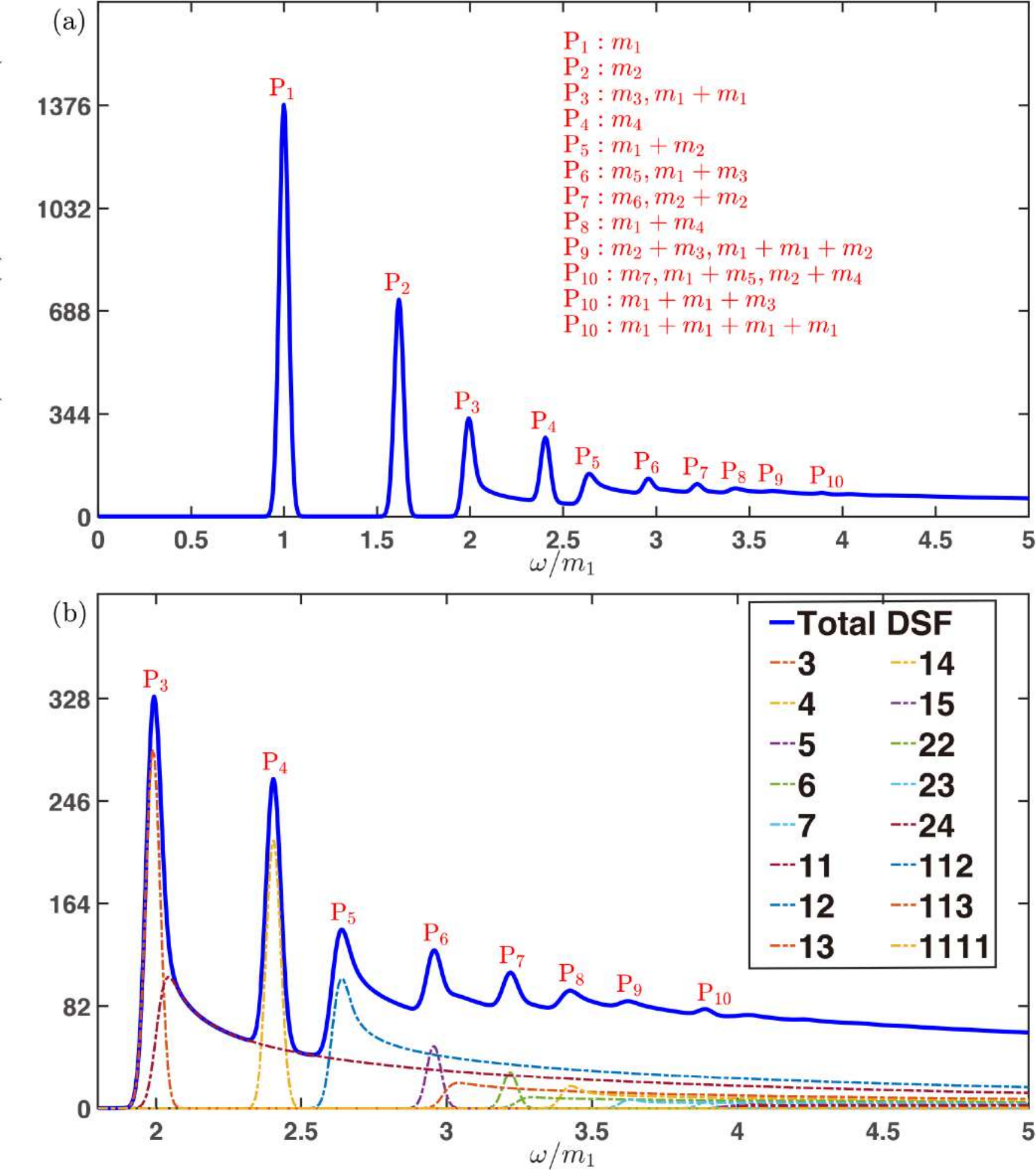}
	\centering
\caption{The DSF of the magnetisation field ${\cal S}^{\sigma\sigma}(\omega,q=0)$ in the $E_8$ model as a function of the frequency $\omega$ \cite{2021PhRvB.103w5117W}. The result from the spectral expansion was convolved with a Gauss profile of width $0.05 m_1$ which regularises the one-particle $\delta$-peaks.\\ 
(a) The full result displays several peaks $P_i$ associated with single or multi-particle excitations.\\
(b) The details of the DSF  for $\omega\gtrsim 2 m_1$, with dashed curves showing the contribution from individual channels, which are labeled according to their particle contents, e.g. ``11" stands for $m_1 + m_1$ channel. \\
The units for the value of the DSF correspond to $m_1=1$, with the expectation value of the order parameter $\sigma$ given in \eqref{eq:vev_E8}.}
	\label{fig:Ssigmasigma}
\end{figure}

Substituting the form factors determined in Subsection \ref{subsec:e8ff} into the spectral representation presented in Subsection \ref{Structurefactor} leads to the determination of the Dynamical Structure Factor in the $E_8$ field theory. This is a straightforward numerical task that we do not discuss in detail, confining ourselves to displaying in Figure \ref{fig:Ssigmasigma} the results obtained in \cite{2021PhRvB.103w5117W}. 

The experimental realisation of the two-dimensional Ising model at its critical point perturbed by the magnetic field has been the subject of several studies carried out recently by different groups. The first study was performed by Coldea et al. in \cite{2010Sci...327..177C}, where they reported to have realised this system experimentally by using strong transverse magnetic fields to tune the quasi–one–dimensional Ising ferromagnet $\rm{Co Nb_2 O_6}$ (cobalt niobate) through its critical point. In this experiment, the Dynamical Structure Factor (DSF) of the spin-spin correlation function was measured in terms of neutron scattering. Just below the critical field, the spin dynamics showed a fine structure with two sharp modes at low energies, in a ratio that approached the golden mean $(1 + \sqrt{5})/2$ predicted for the first two meson particles of the $E_8$ spectrum. Although this was a very significant and influential experiment, the energy resolution of this experiment was not enough to distinguish the higher masses and provide a precise determination of the DSF itself.

\begin{figure}[th!]
	\centering
	\includegraphics[width=0.9\textwidth]{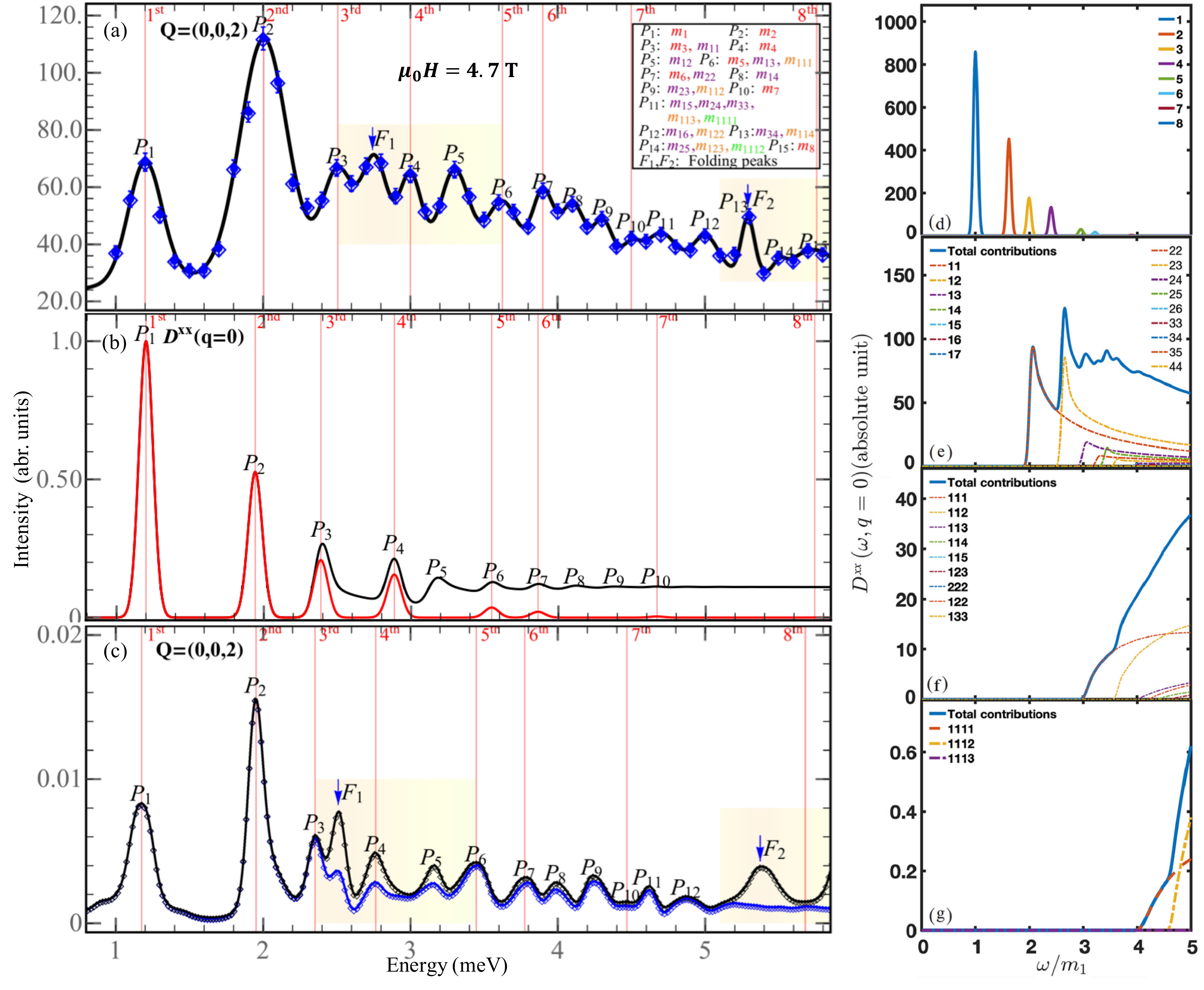}
	\centering
\caption{Figure from ref. \cite{2020arXiv200513302Z} showing experimental results vs. theoretical predictions for the $E_8$ excitation spectrum near the 1D Quantum Critical Point of the quasi-1D antiferromagnet $\rm{Ba Co_2 V_2 O_8}$.  (a) The top panel shows the inelastic neutron scattering intensity along the tridimensional axis $Q=(0,0,2)$. Blue diamonds with error bars correspond to experimental data and black lines are fits with Gaussian functions. The red vertical lines at eight peaks correspond to the eight single $E_8$ particles. Other peaks come from multi-particle excitations and zone-folding effects of the lattice. The peak with mass $m_{i_1 i_2...i_n}$ labels multi-particle channel with particle masses $m_{i_1} m_{i_2} ... m_{i_n}$.(b) The analytical dynamic structure factor $D^{xx}$ (corresponding to ${\cal S}^{\sigma\sigma}$ of Fig. \ref{fig:Ssigmasigma}) calculated from quantum $E_8$ integrable field theory. Red curve stands for single-particle spectra, while black curve is obtained after including multi-particle contributions. In accord with the experiment, $m_1$ is set as 1.2 meV and the analytical data are broadened in a Lorentzian fashion with full-width at half-maximum fixed at $0.08m_1$. (c) Neutron scattering intensity from iTEBD calculations at the zone center. The black and blue curves are results with and without zone-folding effect.
The side panels show DSF spectra for individual scattering channel: (d) single-, (e) two-, (f) three-, and (g) four-particles contributions, where $ijkl$ refer to excitations with combined mass modes of $m_im_jm_km_l$.
}
	\label{fig:Ssigmasigmaexp}
\end{figure}

More refined experiments have recently been carried out, as reported in the papers \cite{2020PhRvB.102j4431A,2020arXiv200513302Z,2020PhRvB.101v0411Z}. The experiments of \cite{2020arXiv200513302Z,2020PhRvB.101v0411Z} used terahertz (THz) spectroscopy to resolve the $E_8$ particles in an antiferromagnetic Ising spin-chain material, namely the quasi 1-D $\rm{Ba Co_2 V_2 O_8}$.  In \cite{2020arXiv200513302Z} the $E_8$ particles and the Dynamical Structure Factor of the spin-spin correlation function were determined by means of nuclear magnetic resonance and inelastic neutron scattering measurements on the same quasi-1D antiferromagnet $\rm{Ba Co_2 V_2 O_8}$. In Fig.~\ref{fig:Ssigmasigmaexp} we report the inelastic neutron scattering intensity along a particular crystal direction (here denoted $Q=(0,0,2)$) together with the theoretical prediction which comes from the result in Fig.\,\ref{fig:Ssigmasigma}. It is worth noting that the experimental neutron dynamic spectrum shows an excellent match with the analytical prediction for the peak positions, and also (with the exception of the first peak) with the spectral weights coming from the $E_8$ theory.  

\subsection{Universal ratios of the Ising model}

The form factors of the Ising model allow the evaluation of the universal amplitudes ratios in the Ising model \cite{DelfinoIsing}; we note that similar results can be obtained for its generalisation to the $q$-state Potts model \cite{DelfinoCardy1}. Let's first provide some notation.

\EQ
 \xi = \left\{\begin{array}{lll}
f_{\pm} |\tau|^{-\nu} \,,&\hspace{3mm}& \tau \rightarrow 0^{\pm} \,,\,\, h=0 \\
f_c |h|^{-\nu_c} \, , & \hspace{3mm} & \tau = 0 \, \,\, h \rightarrow 0
\end{array}
\right.
\EN
$\xi$ is the correlation length, and we also have
\begin{eqnarray}
&&
C = \int d^2 x \,\langle \epsilon(x) \epsilon(0)\rangle = \left\{\begin{array}{lll}
(A_{\pm}/\alpha) |\tau|^{-\alpha} \,,&\hspace{3mm}& \tau \rightarrow 0^{\pm} \,,\,\, h=0 \\
(A_c/\alpha_c) |h|^{-\alpha_c} \, , & \hspace{3mm} & \tau = 0 \, \,\, h \rightarrow 0
\end{array}
\right.
\\
&&
|M| = \langle \sigma \rangle =
\left\{\begin{array}{lll}
B (-\tau)^{\beta} \,,&\hspace{3mm}& \tau \rightarrow 0^{-} \,,\,\, h=0 \\
(|h|/D)^{1/\delta} \, , & \hspace{3mm} & \tau = 0 \, \,\, h \rightarrow 0
\end{array}
\right.\label{VVVVV}
\\
&&
\chi = \int d^2x \,\langle \sigma(x) \sigma(0)\rangle = 
\left\{\begin{array}{lll}
\Gamma_{\pm} |\tau|^{-\gamma} \,,&\hspace{3mm}& \tau \rightarrow 0^{\pm} \,,\,\, h=0 \\
\Gamma_c |h|^{-\gamma_c} \, , & \hspace{3mm} & \tau = 0 \, \,\, h \rightarrow 0
\end{array}
\right.
\end{eqnarray}
where $C$ is the specific heat, $M$ is the magnetisation and $\chi$ is the susceptibility.  For convenience, we report here the list of critical exponents of the model
\begin{eqnarray}
&& \nu = 1 \nonumber\\
&&\nu_c = 8/15 \nonumber\\
&&\alpha = 0  \nonumber\\
&&\alpha_c = 0 \nonumber\\
&&\beta = 1/8  \nonumber\\
&&\delta = 15  \nonumber\\
&&\gamma = 7/4  \nonumber\\
&& \gamma_c = 14/15 
\end{eqnarray}

\subsubsection{Thermal deformation}

First, we consider the deformation in the thermal direction \eqref{eq:ising_thermal}. Using the theoretical results on the Form Factors of this deformation is easy to see that  
we have the exact universal ratio $f_+/f_- = 2$. Moreover evaluating the correlation functions including only states with no more than two particles gives rise to the following universal ratios \cite{DelfinoCardy1}
\begin{align}
   & \frac{A_+}{A_-}=1 && \frac{\Gamma_+}{\Gamma_-}=37.699\dots \nonumber\\
   & R_C=\frac{A_+\Gamma_+}{B^2}=0.3183\dots 
   && R^+_\xi=A_+^{1/2}\xi^0_+=0.3989\dots & \,.
\end{align}
which are remarkably close to the exact results \cite{DelfinoIsing}
\begin{align}
   & \frac{A_+}{A_-}=1 && \frac{\Gamma_+}{\Gamma_-}=37.6936520\dots  \nonumber\\
   & R_C=\frac{A_+\Gamma_+}{B^2}=0.318569391\dots 
   && R^+_\xi=A_+^{1/2}\xi^0_+=\frac{1}{\sqrt{2\pi}} \,=\,0.3989..& \,.
\end{align}

\subsubsection{Magnetic deformation}

Turning to the $E_8$ model defined by the action \eqref{eq:action_E8}, it is possible to determine exactly the coefficient $D$ entering 
Eq. \eqref{VVVVV} since the spontaneous magnetisation is proportional in this case to the trace of the stress-energy tensor which can be computed using the
Thermodynamic Bethe Ansatz (see, for instance, \cite{Mussardo:2020rxh})
\EQ
D = 0.0253610264...
\EN
For the same reason, one can also compute exactly $\Gamma_c$, given by 
\EQ
\Gamma_c = 0.0851721517...
\EN
The exact FF of the $E_8$ model give the following theoretical predictions for universal ratios:
\begin{eqnarray}
&& R_{\chi} = \Gamma_+ D B^{\delta -1} = 6.77828502...
 \nonumber\\
&&R_A = A_c D^{-(1 + \alpha_c)} B^{-2/\beta} = 0.0250658794...
 \nonumber\\
&& Q_2 =(\Gamma_+/\Gamma_c) (f_c/f_+)^{\gamma/\nu} = 3.23513834...
\end{eqnarray} 
These can be compared to results of a power-series expansion \cite{PhysRevB.11.1217} yielding $R_{\chi} \sim 6.78$, and of a transfer matrix computation \cite{Caselle:1999bx} giving  
$R_{\chi} \sim 6.78$, $R_{\chi} = 6.7782(8)$ and $Q_2 = 3.233(4)$. The agreement between the FF theoretical predictions and the numerical determinations of these quantities can be regarded quite satisfactory. 

\section{Tricritical Ising Model}

In this section, we are going to discuss the universality class of the tricritical Ising Model. The Euclidean field theory describes the vicinity of the critical point of the two-dimensional Blume-Capel model \cite{BC1,BC2} which is defined by the partition function
\begin{align}
    Z_\text{BCM}&=\sum_{s_i=\pm 1,t_i=0,1} \exp\left\{-\frac{1}{T}\mathcal{H}_\text{BCM}(\{s_i,t_i\})\right\}\label{eq:2dBCM}\\
    & \mathcal{H}_\text{BCM}(\{s_i,t_i\})=-J\sum_{\langle i,j\rangle}s_i s_j t_i t_j+\Omega\sum_i t_i+K \sum_{\langle i,j\rangle} t_i t_j
    -H\sum_i s_i t_i-H'\sum_{\langle i,j\rangle} \left(s_i t_i t_j+s_j t_j t_i\right)\nonumber
\end{align}
where the indices $i,j$ run over the two-dimensional lattice and $\langle i,j\rangle$ denotes pairs of nearest neighbour sites. The parameter $\Omega$ corresponds to a chemical potential for the vacancy variables $t_i$, $H$ and $H'$ are two external magnetic fields coupling to two relevant order parameter fields, while $K$ is an (irrelevant) nearest neighbour interaction between the vacancies.

\begin{figure}
\centering
		\includegraphics[width=0.5\textwidth]{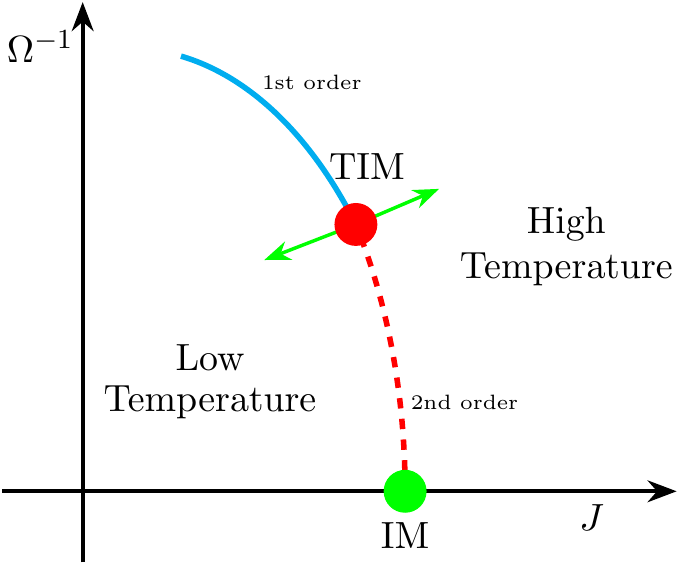}
		\caption{Qualitative phase diagram of the TIM in the plane of the two variables $J$ and $\Omega$. The green arrows show the direction of the thermal deformation by the relevant field $\epsilon$ leading to the $E_7$ model (c.f. Subsection \ref{subsec:TIM_thermal}).}
		\label{fig:phasediagram}
\end{figure}

The phase diagram of this model is much richer than that of the Ising model and is illustrated in Fig. \ref{fig:phasediagram}. Setting $H=H'=0$ and varying the temperature and the vacancy chemical potential consists of a ferromagnetic phase and a paramagnetic phase, which are separated by a curve that consists of two parts. One part corresponds to a first-order, while the other one to a second-order phase transition in the Ising universality class. The point where the two lines meet is a tricritical point which is in a universality class described by a conformal field theory of central charge $7/10$. 

When continued to real-time, the corresponding Minkowski field theory describes the dynamics of the one-dimensional quantum tricritical Ising spin chain governed by the Hamiltonian \cite{vonGehlen:1989yn}
\begin{equation}
    \hat{H}_\text{TIC}=
    -\mathcal{J}\sum_{i} \left\{S_i^z S_{i+1}^z - \alpha (S_i^z)^2 -\beta S_i^x - \gamma  (S_i^x)^2 -h S_i^z \right \}
    \label{eq:TIMC}
\end{equation}
where the spin-$1$ operators are given by
\begin{equation}
    S^z=
    \begin{pmatrix}
    1 & 0 & 0 \\ 0 & 0 & 0 \\ 0 & 0 & -1
    \end{pmatrix}\quad
    S^x=\frac{1}{\sqrt{2}}
    \begin{pmatrix}
    0 & 1 & 0 \\ 1 & 0 & 1 \\ 0 & 1 & 0
    \end{pmatrix}\,.
\end{equation}

\subsection{Universality class and the Kramers--Wannier duality}
\label{subsec:TIM_CFT}

Let us now briefly discuss the universality class and the duality symmetry of the tricritical Ising model (TIM) associated with the second unitary minimal model: the central charge is $c = \tfrac{7}{10}$. We are mainly concerned here with two equivalent quantum field theory formulations of the TIM, namely one based on a Landau-Ginzburg field theory, which explicates the $\mathbb{Z}_2$ spin symmetry, while the other based on supersymmetry, which clarifies the origin of the $\mathbb{Z}_2$ duality symmetry of the model. For a slightly more detailed discussion, we refer to~\cite{2022ScPP...12..162C}.

The lattice Hamiltonian of the Blume-Capel model \eqref{eq:2dBCM} can be put in a one-to-one correspondence with a $\varphi^6$ Landau-Ginzburg Lagrangian based on a scalar field $\varphi$ and the Euclidean action given by: 
\EQ
{\mathcal S} = \int d^2x \left[\frac{1}{2} (\partial_{\mu} \varphi)^2 + 
g_1 \varphi + g_2 \varphi^2 + g_3 \varphi^3 + g_4 \varphi^4 + 
\varphi^6 \right]\, ,
\label{LG}
\EN 
where the tricritical point\footnote{A tricritical point occurs when a second-order phase transition line meets a first-order phase transition line. For the Lagrangian (\ref{LG}), at the tree level the curve which describes the second order phase transition is identified by $g_1 = g_2 = g_3 =0$, but $g_4 >0$; the curve which describes the first order phase transition is given instead by $g_1 = g_3 =0$, with $g_2 >0$ and $g_4 = - 2 \sqrt{g_2}$.} 
 is identified by the condition $g_1=g_2=g_3=g_4=0$ \cite{Zamolodchikov:1986db}. The statistical interpretation of the coupling constants is the following: $g_1$ plays the role of an external magnetic field $h$, $g_2$ measures the displacement of the temperature from its critical value, i.e. $g_2 \sim (T-T_c)$, $g_3$ may be regarded as a sub-leading magnetic field $h'$ and, finally, $g_4$ may be interpreted as a chemical potential for the vacancies. Switching on and tuning the various coupling constants, the model changes its spectrum and its dynamics, as studied earlier in a series of papers 
\cite{Kastor,PLB151:37,Qiu,MSS,MC,FZ,LMC,Zamolodchikov:1989rd,CKM,Zammassless,Lepori:2008et,GM2,AMV,prlFMS,Fioravanti:2000xz,DMSmassless} and recently by the authors in~\cite{2022ScPP...12..162C,2022PhLB..82837008L,2022PhRvD.106j5003L,2022arXiv221101123L}. 

The exact conformal weights of the scaling fields of the model are given by  
\EQ
\Delta_{r,s} = \frac{(5 r - 4 s)^2 -1}{80} \,,
\,
\begin{array}{c}
1 \leq r \leq 3, \\
1 \leq s \leq 4 ,
\end{array}
\label{Kactable}
\EN
The six scalar primary fields of the TIM perfectly match the identification provided by the composite fields of the Landau--Ginzburg theory and by the symmetries of the model. In fact, with respect to the $\mathbb{Z}_2$ spin symmetry of the model $\varphi \rightarrow - \varphi$, the fields are classified as follows\footnote{For simplicity, in the following we use for the fermion fields and the leading order/disorder operators of the TIM the same notations as in the IM, even though it is evident that they are different fields.} (c.f. Table \ref{tab:TIM}):
 \begin{table}
 \bgroup
 \setlength{\tabcolsep}{0.5em}
 \begin{center}
 \begin{tabular}{c c l l c}
     \hline\hline
     \multicolumn{1}{l}{conformal} & $(r,s)$ & field  & physical~role & Landau-Ginzburg \\[-0.5em]
     \multicolumn{1}{l}{weights}   &        & &               & field \\
     \hline
    $(0,0)$ &$(1,1)$ or $(3,4) $& $\mathbb{I}$ & & identity \\[3pt]
    $(\tfrac{3}{80},\tfrac{3}{80})$ & $(2,2)$ or $(2,3)$ & $\sigma$  & magnetisation & \,$\varphi$ \\[3pt]
    $(\tfrac{1}{10},\tfrac{1}{10})$ & $(1,2)$ or $(3,3)$ & $\epsilon$ &  energy & $:\varphi^2:$ \\[3pt]
    $(\tfrac{7}{16},\tfrac{7}{16})$ & $(2,1)$ or $(2,4)$ & $\sigma'$ & submagnetisation & $:\varphi^3:$ \\[3pt]
    $(\tfrac{3}{5},\tfrac{3}{5})$ & $(1,3)$ or $(3,2)$ & $t$ &  chemical potential & $:\varphi^4:$ \\[3pt]
    $(\tfrac{3}{2},\tfrac{3}{2})$ & $(1,4)$ or $(3,1)$ & $\epsilon''$ &  (irrelevant) & $:\varphi^6:$ \\[3pt]
    \hline\hline
  \end{tabular} 
  \end{center}
  \egroup
  \caption{Primary fields of the TIM and their Landau-Ginzburg identifications} 
  \label{tab:TIM}
\end{table} 

\begin{enumerate}
\item Two odd fields: the magnetisation operator
$\sigma  \equiv \varphi$ and the sub-leading magnetic operator $\sigma'  \equiv : \varphi^3:$; 
\item Four even fields: the identity operator ${\bf 1}$, the energy operator $\epsilon  \equiv :\varphi^2:$, and the density operator 
$t \equiv :\varphi^4: $, associated to the vacancies. Finally, there is also the irrelevant field $\epsilon''$. The operator product expansion of these fields gives rise to a sub-algebra of the fusion rules. 
\end{enumerate}

The Kramers--Wannier duality of the tricritical Ising model can be explained most conveniently using the supersymmetric formulation~\cite{PLB151:37,Qiu,MSS}. In two dimensions, super-conformal invariance is associated with two super-currents, $G(z)$ and $\bar G(\bar z)$, where the former is a purely analytic field while the latter is a purely anti-analytic one. They are both fermionic fields, with conformal weights $(\tfrac{3}{2},0)$ and $(0,\tfrac{3}{2})$ respectively. Notice that $G$ corresponds to the $(r,s)=(1,4)$ field in the Kac table.
The OPE of the field $G(z)$ with itself reads  
\EQ
G(z_1) \, G(z_2) =\frac{2 c}{3 (z_1-z_2)^3} + \frac{2}{z_1-z_2} T(z_2) + \cdots,
\label{supegg}
\EN 
where the parameter $c$ is the same central charge that enters the operator expansion of $T(z)$: 
\EQ
T(z_1) T(z_2) =\frac{c}{2 (z_1-z_2)^2} + \frac{2}{(z_1-z_2)^2} T(z_2) + 
\frac{1}{z_1-z_2} \partial T(z_2) + \cdots,
\label{TTTTTT}
\EN
Since $G(z)$ is also a primary field, it satisfies the operator product expansion:
\EQ
T(z_1) G(z_2) =\frac{3}{2 (z_1-z_2)^2} G(z_2) + \frac{1}{z_1-z_2} \partial G(z_2) 
+ \cdots .
\label{primG}
\EN
Let us define the generators $L_n$ and  $G_n$ through the expansions  
\EQ
T(z) =\sum_{n=-\infty}^{\infty}\frac{L_n}{z^{2+n}}
\,
;
\qquad
G(z) =\sum_{m=-\infty}^{\infty} \frac{G_m}{z^{3/2 + m}}.
\label{modiGT}
\EN 
Note that, in the expansion of the field $G(z)$, the indices can assume either integer or half-integer value. In fact, $G(z)$ is a fermionic field defined on the double covering of the plane with a branch cut starting from the origin, the same way as the free fermionic field $\psi$ in the Ising case discussed in subsection~\ref{subsec:IM_CFT} for $\psi$. Making the analytic continuation, $z \rightarrow e^{2\pi i}\,z$, two possible boundary conditions are permitted:
\EQ 
G(e^{2\pi i}\,z) =\pm G(z) \,.
\EN 
In the periodic case ($+$), termed the Neveu-Schwarz (NS) sector, the indices $m$ are half-integers, $m \in \mathbb{Z} + \tfrac{1}{2}$. In the anti-periodic case ($-$), termed the Ramond (R) sector, the indices $m$ are instead integer numbers, $m \in \mathbb{Z}$. 

\begin{table} 
%\begin{indented}  
    \bgroup
    \setlength{\tabcolsep}{0.5em}
    \begin{center}
    \begin{tabular}{c l l}
     \hline\hline
     conformal & field & physical~role \\[-.5em]
     weights &       &               \\
     \hline
    $(\tfrac{3}{80},\tfrac{3}{80})$ & $\mu$ & disorder field \\[3pt]
    $(\tfrac{7}{16},\tfrac{7}{16})$ & $\mu'$ & subleading disorder field  \\[3pt]
    $(\tfrac{3}{5},\tfrac{1}{10})$ & $\psi$ &  fermion  \\[3pt]
    $(\tfrac{1}{10},\tfrac{3}{5})$ & $\bar{\psi}$ & anti-fermion  \\[3pt]
    $(\tfrac{3}{2},0)$ & $G$ & holomorphic supersymmetry current \\[3pt]
    $(0,\tfrac{3}{2})$ & $\bar{G}$ & anti-holomorphic supersymmetry current \\[3pt]
    \hline\hline
  \end{tabular}
  \end{center}
  \egroup
  %\end{indented}
  \caption{Additional operators of the TIM originating from the supersymmetry of the model.}
  \label{tab:twistTIM}
\end{table} 

It is clear from the conformal operator product algebra that the NS/R sector corresponds to even/odd fields respectively. As in the Ising case, the order operators $\sigma$ and $\sigma'$ are accompanied by the Kramers-Wannier dual disorder operators $\mu$ and $\mu'$, forming irreducible representations of the algebra of the zero mode $G_0$ and the fermionic number operator $(-1)^F$:
\EQ
\left\{ (-1)^F, G_n\right\}=0 \hspace{3mm},\, \forall n .
\EN
In the presence of the order/disorder fields, the OPE of the fermionic field $G(z)$ is    
\EQ
G(z) \sigma(w) \,= \,(z-w)^{-3/2} \,\mu(w) + \cdots 
\,
;
\,
G(z) \mu(w) \,= \,\left(\Delta - \frac{c}{24}\right) (z-w)^{-3/2} \,\sigma(w) + \cdots 
\label{cutG}
\EN
(with $\Delta = 3/80$ and $c = 7/10$)
and the same holds replacing $\sigma$ and $\mu$ with $\sigma'$ and $\mu'$ (in the latter case, using in the formula above $\Delta = 7/16$). Similarly to the case of the Ising model, the fields $\sigma$ and $\mu$ are defect operators changing the boundary conditions for fermionic fields, c.f. Eq. \eqref{cutpsi}. 

We introduce a formal operator $D$ that implements the Kramers-Wannier duality, which allows the duality properties of the various fields of the TIM to be summarised as follows: 
\begin{itemize} 
\item 
the magnetisation order parameters change into the disorder operators:
\EQ
\mu =  D^{-1} \sigma  D \, , 
\,  
\mu' =  D^{-1} \sigma' D 
\,.
\label{sigmadual}
\EN 
\item 
the even fields transform instead in themselves:
\EQ
D^{-1} \epsilon D =-\epsilon \hspace{3mm},
\hspace{3mm} 
D^{-1} t D = t  \hspace{3mm}, \hspace{3mm} 
D^{-1} \epsilon'' D = - \epsilon'' , 
\EN
i.e., $\epsilon$ and $\epsilon''$ are odd fields, while $t$ is an even field under the duality transformation .  
\end{itemize}
The transformation properties of the various fields of the TIM under the spin-reversal and duality are shown in Table \ref{tab:discrsym}.

\begin{table}
    \bgroup
    \setlength{\tabcolsep}{0.5em}
  \begin{center}
  \begin{tabular}{| c| c | c|}
  \hline\hline
    field & spin-reversal & Kramers--Wannier  \\[3pt] 
    \hline
    $\epsilon$ & $\epsilon$ & $-\epsilon$ \\
    $t$ & $t$ & $t$ \\
    $\epsilon''$ & $\epsilon''$ & $-\epsilon''$ \\
    $\sigma$ & $-\sigma$ & $\mu$ \\
    $\sigma'$ & $-\sigma'$ & $\mu'$ \\
    \hline\hline        
  \end{tabular}
  \end{center}
    \egroup
\caption{Discrete symmetries of TIM.} \label{tab:discrsym} 
\end{table}

\subsection{Thermal deformation of the tricritical Ising model}
\label{subsec:TIM_thermal}

In this section we consider the perturbation of the tricritical Ising model its energy operator $\epsilon(x)$, with the corresponding action given by 
\begin{eqnarray}
  \mathcal{A}=\mathcal{A}_{TIM} + g \, \int d^2x\, \epsilon(x) \,, 
  \label{ttttherm}
\end{eqnarray} 
where $\mathcal{A}_{TIM}$ is the action of the fixed point. 
The perturbation with positive/negative coupling constant drives the system into its high/low-temperature phases. While in the latter phase, the spin-reversal $\mathbb{Z}_2$ symmetry of the system is spontaneously broken, in the former phase the $\mathbb{Z}_2$ symmetry is unbroken and therefore the corresponding quantum number can be used to label the states. In the low–temperature phase, the massive excitations are given by topologically charged kink states and their neutral bound states, while in the high–temperature phase all excitations are ordinary particle (i.e. topologically trivial) excitations. The two phases are related by a duality transformation and therefore we can restrict our attention to only one of them, which we choose to be the high–temperature
phase. The off-critical theory was shown to be integrable and related to the Toda field theory based on the exceptional algebra ${\rm E}_7$ \cite{MC,FZ}: the conserved charges have spins
\begin{eqnarray}
  s=1,5,7,9,11,13,17 \pmod{18} \,,
\end{eqnarray}  
(these numbers are the Coxeter exponents of the exceptional algebra ${E}_7$) and the set of all $S$-matrix amplitudes are reported in Appendix \ref{AppendixS-matrix}. 

The exact mass spectrum of excitations can be extracted from the pole structure of the $S$ matrices: with respect to the $\mathbb{Z}_2$ spin symmetry of the model, there are three $\mathbb{Z}_2$ odd particle states $A_1, A_3, A_6$ (with masses $m_1$, $m_3$ and $m_6$) and four $\mathbb{Z}_2$ even $A_2, A_4, A_5, A_7$ (with masses $m_2$, $m_4$, $m_5$ and $m_7$) in the high--temperature phase. In the low--temperature phase, the above three $\mathbb{Z}_2$ odd particles correspond to kink excitations which interpolate between the two degenerate ground states whereas the four $\mathbb{Z}_2$ even ones correspond to kink-antikink bound states (a.k.a. breathers). This leads, in particular, to an interesting prediction on the universal ratio of the correlation lengths {\em above} and {\em below}
 the critical temperature \cite{prlFMS,Fioravanti:2000xz}. In fact, identifying the correlation length 
from the leading exponential asymptotic behaviour of the spin-spin connected correlation function in the long-distance limit   
\EQ
\langle 0 | \sigma(x) \sigma(0) | 0\rangle_c^{\pm} 
\sim \exp\left(-\frac{| x|}{\xi^{\pm}}\right) 
\, ,
\label{falling}
\EN 
(where the indices $\pm$ refer to the high and low temperature 
phases respectively), from the $\mathbb{Z}_2$ symmetry property of the 
$\sigma$ field, the self--duality of the model and the 
spectral representation of the above correlator it follows that\footnote{Note that for the thermal deformation of the IM the same universal ratio takes instead the value 
$\xi^+/\xi^- = 2$, since the lowest mass state coupled to the disorder operator is in the IM the two-particle state $A A$.}
\EQ
\frac{\xi^+}{\xi^-} = \frac{m_2}{m_1} = 2 \cos\frac{5\pi}{18} =
1.28557...
\label{xiuniv}
\EN  
In fact, in the high-temperature phase the lowest energy state to which the order parameter $\sigma$ couples to is given by the particle $A_1$. To determine the correlation length in the low-temperature phase, using the self-duality of the model, we can consider the correlator of the disorder operator $\mu$ in the high-temperature phase. However, the lowest energy state to which $\mu$ couples is $A_2$, which leads to the non-trivial universal ratio (\ref{xiuniv}) of the correlation lengths above and below the critical temperature. 
\begin{table} 
    \bgroup
    \setlength{\tabcolsep}{0.5em}
  \begin{center}
  \begin{tabular}{@{}l l l l}
  \hline
  exact & numerical & parity & excitation\\
  \hline
   $m_1$ & 1 & \red{odd}  & kink\\
   $m_2=2m\cos(5\pi/18)$ & 1.285(6) & even  & particle\\
   $m_3=2m_1\cos(\pi/9)$ & 1.879(4) & \red{odd} & kink \\
   $m_4=2m_1\cos(\pi/18)$ & 1.969(6) & even & particle\\
   $m_5=4m_1\cos(\pi/18)\cos(5\pi/18)$ & 2.532(1) & even  & particle\\
   $m_6=4m_1\cos(2\pi/9)\cos(\pi/9)$ & 2.879(4) & \red{odd} & kink\\
   $m_7=4m_1\cos(\pi/18)\cos(\pi/9)$ & 3.701(7)& even & particle\\
  \hline
  \end{tabular}
  \end{center}
    \egroup
 \caption{Spectrum of the thermal deformation of the tricritical Ising Model.} \label{tab:e7} 
\end{table}
It is also worth reminding that in the TIM the relationship between the mass gap and the coupling constant is known exactly and given by \cite{FATEEV199445}:
\begin{eqnarray}
  m_1&=&\left(\frac{2\Gamma\left(\frac{2}{9}\right)}{\Gamma\left(\frac{2}{3}\right)\Gamma\left(\frac{5}{9}\right)}\right)
  \left(\frac{4\pi^2\Gamma\left(\frac{2}{5}\right)\Gamma^3\left(\frac{4}{5}\right)}{\Gamma^3\left(\frac{1}{5}\right)\Gamma\left(\frac{3}{5}\right)}\right)^{5/18} |g|^{5/9} \\ 
  &=&3.7453728362\dots |g|^{5/9}  \,. \nonumber
  \label{masscouplingrelation}
\end{eqnarray}
This relation turns out to be useful in our numerical TCSA studies, where it can be used to normalise all units in terms of the lowest mass gap $m_1$ of the theory.

\subsubsection{Form Factors of the thermal field}\label{subsec:ffe7thermal}

Since the thermal deformation of the TIM corresponds to an integrable quantum field theory, the form factor expansion can be exploited in order to compute the correlation functions of the model. The operator $\epsilon(x)$ is proportional to the trace of the stress-energy tensor $\Theta(x)$:
\EQ
\Theta(x)=2\pi g\left( 2-2\Delta_\epsilon\right)\epsilon(x).
\EN
The conservation law of the stress-energy tensor implies that the polynomial $Q_{ab}(\th)$ can be factorised as 
\EQ
Q^\Theta_{ab}(\th) = \left(\cosh\th +
\frac{m_a^2+m_b^2}{2m_am_b}\right)^{1-\delta_{ab}} P_{ab}(\th)\,,
\EN
where
\EQ
P_{ab}(\th)\equiv\sum_{k=0}^{N_{ab}'} a^k_{ab}\,\cosh^k\th\,.
\EN
Moreover, for the diagonal elements $F^\Theta_{aa}$ we have the normalisation
\EQ
F_{aa}^\Theta(i\pi) =\langle A_a(\th_a)|\Theta(0)|A_a(\th_a) \rangle = 2\pi m^2_a\,.
\label{ipi}
\EN
The above properties determine the form factors of $\Theta$ uniquely, and its FF were computed in \cite{AMV}, to which the interested reader is referred for more details.

\subsubsection{Form Factors of the order and disorder operators}\label{sec:TIMFF}

It was observed in~\cite{Fioravanti:2000xz} that in the TIM there is an important novelty compared to the IM, i.e. there are {\em two} odd spin operators $\sigma$ and $\sigma'$, whose conformal dimensions differ less than 1 (c.f. Section~\ref{subsec:TIM_CFT}). This circumstance has a drastic consequence on the resulting computation of the FF of these operators (similarly for their duals). Namely, all the form factor equations (c.f. Subsection~\ref{subsec:FFeqs}) that can be written down for the FF of these operators, including those involving their asymptotic behaviour, are exactly the {\em same}! This implies that these operators cannot be distinguished from the form factor  equations they satisfy, just as we have seen for the case of the energy and magnetisation operators in the magnetic formation of the Ising model in Subsection \ref{subsec:e8ff}. None of the order/disorder operators corresponds to the perturbation either, so this shortcut which was exploited for the magnetisation operator in the Ising $E_8$ case and also for the thermal field in Subsection \ref{subsec:ffe7thermal} above,  is not available here either. 

Correspondingly, we expect to find a linear system of $n$ equations for the $n$ unknown parameters entering the FF of these operators which {\em necessarily} must have rank $(n-1)$, i.e. there must be two free parameters to accommodate the two different magnetisation operators of the model. This is indeed the case, and form factor equations only allow the FF of $\sigma(x)$ and $\sigma'(x)$ to be fixed in terms of their one-particle FF $F_1$ and $F_3$ on the particle excitations $A_1$ and $A_3$. Interestingly enough, these two parameters can then be determined for both operators by exploiting the self-duality of the TIM and the cluster properties of the FF. Just like for the Ising model, the clustering of the FF of the order and disorder operators can be fixed in terms of the parity of the number of particles, in analogy with Eqs. \eqref{aympfact1} and \eqref{aympfact2}, which are valid separately for the leading $\sigma$/$\mu$ and the subleading $\sigma'$/$\mu'$.

In the following, we collect the results of~\cite{2022ScPP...12..162C} for the form factors in the high-temperature phase of the TIM, up to certain two particle form factors. Similarly to the Ising model, the FF in the low-temperature phase can be obtained by application of the Kramers--Wannier duality which swaps the roles of the order fields $\sigma$ and $\sigma'$ with the disorder fields $\mu$ and $\mu'$.  

The multi-particle states occurring in the FF can be classified according to their threshold energy and their $\mathbb{Z}_2$ parity, as shown in Table \ref{multiparticlesss}. This classification is convenient as their threshold energy specifies the importance of their contributions in the spectral expansion of correlation functions and dynamical structure factors, while their parity determines their occurrence via superselection rules implied by the $\mathbb{Z}_2$ spin reversal symmetry.

The building blocks necessary for constructing the Form Factor parameterisation \eqref{FF_Ansatz} are collected in Appendix \ref{Appendix:FF2ptE7}. Using them, the general two-particle Form Factor $F^{\Phi}_{ab}(\th)$ can be written as 
\EQ
F^{\Phi}_{ab}(\th)=\frac{Q^{\Phi}_{ab}(\th)}{D_{ab}(\th)}
F^{min}_{ab}(\th)\,,
\EN
The polynomial $Q^{\Phi}_{ab}(\th)$ is the term that contains the information about the 
operator $\Phi$, and can be written as 
\EQ
Q_{ab}^{\Phi}(\th)\equiv\sum_{k=0}^{N_{ab}} a^k_{ab}\,\cosh^k\th\,.
\label{polyy}
\EN
Its degree $N_{ab}$ is determined by the upper bound \eqref{bound} on the asymptotic behaviour of the FF of the operator $\Phi(x)$.

For excitations that are non-local with respect to the operator $\Phi$, the FF Ansatz includes an extra term $\cosh\theta/2$ (either in the numerator or in the denominator, according to the asymptotic behaviour in $\theta$ of the form factor) which is even under $\theta \rightarrow -\theta$ but changes sign under the transformation $\theta \rightarrow \theta + 2 \pi i$ which probes the (non-)locality of $\Phi$ with respect to the excitations. 

Since each polynomial ${\cal P}_{\alpha}(\th)$ grows asymptotically as $e^\theta$, while the minimal form factor behaves as shown in Eq.\,(\ref{dec}), it is then easy to determine from the bound (\ref{bound}) the maximum degree $N_{ab}$ of the polynomial $Q_{ab}(\th)$. The task is then to extract the one-particle FF $F_i$ and the coefficients $a_{ij}^n$ in the polynomial part~\eqref{polyy} of the two-particle form factor Ansatz.

\begin{table}
\begin{center}
\bgroup
\setlength{\tabcolsep}{0.5em}
\begin{tabular}{||c|r|c||c|r|c||} \hline
{state}\rule[-2mm]{0mm}{7mm} &  $\omega/m_1$ \rule[-2mm]{0mm}{7mm} &
{parity} \rule[-2mm]{0mm}{7mm}  
& {state}\rule[-2mm]{0mm}{7mm} &  $\omega/m_1$ \rule[-2mm]{0mm}{7mm} &
{parity} \rule[-2mm]{0mm}{7mm}   \\ \hline\hline
\rule[-2mm]{0mm}{7mm}$A_2$ &1.28558 & even & % \\ \hline
\rule[-2mm]{0mm}{7mm}$A_1$ &1.00000 & \red{odd} \\ \hline
\rule[-2mm]{0mm}{7mm}$A_4$ &1.96962 & even  &  %\\ \hline
\rule[-2mm]{0mm}{7mm}$A_3$ &1.87939 & \red{odd} \\ \hline
\rule[-2mm]{0mm}{7mm}$A_1$ $A_1$ & $\geq$ 2.00000 & even  & %\\ \hline
\rule[-2mm]{0mm}{7mm}$A_1$ $A_2$ & $\geq$ 2.28558 & \red{odd} \\ \hline
\rule[-2mm]{0mm}{7mm}$A_2$ $A_2$ & $\geq$ 2.57115 & even  & % \\ \hline
\rule[-2mm]{0mm}{7mm}$A_6$ & $\geq$  2.87939 & \red{odd} \\ \hline
\rule[-2mm]{0mm}{7mm}$A_1$ $A_3$ & $\geq$  2.87939 & even & %\\ \hline
\rule[-2mm]{0mm}{7mm}$A_1$ $A_4$ & $\geq$  2.96952 & \red{odd} \\ \hline
\rule[-2mm]{0mm}{7mm}$A_5$  & 2.53209 & even  & %\\ \hline
\rule[-2mm]{0mm}{7mm}$A_2$ $A_3$ & $\geq$  3.16496 & \red{odd} \\ \hline
\rule[-2mm]{0mm}{7mm}$A_2$ $A_4$ & $\geq$ 3.25519 & even & %\\ \hline
\rule[-2mm]{0mm}{7mm}$A_1$ $A_5$ & $\geq$ 3.53209 &\red{odd}  \\ \hline
\rule[-2mm]{0mm}{7mm}$A_7$ &  $3.70167$ & even & %\\ \hline
\rule[-2mm]{0mm}{7mm}$A_3$ $A_4$ & $\geq$ 3.84901 & \red{odd} \\ \hline
\rule[-2mm]{0mm}{7mm}$A_3$ $A_3$ & $\geq$ 3.75877 & even    \\ \cline{1-3}
\rule[-2mm]{0mm}{7mm}$A_2$ $A_5$ & $\geq$ 3.81766  & even   \\ \cline{1-3}
\rule[-2mm]{0mm}{7mm}$A_1$ $A_6$ & $\geq$ 3.87939 & even   \\ \cline{1-3}
\end{tabular}
\egroup
\end{center}
\caption{Lowest energy states sorted according to the energy of their threshold and their $\mathbb{Z}_2$ parity.}
\label{multiparticlesss}
\end{table}

\vspace{3mm}
\noindent
{\bf Order operator:} Anticipating that the FF of $\sigma$ and $\sigma'$ satisfy the same set of equations, in the following we denote  a generic $\mathbb{Z}_2$ odd (order) operator by $\Phi$ 
and the corresponding dual (disorder) operator by $\tilde\Phi$. The order operators have non-vanishing matrix elements between states with different $\mathbb{Z}_2$ parity. In particular, this means that their vacuum expectation value (in the high-temperature phase) is zero. According to the $\mathbb{Z}_2$ parity of the various particles, the non-vanishing one-particle FF are 
\beq
F^\Phi_1, F^\Phi_3, F^\Phi_6
\,.
\label{OONNEE}
\eeq
Given the presence of two order operators, we expect to find solutions of the form factor  equations which depend on two free parameters, which we choose to be $F^\Phi_1$ and $F^\Phi_3$. This means that it should be possible to fix $F^\Phi_6$ in terms of the previous quantities $F^\Phi_1$ and $F^\Phi_3$, as shown below in \eqref{f6}.

Labelling the multi-particle states in terms of the increasing value of their rest energy, 
the first non-vanishing two-particle FF of the order operators is
\beq
F^\Phi_{12}(\theta)\,=\,Q^\Phi_{12}(\theta)\frac{F^{ min}_{12}(\theta)}{D_{12}(\theta)}\,\,\,,\label{onetwo}
\eeq
where
\beq
F_{12}^{\rm min}(\theta)\,=\,g_{\tfrac{7}{18}}(\theta)g_{\tfrac{13}{18}}(\theta)\quad,\quad
D_{12}(\theta)\,=\,\mathcal{P}_{\tfrac{7}{18}}(\theta)\mathcal{P}_{\tfrac{13}{18}}(\theta)\,\,\,.
\eeq
$Q_{12}(\theta)$ is the polynomial that determines the operator and can be written as
\beq
Q_{12}(\theta)\,=\,\sum_{n=0}^{N_{12}}a_{12}^n\cosh^{n}(\theta)\,\,\,.\label{qpolynomial}
\eeq
The degree of this polynomial is fixed to be $N_{12}=1$, by using Eq. (\ref{bound}) using that $\displaystyle\lim_{\theta\to\infty}g_{\alpha}(\theta)\sim \exp(\vert\theta\vert/2)$ (see Appendix \ref{Appendix:FF2ptE7}) together with $\displaystyle\lim_{\theta\to\infty}\mathcal{P}_\alpha(\theta)\sim\exp(\theta)$. Hence, we need to fix two constants, 
$a_{12}^0$ and $a_{12}^1$, in order to determine the FF $F^\Phi_{12}(\theta)$. Using the bound state\footnote{In the following when implementing the bound state residue equations for the FF we assume that {\em all} the on-shell three-particle coupling $\Gamma_{ab}^c$ are positive, i.e. $\Gamma_{ab}^c > 0$. }
 form factor  equation \eqref{recurb}, these quantities can be expressed in terms of the constant one-particle FF $F^\Phi_1$ and $F^\Phi_3$.  The two linear equations for $a_{12}^0$ and $a_{12}^1$ are then\footnote{The coefficients of $a_{i,j}^k$ are computed by numerical integration of the minimal form factor functions \eqref{fmin}.  Although they can be performed to any desired precision, we only give here the first few non-zero digits to keep the equations simple.}
\begin{eqnarray}
&&  0.113447 \,a_{12}^0  - 0.0729222 \,a_{12}^1 = F^{\Phi}_1 \,\,\,;\label{normalisationodd}\\
&& 0.0328461 \, a_{12}^0 + 0.011234 \, a_{12}^1 = - F^\Phi_3 \,\,\,.\nonumber
\end{eqnarray}
Hence, if we knew the values of both $F^\Phi_1$ and $F^\Phi_3$, for $\Phi=\sigma$ or $\Phi=\sigma'$, we could determine the two coefficients $a_{12}^0$ and $a_{12}^1$ of this form factor for these operators. 

The next two-particle FF to consider for the magnetisation operators is
\beq
F_{14}^\Phi(\theta)\,=\,\frac{Q_{14}^\Phi(\theta)}{D_{14}(\theta)}F_{14}^{\rm min}(\theta)\,\,\,, \label{onefour}
\eeq
 where
\begin{eqnarray}
F_{14}^{\rm min}(\theta)&=&g_{\tfrac{1}{6}}(\theta)\,g_{\tfrac{17}{18}}(\theta)\,g_{\tfrac{11}{18}}(\theta)\,g_{\tfrac{1}{2}}(\theta),\nonumber\\
&&\nonumber\\
D_{14}(\theta)&=&\mathcal{P}_{\tfrac{1}{6}}(\theta)\,\mathcal{P}_{\tfrac{17}{18}}(\theta)\,\mathcal{P}_{\tfrac{11}{18}}(\theta)\,\mathcal{P}_{\tfrac{1}{2}}(\theta),\nonumber
\end{eqnarray}
and
\beq
Q_{14}(\theta)=\sum_{n=0}^{N_{14}}a_{14}^n\cosh^n(\theta).\nonumber
\eeq
By enforcing the asymptotic behaviour (\ref{onefour}) of the FF, we can fix $N_{14}\leq 2$ and therefore there are three new constants to determine, i.e. $a_{14}^0,\,a_{14}^1,\,a_{14}^2$. There are three bound state residue equations, namely those relative to the simple poles coming from the particles $A_1, A_3$ and $A_6$: two of them which lead us back to $F^\Phi_1$, $F^\Phi_3$ while the third one brings in the new one-particle FF $F^\Phi_6$ 
\begin{eqnarray}
&& 0.0738796 \, a_{14}^0  - 0.0727572 \, a_{14}^1 + 0.071651\, a_{14}^2 = F^\Phi_1 \nonumber \\
&& - 0.00137278 \,a_{14}^0  + 0.000469519 a_{14}^1  - 0.000160585 \, a_{14}^2 = F^\Phi_3 \label{onetwoonefour}\\
&& 0.000110665 \, a_{14}^0  + 0.0000958386 \,a_{14}^1 + 0.0000829987 \,a_{14}^2 = - F^\Phi_6 \nonumber 
\end{eqnarray}
It is easy to see that, in absence of the values of $F^\Phi_1, F^\Phi_3$ and $F^\Phi_6$, the linear equations (\ref{normalisationodd}) and (\ref{onetwoonefour}) written so far are not enough to find the five unknown constants $(a_{12}^0, a_{12}^1, a_{14}^0, a_{14}^1, a_{14}^2)$. Hence, our strategy consists of considering more form factors, until there are enough linear equations to fix all the necessary constants.

The next FF to consider, i.e. $F_{23}^\Phi(\theta)$, brings in three new unknown constants, $a_{23}^0,\,a_{23}^1,\,a_{23}^2$ but there are also three bound state residue equations 
relative to the particles $A_1$, $A_3$ and $A_6$. They yield the linear equations
\begin{eqnarray}
&& 0.0517869\, a_{23}^ 0  - 0.0448488 \, a_{23}^1 + 0.0388402\, a_{23}^2 = F^\Phi_1\nonumber \\
&& -0.00638769 \,a_{23}^0 + 0.00218472 \, a_{23}^1 - 0.000747218 \, a_{23}^2 = F^\Phi_3 \\
&& 0.000693771 \,a_{23}^0  + 0.000445947 \, a_{23}^1 + 0.000286649\, a_{23}^2 = - F^\Phi_6 \nonumber
\end{eqnarray}
For the next two-particle FF given by  $F_{15}^\sigma(\theta)$, there are three new unknown constants, $a_{15}^0,\,a_{15}^1,\,a_{15}^2$. The bound state axiom for the two poles related to the particles $A_3$ and $A_6$ imply the equations
\begin{eqnarray}
&& 0.0439601 \, a_{15}^0  - 0.0336754\, a_{15}^1 + 0.0257969\, a_{15}^2 \, =\, F^\Phi_3 \\
&& 0.00254047 \, a_{15}^0 + 0.000441149\, a_{15}^1 + 0.0000766047\, a_{15}^2 = F^\Phi_6 \nonumber
\end{eqnarray}
Since the S-matrix amplitude $S_{15}(\theta)$ also has a double pole at $\theta=i\tfrac{\pi}{3}$, it can be exploited to give a corresponding residue equation according to \eqref{multiparticlepole} which yields the linear equations
\beq
\lim_{\theta\to i \tfrac{2\pi}{3}}\frac{F_{15}^\Phi(\theta)}{\Gamma_{12}^1\Gamma_{22}^5}\,=\,F_{12}^\Phi\left(i\frac{\pi}{6}\right),\label{doubleonefiveonetwo}
\eeq
and
\beq
\lim_{\theta\to i \tfrac{\pi}{3}}\frac{F_{15}^\Phi(\theta)}{\Gamma_{23}^1\Gamma_{22}^5}\,=\,F_{23}^\Phi\left(i\frac{\pi}{18}\right).\label{doubleonefivetwothree}
\eeq
These two conditions give rise to the additional two equations 
\begin{eqnarray}
&& -0.0961442 \,a_{15}^0 + 0.0480721\, a_{15}^1 - 0.024036 \,a_{15}^2 = 0.126345\, a_{12}^0 + 0.109418\, a_{12}^1 \nonumber \\ 
&& -0.00651126 \,a_{15}^0 - 0.00325563\, a_{15}^1 - 0.00162782 \,a_{15}^2 = 0.00189226\, a_{23}^0 +  \nonumber \\
&& \quad 0.00186351\, a_{23}^1 + 0.0018352\, a_{23}^2 
\end{eqnarray}
The FF $F^\Phi_{34}$ introduces four new constants $a_{34}^0,a_{34}^1,a_{34}^2,a_{34}^3$ and satisfies a bound state residue equation and four double pole residue equations.

Putting together all the equations collected, including the one-particle FF $F^\Phi_1, F^\Phi_3$ and $F^\Phi_6$ involved in the computation, there are $18$ unknown constants to fix. Altogether, for the above set of form factors, there are $17$ equations ($11$ from simple poles and $6$ from double ones), but it turns out that only $16$ of them are linearly independent. Therefore there exists a two-parameter family of solutions and $F^\Phi_1$ and $F^\Phi_3$ can be taken as the two parameters to label them. First of all, we indeed find that $F_6$ is no longer a free parameter since it is fixed in terms of  $F^\Phi_1$ and $F^\Phi_3$
\EQ
F^\Phi_6  =  0.115722 F^\Phi_1+0.587743 F^\Phi_3\,.
\label{f6}
\EN 
Secondly, the coefficients entering the various FF are determined as  
\EQ
\label{eq:FForder}
\begin{tabular}{lll}
%\hline
$F_{12}$: & \hspace{2mm} $a^0_{12}=3.06131 F^\Phi_1-19.8715 F^\Phi_3$ &\hspace{2mm} $a^1_{12} =-8.95069 F^\Phi_1-30.9146 F^\Phi_3$ \\
%\hline
$F_{14}$: & \hspace{2mm} $a^0_{14}=-160.899 F^\Phi_1-1600.15 F^\Phi_3$ &\hspace{2mm} $a^1_{14}=-626.504 F^\Phi_1-3040.25 F^\Phi_3$ \\
& \hspace{2mm} $a^2_{14}=-456.311 F^\Phi_1-1437.25 F^\Phi_3$ &  \\
%\hline
$F_{15}$: & \hspace{2mm} $a^0_{15}=32.1365 F^\Phi_1+177.579 F^\Phi_3$ & \hspace{2mm} $a^1_{15}=70.7301 F^\Phi_1+289.789 F^\Phi_3$ \\
&\hspace{2mm} $a^2_{15}=37.5681 F^\Phi_1+114.447 F^\Phi_3$ & \\
%\hline
$F_{32}$: & \hspace{2mm} $a^0_{23}=-38.6198 F^\Phi_1-337.751 F^\Phi_3$ &\hspace{2mm} $a^1_{23}=-142.958 F^\Phi_1-621.037 F^\Phi_3$ \\
&\hspace{2mm} $a^2_{23}=-87.8337 F^\Phi_1-266.777 F^\Phi_3$ & \\
%\hline
$F_{34}$: & \hspace{2mm} $a^0_{34}=-493.626 F^\Phi_1-2722.44 F^\Phi_3$ &\hspace{2mm} $a^1_{34}=-1617.98 F^\Phi_1-6682.94 F^\Phi_3$ \\
&\hspace{2mm} $a^2_{34}=-1495.64 F^\Phi_1-5104.39 F^\Phi_3$ 
&\hspace{2mm}$a^3_{34}=-399.244 F^\Phi_1-1174.92 F^\Phi_3$  
\end{tabular}
\EN

\vspace{3mm}
\noindent
{\bf Disorder operator:} Let's now focus the attention on the FF's of the dual disorder operators $\mu(x)$ and $\mu'(x)$ (hereafter collectively denoted as $\tilde\Phi$). In order to write down and find the solution of the FF equations for the matrix elements of these operators, it is necessary to consider that   
\begin{itemize}
    \item these operators couple to the $\mathbb{Z}_2$ even multi-particle states.
    \item moreover, 
    their vacuum expectation values $F_0^{\tilde\Phi} = \langle 0 | \tilde\Phi | 0 \rangle$ (in the conformal normalisation of both operators) are exactly known \cite{FATEEV1998652}.
    \end{itemize}
    
Taking into account the parity of the particles, the asymptotic behaviour of the FF, and the semi-locality of the dual operators with respect to the kink states (but not with respect to the pure particle excitations), we have the following Ansatz for the corresponding lowest FF's of the dual operators 
  
\begin{alignat}{4}
    \label{eq:ffevenans}
    F^{\tilde\Phi}_{11}(\theta) &=\frac{1}{\cosh\frac{\theta}{2}}\frac{Q_{11}^{\tilde\Phi}(\theta)}{D_{11}(\theta)}F_{11}^{\rm min}(\theta); &&\quad N_{11}=1 \nonumber\\
    F^{\tilde\Phi}_{13}(\theta)  &= \cosh{\frac{\theta}{2}}\frac{Q_{13}^{\tilde\Phi}(\theta)}{D_{13}(\theta)}F_{13}^{\rm min}(\theta);&&\quad  N_{13}=1 \nonumber\\
    F^{\tilde\Phi}_{22}(\theta)  &= \frac{Q_{22}^{\tilde\Phi}(\theta)}{D_{22}(\theta)}F_{22}^{\rm min}(\theta);&&\quad N_{22}=1 \\
   F^{\tilde\Phi}_{24}(\theta)  &= \frac{Q_{24}^{\tilde\Phi}(\theta)}{D_{24}(\theta)}F_{24}^{\rm min}(\theta);&&\quad N_{24}=2 \nonumber\\
   F^{\tilde\Phi}_{33}(\theta)  &= \frac{1}{\cosh\frac{\theta}{2}}\frac{Q_{33
   }^{\tilde\Phi}(\theta)}{D_{33}(\theta)}F_{33}^{\rm min}(\theta); &&\quad N_{33}=3\nonumber
\end{alignat}
Notice that we included an extra factor  $1/\cosh\theta/2$ (which induces an annihilation pole present for equal particles at $\theta = i \pi$) in the expressions $F^{\tilde\Phi}_{11}(\theta)$ and $F^{\tilde\Phi}_{33}(\theta)$ in view of the kink nature of these excitations in the low-temperature phase, which via KW duality implies  the non-locality of both excitations $A_1$ and $A_3$ with respect to the disorder operators in the high-temperature phase. For the same reason we also introduced an extra term $\cosh \theta/2$ in $F^{\tilde\Phi}_{13}(\theta)$, which also guarantees that this FF (similarly to all the others) has the expected constant asymptotic behaviour at $|\theta| \rightarrow \infty$. Taking into account the known values of the vacuum expectation values $F^{\tilde\Phi}_0$,  in this case, we have $19$ unknowns (including in this counting, at this stage also the VEVs of the two disorder operators, even though their values are eventually known by other means) and $20$ equations: among them
\begin{itemize}
\item 2 equations comes from the kinematical pole of $F^{\tilde\Phi}_{11}$ and $F^{\tilde\Phi}_{33}$  since the particles $A_1$ and $A_3$ (being a kink in the low-temperature phase) are non-local with respect the disorder operators $\tilde\Phi$ in the high-temperature phase by KW duality. 
\item $12$ equations come from the bound state pole residue equations. 
\item $6$ equations come from the double pole residue equations. 
\end{itemize} 
However, only $18$ of these linear equations are linearly independent, leading to a solution which can be parameterised in terms of $F^{\tilde\Phi}_0$ and $F^{\tilde\Phi}_2$ as follows:
first of all, the one-particle FF's $F^{\tilde\Phi}_4, F^{\tilde\Phi}_5$ and $F^{\tilde\Phi}_7$ are given in terms of $F^{\tilde\Phi}_0$ and $F^{\tilde\Phi}_2$ as follows 
\begin{eqnarray}
F^{\tilde\Phi}_4 & = & 0.0979846 F^{\tilde\Phi}_0-0.71782 F^{\tilde\Phi}_2 \nonumber \\
F^{\tilde\Phi}_5  & =  & 0.529952 F^{\tilde\Phi}_2-0.100203 F^{\tilde\Phi}_0 \\
F^{\tilde\Phi}_7  & = &  0.037784 F^{\tilde\Phi}_0-0.157422 F^{\tilde\Phi}_2 \nonumber 
\end{eqnarray}
while the remaining coefficients of the various FF's are fixed as
\EQ
\begin{tabular}{lll}
%\hline
$F^{\tilde\Phi}_{11}$: & \hspace{2mm} $a^0_{11}=5.08848 F^{\tilde\Phi}_2-0.21014 F^{\tilde\Phi}_0$ &\hspace{2mm} $a^1_{11}=5.08848 F^{\tilde\Phi}_2-1.21014 F^{\tilde\Phi}_0$  \\
%\hline
$F^{\tilde\Phi}_{13}$: & \hspace{2mm} $a^0_{13}=89.6538 F^{\tilde\Phi}_2-13.8826 F^{\tilde\Phi}_0$ &\hspace{2mm} $a^1_{13}=63.3814 F^{\tilde\Phi}_2-18.1224 F^{\tilde\Phi}_0$ \\
%\hline
$F^{\tilde\Phi}_{22}$: & \hspace{2mm} $a^0_{22}=22.2545 F^{\tilde\Phi}_2-2.57115 F^{\tilde\Phi}_0$ &\hspace{2mm} $a^1_{22}=17.9847 F^{\tilde\Phi}_2-5.1423 F^{\tilde\Phi}_0$  \\
%\hline
$F^{\tilde\Phi}_{24}$: & \hspace{2mm} $a^0_{24}=162.262 F^{\tilde\Phi}_2-28.5588 F^{\tilde\Phi}_0$ &\hspace{2mm} $a^1_{24}=251.212 F^{\tilde\Phi}_2-57.9849 F^{\tilde\Phi}_0$ \\
&\hspace{2mm} $a^2_{24}=90.1906 F^{\tilde\Phi}_2-27.0272 F^{\tilde\Phi}_0$ & \\
$F^{\tilde\Phi}_{33}$: & \hspace{2mm} $a^0_{33}=438.962 F^{\tilde\Phi}_2-86.4572 F^{\tilde\Phi}_0$ &\hspace{2mm} $a^1_{33}=1008.33 F^{\tilde\Phi}_2-233.757 F^{\tilde\Phi}_0$\\ 
&\hspace{2mm} $a^2_{33}=730.328 F^{\tilde\Phi}_2-195.529 F^{\tilde\Phi}_0$
&\hspace{2mm} $a^3_{33}=160.963 F^{\tilde\Phi}_2-49.2296 F^{\tilde\Phi}_0$ 
\end{tabular}
\label{FFFFFFF}
\EN
The expectation values of the disorder fields $\mu$ and $\mu'$ were obtained in \cite{FATEEV1998652}:
\EQ
\begin{array}{lllll}
F^{\mu}_0 &=& 1.59427 \ldots\, |g|^{ 1/24}  &= &  1.44394 \ldots \, m_1^{3/40}\\
F^{\mu'}_0 &=& 2.45205 \ldots\, |g|^{35/72}. &= & 0.772185 \ldots   \, m_1^{7/8} 
\end{array}
\label{eq:tim_exact_vevs}
\EN
The rest of the ingredients can be extracted using the cluster property (c.f. Subsection ~\ref{subsec:FFeqs}) together with the Kramers--Wannier duality:
\begin{eqnarray}
	|F^{\mu}_2|&=&0.462658\dots\nonumber\\
	|F^{\mu'}_2|&=&2.05131\dots\,.
\end{eqnarray}
and
\begin{eqnarray}
	|F^{\sigma}_1| & = & 0.710426\dots \nonumber\\
	\label{eq:FF1ptsol}|F^{\sigma}_3| & = &0.252315\dots\nonumber\\
	|F^{\sigma'}_1 |& = & 2.05592\dots \nonumber\\
	|F^{\sigma'}_3| & = &1.71395\dots 
\end{eqnarray}
(written in units $m_1=1$), where all the amplitudes can be chosen to be real and the sign of $F_3^{\Phi}$ must be opposite to $F_1^{\Phi}$ when substituting to Eqs. \eqref{eq:FForder}.\footnote{Note that this determination of the relative sign corresponds to fixing some of the signs of the three-particle couplings listed in Table \ref{tab:Sgammas}.}

\subsection{Dynamical Structure Factors of the thermal deformation of the Tricritical Ising Model}\label{DSFTIM}

In any lattice realisation of the tricritical Ising model, such as the Blume--Capel model \eqref{eq:2dBCM} or the quantum spin chain \eqref{eq:TIMC}, it is necessary to specify the relation between the spin operator\footnote{This is the actual operator that couples to the neutrons in the scattering experiments.} 
 $S(x,t)$ on the lattice and the relevant spin fields which are present in the low energy effective continuum $E_7$ quantum field theory. On a general ground, based on 
 the spin $Z_2$ symmetry of the various fields involved, this relationship is of the form:
\begin{equation}
    {\cal S}(x,t) = a_\sigma\sigma(x,t) + a_{\sigma'}\sigma'(x,t) + \cdots ...
\end{equation}
where the constants $a_{\sigma,\sigma'}$ are specific to the lattice model realisation.
The dynamical spin response, ${\cal S}(\omega,q)$ that would be measured in a neutron scattering experiment can be expressed in terms of the imaginary part of a retarded spin-spin correlation function:
\begin{eqnarray}
    S(\omega,q) &=& \int dxdt e^{i\omega t-iqx}
    \langle {\cal S}(x,t){\cal S}(0,0)\rangle
    = \int dxdt e^{i\omega t-iqx}\sum_{i,j=\sigma,\sigma'} a_i a_j
    \langle\sigma_i(x,t)\sigma_j(0,0)\rangle\nonumber\\
    &\equiv& 
    \sum_{i,j=\sigma,\sigma'} a_i a_j {\cal S}^{ij}(\omega,q)
\end{eqnarray}
The details of the calculation of ${\cal S}^{ij}(\omega,q)$ in terms of the FF of the various fields entering these quantities are presented in Section~\ref{Structurefactor}.
We restrict ourselves here to just a few comments, focusing our attention on the high-temperature phase (as usual, the analysis of  the low-temperature is done by interchanging the role of order and disorder operators); notice that contrary to the Ising model, in the TIM all operators have one- and two-particle contributions. The order parameters have contributions from odd states while the disorder operators from the even particle states. These terms in the dynamical structure factors can be evaluated by combining the results of the previous sections and Subsection~\ref{Structurefactor}.
\begin{table}
	\centering
	\bgroup
    \setlength{\tabcolsep}{0.5em}
	\begin{tabular}{|c|c|c|c|}
		\hline
		& & &\\[-1.3em]
		 particle & 
		 $2\pi \frac{|F_i^{\sigma}|^2}{m_i}$ & $2\pi \frac{|F_i^{\sigma'}|^2}{m_i}$ & 
		 $2\pi \frac{F_i^{\sigma}F_i^{\sigma'*}}{m_i}$\\
		 & & &\\[-1.3em]
		\hline
		$1$ & $3.17116$  & $26.5578$ &$9.1771$\\
		$3$ & $0.212838$  &$9.82114$ & $1.44579$\\
		$6$ & $0.0095296$ & $1.29193$& $0.110957$\\
		\hline
	\end{tabular}
	\egroup
	\caption{One-particle weights in the contributions to the DSF ${\cal S}^{ij}$, $i,j=\sigma,\sigma'$.}
	\label{tab:1ptsigma}
\end{table}

\begin{table}
	\centering
    \bgroup
    \setlength{\tabcolsep}{0.5em}
	\begin{tabular}{|c|c|c|c|}
		\hline
		& & &\\[-1.3em]
		particle & 
		$2\pi \frac{|F_i^{\mu}|^2}{m_i}$ & $2\pi \frac{|F_i^{\mu'}|^2}{m_i}$ & 
		$2\pi \frac{F_i^{\mu}F_i^{\mu'*}}{m_i}$\\
		 & & &\\[-1.3em]
		\hline
		$2$ & $1.04617$  & $20.5658$ &$4.63846$\\
		$4$ & $0.115915$  &$6.22406$ & $0.849389$\\
		$5$ & $0.0250625$ & $2.52991$ & $0.25805$\\
		$7$ & $0.000566863$ & $0.146462$ & $0.00911174$\\
		\hline
	\end{tabular}
	\egroup
	\caption{One-particle weights in the contributions to the DSF ${\cal S}^{ij}$, $i,j=\mu,\mu'$.}
	\label{tab:1ptmu}
\end{table}

\begin{figure}
	\centering
	\begin{subfigure}{0.45\textwidth}
		\includegraphics{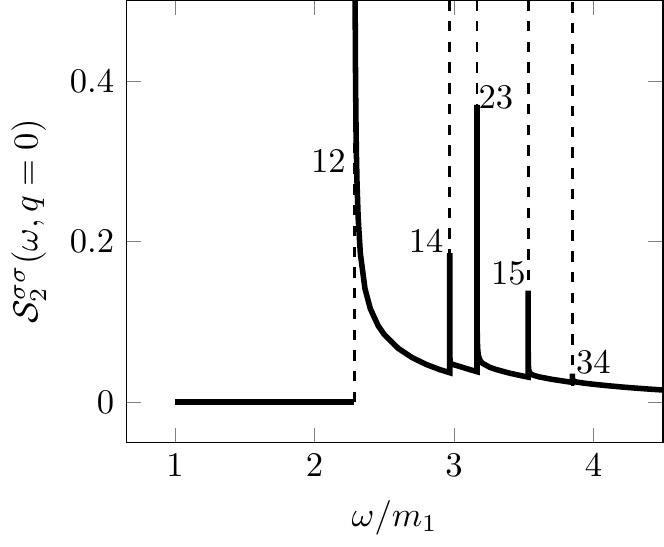}
		\caption{Two-particle contribution to $\mathcal{S}^{\sigma\sigma}$}
			\end{subfigure}
	\begin{subfigure}{0.45\textwidth}
		\includegraphics{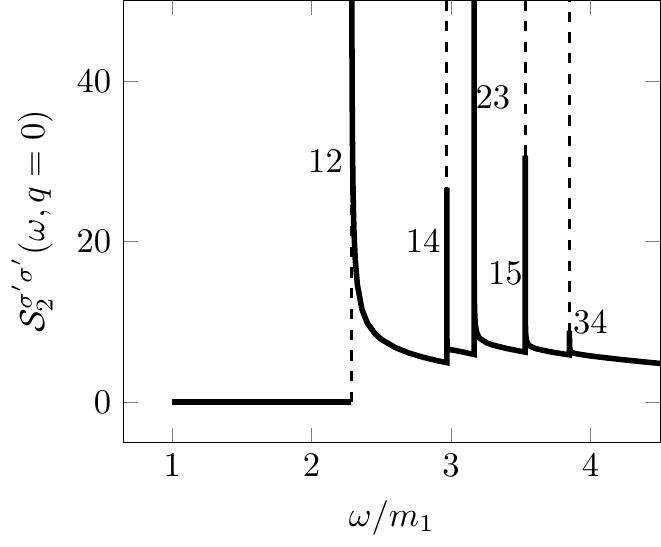}
		\caption{Two-particle contribution to $\mathcal{S}^{\sigma'\sigma'}$}
			\end{subfigure}
	\caption{\label{Dss} Two-particle contributions to the dynamical structure factor  of the leading and subleading magnetisation operators. The values of the dynamical structure factors are shown in units of the massgap $m_1$ using the exact VEV \eqref{eq:tim_exact_vevs}.}
\end{figure}

\begin{figure}
	\centering
	\begin{subfigure}{0.45\textwidth}
		\includegraphics{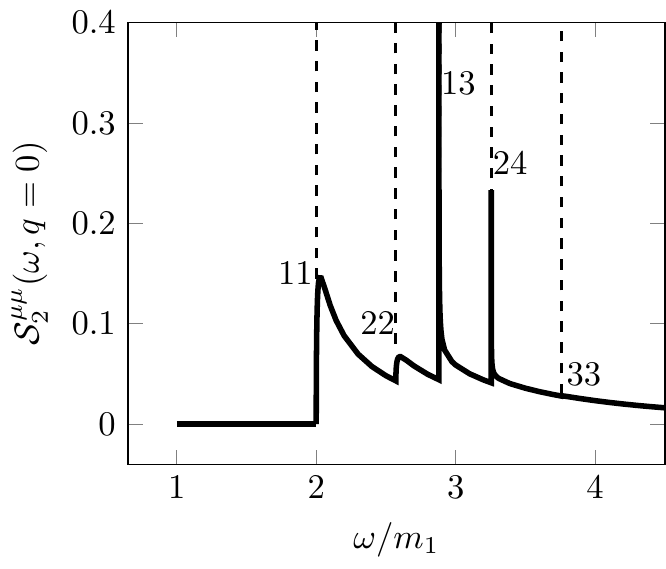}
		\caption{Two-particle contribution to $\mathcal{S}^{\mu\mu}$}
	\end{subfigure}
	\begin{subfigure}{0.45\textwidth}
		\includegraphics{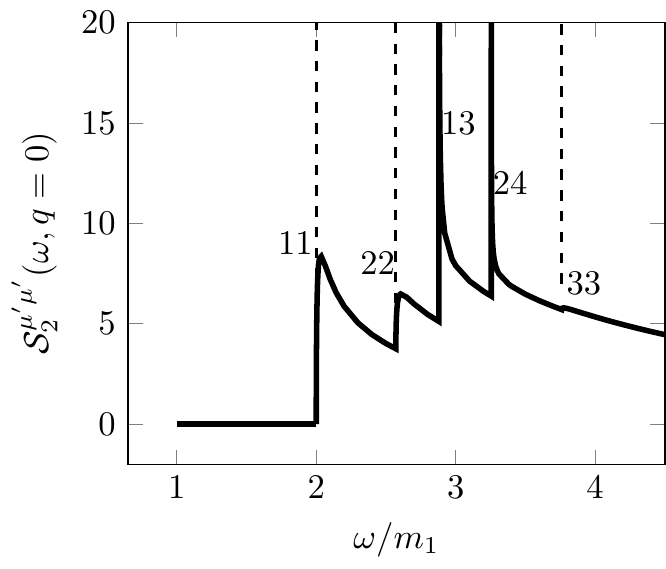}
		\caption{Two-particle contribution to $\mathcal{S}^{\mu'\mu'}$}
	\end{subfigure}
	\caption{\label{Dmm} Two-particle contributions to the dynamical structure factor  of the leading and subleading disorder operators. The values of the dynamical structure factors are shown in units of the mass gap $m_1$ using the exact VEV \eqref{eq:tim_exact_vevs}.}
\end{figure}

\begin{figure}
	\centering
	\begin{subfigure}{0.45\textwidth}
		\includegraphics{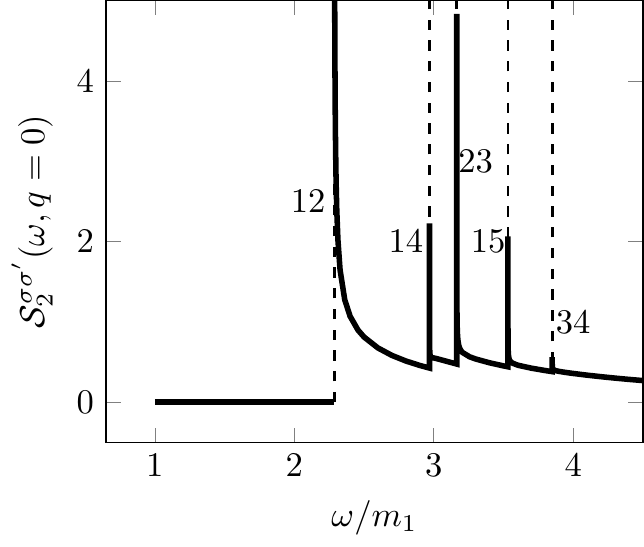}
		\caption{Two-particle contribution to $\mathcal{S}^{\sigma\sigma'}$}
	\end{subfigure}
	\begin{subfigure}{0.45\textwidth}
		\includegraphics{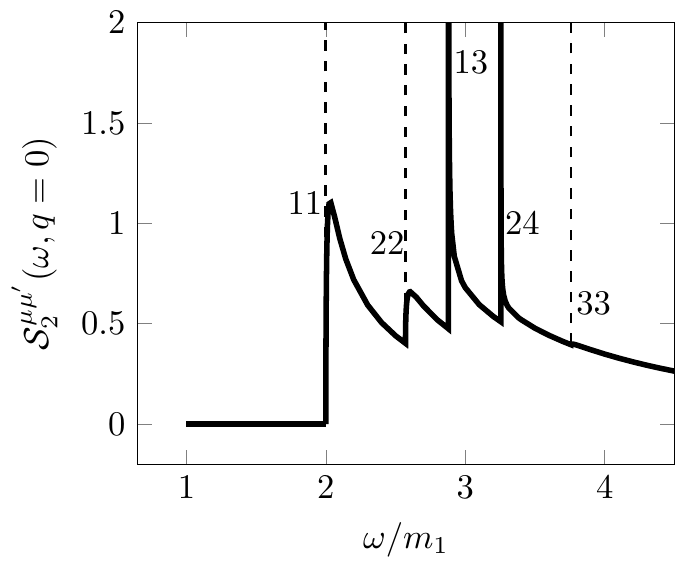}
		\caption{Two-particle contribution to $\mathcal{S}^{\mu\mu'}$}
	\end{subfigure}
	\caption{\label{Dmixed} Two-particle contributions to the mixed dynamical structure factors of the leading and subleading magnetisation and disorder operators.}
\end{figure}

The weights of one-particle Dirac-$\delta$ contributions for the order and disorder operators are listed in Tables~\ref{tab:1ptsigma} and~\ref{tab:1ptmu}, respectively. The various two-particle contributions can also be evaluated using the formula (\ref{eq:S2}) presented in Subsection \ref{Structurefactor}, and are plotted in Figures~\ref{Dss}, \ref{Dmm} and~\ref{Dmixed}, while their numerical values are collected in Table~\ref{tab:D2pt}. The behaviour of the contribution ${\cal S}^{ij}_{2}$ just above the threshold $\omega=m_i+m_j$ is given by 
\begin{equation}
    S^{ij}_{2}(\omega,q=0)=
    {\cal C}_{ij}\left[\omega-(m_i+m_j)\right]^{1/2-\delta_{i,j}}
    +O(\left[\omega-(m_i+m_j)\right]^{3/2-\delta_{i,j}})
\end{equation}
where ${\cal C}_{ij}$ is a constant.

Comparing these results to the DSF in the Ising model discussed in Section \ref{DSFIM}, we can see significant differences which can be used to distinguish them in experimental signatures. First, the Ising DSF shows a signal from a single excitation, while in the TIM there is a complicated spectrum corresponding to different masses and their combinations, with their ratios predicted by the spectrum $E_7$ scattering theory (cf. Table \ref{tab:e7}). 

The threshold behaviour is also significantly different. For the Ising model, the contributions coming from multi-particle continua vanishes as a power at the threshold. However, in the TIM there are two types of two-particle thresholds:

\begin{itemize}
    \item For thresholds with identical particles $A_aA_a$, the behaviour is similar to the Ising case. This is due to the behaviour $S_{aa}(\theta=0)=-1$ of the scattering amplitude, which due to \eqref{w1} implies that the two-particle form factor vanishes at the threshold:
    \begin{equation}
        F_{aa}^{min}(\theta=0)=0\,.
    \end{equation}
    From \eqref{eq:S2} this leads to the two-particle contribution vanishing at the threshold.
    \item For thresholds with different particles $A_aA_b$, $a\neq b$, the two-particle contribution diverges at the threshold. In this case $S_{ab}(\theta=0)=+1$ and $F_{ab}^{min}(\theta=0)$ has a finite value. As a result, the zero of the denominator $|\sinh(\theta_1-\theta_2)|$ in \eqref{eq:S2} leads to singular behaviour at the threshold.
\end{itemize}

In any experimental realisation of the spin system, the large-scale dynamics depends on the relation between the temperature and vacancy couplings, which can be tuned using suitable experimental parameters. As a function of the parameters, the system is then expected to exhibit a crossover from the simple Ising DSF computed in Subsection \ref{DSFIM} to the DSF characteristic of the TIM. When the parameters are tuned so that the dynamics corresponds to that of the $E_7$ model, the operator describing the response in an inelastic neutron scattering experiment is a combination of $\sigma$ and $\sigma'$, when the system is in the high-temperature phase, or $\mu$ and $\mu'$ in the low-temperature phase. These two phases are then clearly distinguishable by their different threshold structures. Note that while the specific amplitudes depend on the particular combination of operators, the locations of one-particle peaks and two-particle thresholds are fixed by the spectrum of the TIM.  In moving from an $E_7$ DSF to that of an Ising DSF due to perturbations that break the $E_7$ integrability, we expect that at weak integrability breaking the various delta-function peaks and threshold singularities will shift (if they are at a frequency $\omega < 2m_1$) or broaden (if $\omega > 2m_1$), reflecting the shifts in masses and decay processes due to weak integrability breaking, respectively.  

\begin{table}
    \centering
    \bgroup
    \setlength{\tabcolsep}{0.5em}
    \begin{tabular}{|c|c|c|c|c|c|c|}
    \hline
    $\omega$ & $\mathcal{S}_2^{\sigma \sigma}$ & $\mathcal{S}_2^{\sigma' \sigma'}$ & $\mathcal{S}_2^{\mu \mu}$ & $\mathcal{S}_2^{\mu' \mu'}$ & $\mathcal{S}_2^{\sigma \sigma'}$ & $\mathcal{S}_2^{\mu \mu'}$ \\
    \hline
 $2.1$  &  $0.$        &  $0.$       &  $0.118391$  &  $7.22124$  &  $0$        &  $0.924625$ \\
 $2.4$  &  $0.116344$  &  $9.84403$  &  $0.057359$  &  $4.45951$  &  $1.07018$  &  $0.505759$ \\
 $2.7$  &  $0.055504$  &  $6.11336$  &  $0.058118$  &  $5.99721$  &  $0.58250$  &  $0.589421$ \\
 $3.$   &  $0.046533$  &  $6.51685$  &  $0.058922$  &  $7.89907$  &  $0.55057$  &  $0.679920$ \\
 $3.3$  &  $0.040778$  &  $7.11038$  &  $0.046481$  &  $7.62455$  &  $0.53811$  &  $0.591966$ \\
 $3.6$  &  $0.031957$  &  $6.71276$  &  $0.032408$  &  $6.15006$  &  $0.46253$  &  $0.442530$ \\
 $3.9$  &  $0.024489$  &  $6.03023$  &  $0.025383$  &  $5.55370$  &  $0.38338$  &  $0.370885$ \\
 $4.2$  &  $0.018951$  &  $5.34815$  &  $0.020103$  &  $4.95863$  &  $0.31722$  &  $0.310682$ \\
 $4.5$  &  $0.015140$  &  $4.82431$  &  $0.016282$  &  $4.45800$  &  $0.26889$  &  $0.264021$ \\
 $4.8$  &  $0.012338$  &  $4.37690$  &  $0.013441$  &  $4.03219$  &  $0.23081$  &  $0.227167$ \\
 $5.1$  &  $0.010227$  &  $3.99007$  &  $0.011279$  &  $3.66613$  &  $0.20027$  &  $0.197559$ \\
 $5.4$  &  $0.008604$  &  $3.65296$  &  $0.009598$  &  $3.34868$  &  $0.17540$  &  $0.173419$ \\
 $5.7$  &  $0.007333$  &  $3.35720$  &  $0.008268$  &  $3.07131$  &  $0.15491$  &  $0.153482$ \\
 $6.$   &  $0.006322$  &  $3.09619$  &  $0.007199$  &  $2.82736$  &  $0.13782$  &  $0.136829$ \\
	    \hline
    \end{tabular}
    \egroup
    \caption{Two-particle contributions to various dynamical structure factors calculated from the form-factor  solution presented in the previous subsections.}
    \label{tab:D2pt}
\end{table}

\subsection{Generalised susceptibilities in the tricritical Ising model}\label{subsec:gen_susc_TIM}

The universal ratios in the scaling region of the tricritical Ising model were obtained in~\cite{prlFMS,Fioravanti:2000xz}. The one-point functions and the generalised susceptibilities were extracted using a combination of the Truncated Conformal Space Approach~\cite{YZ} and integrability techniques. One-point functions were known at the time~\cite{russian2} for the thermal and the subleading magnetic (integrable) deformations. In the thermal deformation, the FF of the thermal field up to two particles were available~\cite{AMV}. One-particle FF were also extracted from the TCSA for the general case. Together with short-distance expansions the authors of~\cite{prlFMS,Fioravanti:2000xz} managed to extract a large variety of universal ratios, which we refrain from replicating here, and instead refer the interested reader to the original works.

However, the recent construction of the FF of the magnetisation operators in~\cite{2022ScPP...12..162C} allows for a novel way of determining the generalised susceptibilities using a form factor expansion for the connected correlation functions. The results are reported in Table~\ref{tab:GenSuscTIM} together with a comparison to those obtained in~\cite{Fioravanti:2000xz} by different means. In certain cases it is also possible to obtain the exact value of the generalised susceptibilities by directly exploiting the $\Delta$-theorem \cite{DSC}:
\begin{equation}
    \Gamma^i_{ik} = - \frac{\Delta_k}{1-\Delta_i} B_{ik}\,.
\end{equation}
We note that in most cases the agreement between the different approaches is excellent, with the exception of some susceptibilities involving the subleading magnetisation $\sigma'$. This latter finding is not surprising since it was already found in~\cite{2022ScPP...12..162C} that spectral sums involving $\sigma'$ (such as the $\Delta$-theorem sum rule) converge much slower than those involving $\sigma$ and $\epsilon$, due to $\sigma'$ being much less relevant than the other two fields.

\begin{table}[t]
    \centering
    \begin{tabular}{|c|c|c|c|c|c|}
     \hline
         Susceptibilities & Form factor approx. & Integration by~\cite{Fioravanti:2000xz} & TCSA by~\cite{Fioravanti:2000xz} & $\Delta$ sum rule \\ \hline
         $\Gamma^{\epsilon-}_{\sigma\sigma}$ & $0.02716$ & $0.026(2)$ & $0.026(7)$ & \\ \hline 
         $\Gamma^{\epsilon-}_{\sigma\epsilon}$ & $-0.06512$ & $\pm0.06(3)$ & $\pm0.06(6)$ & $\pm0.0662\dots$ \\ \hline
         $\Gamma^{\epsilon-}_{\sigma\sigma'}$ & $0.4176$ & $0.4(4)$ & $0.4(2)$ & \\ \hline
         $\Gamma^{\epsilon-}_{\epsilon\epsilon}$ & $0.1578$ & $0.15(8)$ & $0.16(1)$ & $0.16315\dots$ \\ \hline
         $\Gamma^{\epsilon-}_{\epsilon\sigma'}$ & $-1.0353
$ & $\pm 1.1(2)$ & $\pm 1.1(0)$ & $\pm 1.1145\dots$ \\ \hline
         $\Gamma^{\epsilon-}_{\sigma'\sigma'}$ & $7.999$ & $12.(6)$ & & \\ \hline
         $\Gamma^{\epsilon+}_{\sigma\sigma}$ & $0.09318$ & $0.093(9)$ & $0.093(7)$ & \\ \hline
         $\Gamma^{\epsilon+}_{\sigma\sigma'}$ & $0.8676$ & $-0.8(9)$ & $-0.8(8)$ & \\ \hline
         $\Gamma^{\epsilon+}_{\sigma'\sigma'}$ & $9.938$ & $16.(5)$ & & \\ \hline
    \end{tabular}
    \caption{Generalised susceptibilities in the thermal deformations of the tricritical Ising model. The exact results are from the $\Delta$-sum rule. }
    \label{tab:GenSuscTIM}
\end{table}

\section{Summary}

In this work, we reviewed the construction of experimentally relevant quantities such as universal amplitude ratios and dynamical structure functions in low-dimensional models of classical/quantum statistical mechanics, using a description of the vicinity of their critical point by integrable quantum field theories. The crucial tool that allows their computation is the bootstrap approach to scattering amplitudes and form factors.

In the case of the Ising model considered in Section \ref{sec:Ising}, both the thermal and magnetic perturbations are integrable, allowing the extraction of numerous predictions about the scaling region. Besides universal amplitude ratios, we discussed the dynamical structure factors which predict the response of the system to experimental probes. 

In particular, the magnetic deformation of the Ising model leads to a very interesting dynamics with $8$ particle excitations associated with the exceptional Lie algebra $E_8$, originally discovered by Zamolodchikov \cite{1989IJMPA...4.4235Z}. After the first experimental realisation of this model in the ferromagnetic material $\text{Co}\text{Nb}_2\text{O}_6$ \cite{2010Sci...327..177C} (cf. also \cite{2020PhRvB.102j4431A}), it was also realised in the antiferromagnetic system $\text{BaCo}_2\text{V}_2\text{O}_8$ and studied using both inelastic neutron scattering \cite{2020arXiv200513302Z} and terahertz spectroscopy \cite{2020PhRvB.101v0411Z}. The advantage of the antiferromagnetic system is that it allows much more detailed and precise mapping of the spectrum. The results were found to match theoretical calculations whose details were reported in \cite{2021PhRvB.103w5117W}.

After reviewing the recent theoretical and experimental successes in the Ising model, we turned to the universality class of the tricritical Ising model. In terms of two-dimensional classical statistical systems, it corresponds to the Blume-Capel model which was originally proposed to describe magnetic materials with crystal-field coupling \cite{BC1}, but it can also be applied to $^3$He-$^4$He mixtures \cite{PRA4:1071}. The associated quantum spin chain is a spin-$1$ system with the Hamiltonian \eqref{eq:TIMC} \cite{vonGehlen:1989yn}.

Similarly to the $E_8$ case, the thermal deformation of the tricritical Ising model also has very rich dynamics with $7$ particles corresponding to the exceptional Lie algebra $E_7$. We also emphasize that the $E_7$ model has several novelties compared to the Ising $E_8$ case, as a consequence of the presence of topologically charged excitations (kinks) in the spectrum. In terms of determining the dynamical structure factors, the determination of form factors of magnetisation operators turns out to be very nontrivial. As a result, the calculation of DSF potentially relevant for experiments was only accomplished very recently \cite{2022ScPP...12..162C}. We also note that determining these form factors also enabled a novel evaluation of generalised susceptibilities relevant to universal amplitude ratios. This is a new result of this work and was reported in Subsection \ref{subsec:gen_susc_TIM}.

Here we also briefly mention that the presence of kinks in the $E_7$ spectrum results in other very interesting phenomena. In the thermal deformation of the Ising model \cite{1978PhRvD..18.1259M}, breaking integrability by switching on external magnetic fields induces kink confinement which has a very pronounced effect on the non-equilibrium dynamics \cite{2017NatPh..13..246K}. Similar phenomena are expected to occur in the tricritical Ising model. However, the confinement dynamics in the tricritical model is much richer than in the ordinary Ising case, due to the presence of two different species of topological excitations \cite{2022PhLB..82837008L}. In addition, an intricate vacuum structure leads to a fascinating phenomenology of false vacuum decay \cite{2022PhRvD.106j5003L}.

The above considerations demonstrate that a physical realisation of the tricritical Ising spin chain is a very interesting problem whose solution is expected to be extremely rewarding. However, presently it remains an open experimental challenge. One possible way towards this goal is indicated by finding Ising tricriticality an interacting Majorana chain proposed as a model of topological materials \cite{affleck_majorana}, while a more recent option is provided by Rydberg atom arrays \cite{fendley_rydberg}. We close by expressing our hope that experimental advances are soon going to make the fascinating phenomenology of the tricritical Ising model accessible to laboratory studies.

\subsection*{Acknowledgments}

GM acknowledges the grant Prin 2017-FISI. The work of ML was supported by the National Research Development and Innovation Office of Hungary under the postdoctoral grant PD-19 No. 132118 and partially by the OTKA Grant K 134946. GT and ML were partially supported by the National Research, Development and Innovation Office (NKFIH) through the OTKA Grant K 138606, and also within the Quantum Information National Laboratory of Hungary (Grant No. 2022-2.1.1-NL-2022-00004). This collaboration was supported in part by the CNR/MTA Italy-Hungary 2023-2025 Joint Project “Effects of strong correlations in interacting many-body systems and  quantum circuits”.

\providecommand{\href}[2]{#2}\begingroup\raggedright\endgroup

\clearpage

\appendix

\section{Scattering amplitudes in the $E_8$ model}\label{Appendix:FF2ptE8}

\begin{table}[b!]
\begin{center}
\begin{tabular}{|c|l|}\hline
$a$  $b$ &
$S_{ab}$ \\ \hline 
1  1 &
$ \st{\bf 1}{(20)} \, \st{\bf 2}{(12)} \, \st{\bf 3}{(2)} $\\ 
1  2 &
$ \st{\bf 1}{(24)} \, \st{\bf 2}{(18)} \, \st{\bf 3}{(14)} \, \st{\bf 4}{(8)}
$\\ 
1  3 &
$ \st{\bf 1}{(29)} \, \st{\bf 2}{(21)} \, \st{\bf 4}{(13)} \,
\st{\bf 5}{(3)} \, (11)^2 $ \\ 
1  4 &
$ \st{\bf 2}{(25)} \, \st{\bf 3}{(21)} \, \st{\bf 4}{(17)} \,
\st{\bf 5}{(11)} \, \st{\bf 6}{(7)} \, (15) $ \\ 
1  5 &
$ \st{\bf 3}{(28)} \, \st{\bf 4}{(22)} \, \st{\bf 6}{(14)} \,
\st{\bf 7}{(4)} \, (10)^2 \, (12)^2 $ \\ 
1  6 &
$ \st{\bf 4}{(25)} \, \st{\bf 5}{(19)} \, \st{\bf 7}{(9)} \,
(7)^2 \, (13)^2 \, (15) $ \\ 
1  7 &
$ \st{\bf 5}{(27)} \, \st{\bf 6}{(23)} \, \st{\bf 8}{(5)} \,
(9)^2 \, (11)^2\, (13)^2 \, (15) $ \\ 
1  8 &
$ \st{\bf 7}{(26)} \, \st{\bf 8}{(16)^3} \, (6)^2 \,
(8)^2 \, (10)^2 \, (12)^2 $ \\ 
2  2 &
$ \st{\bf 1}{(24)} \, \st{\bf 2}{(20)} \, \st{\bf 4}{(14)} \,
\st{\bf 5}{(8)} \,\st{\bf 6}{(2)} \, (12)^2 $ \\ 
2  3 &
$ \st{\bf 1}{(25)} \, \st{\bf 3}{(19)} \, \st{\bf 6}{(9)} \, (7)^2 \,
(13)^2 \, (15) $ \\ 
2  4 &
$ \st{\bf 1}{(27)} \,\st{\bf 2}{(23)} \, \st{\bf 7}{(5)} \,
(9)^2 \, (11)^2 \, (13)^2\,(15)$ \\ 
2  5 &
$ \st{\bf 2}{(26)} \,\st{\bf 6}{(16)^3} \, (6)^2 (8)^2 (10)^2 (12)^2 $
\\ 
2  6 &
$ \st{\bf 2}{(29)} \, \st{\bf 3}{(25)} \, \st{\bf 5}{(19)^3} \,
\st{\bf 7}{(13)^3} \, \st{\bf 8}{(3)} \, (7)^2 (9)^2 (15) $ \\ 
2  7 &
$ \st{\bf 4}{(27)} \, \st{\bf 6}{(21)^3} \, \st{\bf 7}{(17)^3} \,
\st{\bf 8}{(11)^3} \, (5)^2 (7)^2 (15)^2 $ \\ 
2  8 &
$ \st{\bf 6}{(28)} \, \st{\bf 7}{(22)^3} \, (4)^2 (6)^2 (10)^4 (12)^4 (16)^4 $
\\ 
\hline
\end{tabular}
\end{center}
\caption{$S$-matrix of the Ising model in a magnetic field at $T=T_c$. Continued in Table \ref{tab:e8smat2}.}
\label{tab:e8smat1}
\end{table}

%\newpage

%\begin{table}[ht]
\begin{table}
\begin{center}
\begin{tabular}{|c|l|}\hline
3  3 &
$ \st{\bf 2}{(22)} \, \st{\bf 3}{(20)^3} \, \st{\bf 5}{(14)} \,
\st{\bf 6}{(12)^3} \, \st{\bf 7}{(4)} \, (2)^2 $ \\ 
3  4 &
$ \st{\bf 1}{(26)} \, \st{\bf 5}{(16)^3} \, (6)^2 (8)^2 (10)^2 (12)^2 $
\\ 
3  5 &
$ \st{\bf 1}{(29)} \, \st{\bf 3}{(23)} \, \st{\bf 4}{(21)^3} \,
\st{\bf 7}{(13)^3} \, \st{\bf 8}{(5)} \, (3)^2 (11)^4 (15) $ \\ 
3  6 &
$ \st{\bf 2}{(26)} \, \st{\bf 3}{(24)^3} \, \st{\bf 6}{(18)^3} \,
\st{\bf 8}{(8)^3} \, (10)^2 (16)^4 $ \\ 
3  7 &
$ \st{\bf 3}{(28)} \, \st{\bf 5}{(22)^3} \, (4)^2 (6)^2 (10)^4 (12)^4 (16)^4
$ \\ 
3  8 &
$ \st{\bf 5}{(27)} \, \st{\bf 6}{(25)^3} \, \st{\bf 8}{(17)^5} \,
(7)^4 (9)^4 (11)^2 (15)^3 $ \\
4  4 &
$ \st{\bf 1}{(26)} \, \st{\bf 4}{(20)^3} \, \st{\bf 6}{(16)^3} \,
\st{\bf 7}{(12)^3} \, \st{\bf 8}{(2)} \, (6)^2 (8)^2 $ \\ 
4  5 &
$ \st{\bf 1}{(27)} \, \st{\bf 3}{(23)^3} \, \st{\bf 5}{(19)^3} \,
\st{\bf 8}{(9)^3} \, (5)^2 (13)^4 (15)^2 $ \\ 
4  6 &
$ \st{\bf 1}{(28)} \, \st{\bf 4}{(22)^3} (4)^2 (6)^2 (10)^4
(12)^4 (16)^4 $ \\ 
4  7 &
$ \st{\bf 2}{(28)} \, \st{\bf 4}{(24)^3} \, \st{\bf 7}{(18)^5} \,
\st{\bf 8}{(14)^5} \, (4)^2 (8)^4 (10)^4 $ \\ 
4  8 &
$ \st{\bf 4}{(29)} \, \st{\bf 5}{(25)^3} \, \st{\bf 7}{(21)^5} \,
(3)^2 (7)^4 (11)^6 (13)^6 (15)^3 $ \\ 
5  5 &
$ \st{\bf 4}{(22)^3} \, \st{\bf 5}{(20)^5} \, \st{\bf 8}{(12)^5} \,
(2)^2 (4)^2 (6)^2 (16)^4 $ \\ 
5  6 &
$ \st{\bf 1}{(27)} \, \st{\bf 2}{(25)^3} \, \st{\bf 7}{(17)^5} \,
(7)^4 (9)^4 (11)^4 (15)^3 $ \\ 
5  7 &
$ \st{\bf 1}{(29)} \, \st{\bf 3}{(25)^3} \, \st{\bf 6}{(21)^5} \,
(3)^2 (7)^4 (11)^6 (13)^6 (15)^3 $ \\ 
5  8 &
$ \st{\bf 3}{(28)} \, \st{\bf 4}{(26)^3} \, \st{\bf 5}{(24)^5} \,
\st{\bf 8}{(18)^7} \, (8)^6 (10)^6 (16)^8 $ \\ 
6  6 &
$ \st{\bf 3}{(24)^3} \, \st{\bf 6}{(20)^5} \, \st{\bf 8}{(14)^5} \,
(2)^2 (4)^2 (8)^4 (12)^6 $ \\ 
6  7 &
$ \st{\bf 1}{(28)} \, \st{\bf 2}{(26)^3} \, \st{\bf 5}{(22)^5} \,
\st{\bf 8}{(16)^7} \, (6)^4 (10)^6 (12)^6 $ \\ 
6  8 &
$ \st{\bf 2}{(29)} \, \st{\bf 3}{(27)^3} \, \st{\bf 6}{(23)^5} \,
\st{\bf 7}{(21)^7} \, (5)^4 (11)^8 (13)^8 (15)^4 $ \\ 
7  7 &
$ \st{\bf 2}{(26)^3} \, \st{\bf 4}{(24)^5} \, \st{\bf 7}{(20)^7} \,
(2)^2 (8)^6 (12)^8 (16)^8 $ \\ 
7  8 &
$ \st{\bf 1}{(29)} \, \st{\bf 2}{(27)^3} \, \st{\bf 4}{(25)^5} \,
\st{\bf 6}{(23)^7} \, \st{\bf 8}{(19)^9} \, (9)^8 (13)^{10} (15)^ 5 $
\\ 
8  8 &
$ \st{\bf 1}{(28)^3} \, \st{\bf 3}{(26)^5} \, \st{\bf 5}{(24)^7} \,
\st{\bf 7}{(22)^9} \, \st{\bf 8}{(20)^{11}} \, (12)^{12} (16)^{12} $ \\
\hline
\end{tabular}
\end{center}
\caption{Continuation of $S$-matrix of the Ising model in a magnetic field at $T=T_c$.}
\label{tab:e8smat2}
\end{table}

Using the formalism introduced in Section \ref{sec:Smatrix}, the exact two-body $S$-matrix amplitudes of the critical Ising model in a magnetic field can be written as
\EQ
S_{ab}(\theta)=\prod_{\alpha\in\mathcal{A}_{ab}}\left(f_{\alpha}(\theta)\right)^{p_\alpha},
\label{productS1}
\EN
where 
\EQ
f_\alpha(\theta)\,\equiv \,\frac{\tanh\frac{1}{2}\left(\theta + i\pi\frac{\alpha}{30}\right)}{\tanh\frac{1}{2}\left(\theta - i \pi \frac{\alpha}{30}\right)}\,.
\label{building}
\EN
For a given pair $A_a$ and $A_b$ of the asymptotic particles, the various integer numbers $\alpha=1,\dots, 29$ give the location of the various poles of the S-matrix (in units of $\pi/30$), while the integer numbers $p_\alpha$ give the multiplicity of each of these poles. The set of the $\alpha$'s of the various channels is listed in in Tables \ref{tab:e8smat1} and \ref{tab:e8smat2}, where we use the notation 
$$\st{\bf a}{(\alpha)}^{p_\alpha}$$
 to denote the location of the pole $\alpha$, of multiplicity $p_\alpha$, corresponding to the bound state $A_a$.

\section{Scattering amplitudes and building blocks for the form factors in the $E_7$ model}\label{Appendix:FF2ptE7}

Here we collect the explicit expressions for the scattering amplitudes and the building blocks of the FF in the $E_7$ model which are needed in the main text.

\subsection{The $E_7$ scattering amplitudes}\label{AppendixS-matrix}

Using the formalism introduced in Section \ref{sec:Smatrix}, the exact two-body $S$-matrix amplitudes of the high and low-temperature phases of the tricritical Ising model can be written as \cite{MC,FZ}
\EQ
S_{ab}(\theta)=\prod_{\alpha\in\mathcal{A}_{ab}}\left(f_{\alpha}(\theta)\right)^{p_\alpha},
\label{productS2}
\EN
where 
\EQ
f_\alpha(\theta)\,\equiv \,\frac{\tanh\frac{1}{2}\left(\theta + i\pi\frac{\alpha}{18}\right)}{\tanh\frac{1}{2}\left(\theta - i \pi \frac{\alpha}{18}\right)}\,.
\label{building2}
\EN
For a given pair $A_a$ and $A_b$ of the asymptotic particles, the various integer numbers $\alpha=1,\dots, 17$ give the location of the various poles of the S-matrix (in unit of $\pi/18$), while the integer numbers $p_\alpha$ give the multiplicity of each of these poles. The set of the $\alpha$'s of the various channels is given below, where we use the notation 
$$\st{\bf a}{(\alpha)}^{p_\alpha}$$
 to denote the location of the pole $\alpha$, of multiplicity $p_\alpha$, corresponding to the bound state $A_a$. The eventual 
minus signs in some of the amplitudes means that the corresponding product (\ref{productS2}) must be multiplied by a $-$ sign.  

\begin{table}
\begin{centering}
\bgroup
\setlength{\tabcolsep}{0.5em}
\begin{tabular}{|c|c|}
\hline 
$a$ \, $b$  & $S_{ab}$ \tabularnewline
\hline 
\hline 
\rule[-2mm]{0mm}{10mm}1 \, 1  & $-\st{\bf 2}{(10)}\,\st{\bf 4}{(2)}$\tabularnewline
\hline 
\rule[-2mm]{0mm}{10mm}1 \, 2  & $\st{\bf 1}{(13)}\,\st{\bf 3}{(7)}$\tabularnewline
\hline 
\rule[-2mm]{0mm}{10mm}1 \, 3  & $-\st{\bf 2}{(14)}\,\st{\bf 4}{(10)}\,\st{\bf 5}{(6)}$ \tabularnewline
\hline 
\rule[-2mm]{0mm}{10mm}1 \, 4  & $\st{\bf 1}{(17)}\,\st{\bf 3}{(11)}\,\st{\bf 6}{(3)}\,(9)$ \tabularnewline
\hline 
\rule[-2mm]{0mm}{10mm}1 \, 5  & $\st{\bf 3}{(14)}\,\st{\bf 6}{(8)}\,(6)^{2}$ \tabularnewline
\hline 
\rule[-2mm]{0mm}{10mm}1 \, 6  & $-\st{\bf 4}{(16)}\,\st{\bf 5}{(12)}\,\st{\bf 7}{(4)}\,(10)^{2}$ \tabularnewline
\hline 
\rule[-2mm]{0mm}{10mm}1 \, 7  & $\st{\bf 6}{(15)}(9)\,(5)^{2}\,(7)^{2}$ \tabularnewline
\hline 
\rule[-2mm]{0mm}{10mm}2 \, 2  & $\st{\bf 2}{(12)}\,\st{\bf 4}{(8)}\,\st{\bf 5}{(2)}$ \tabularnewline
\hline 
\rule[-2mm]{0mm}{10mm}2 \, 3  & $\st{\bf 1}{(15)}\,\st{\bf 3}{(11)}\,\st{\bf 6}{(5)}\,(9)$ \tabularnewline
\hline 
\rule[-2mm]{0mm}{10mm}2 \, 4  & $\st{\bf 2}{(14)}\,\st{\bf 5}{(8)}\,(6)^{2}$ \tabularnewline
\hline 
\rule[-2mm]{0mm}{10mm}2 \, 5  & $\st{\bf 2}{(17)}\,\st{\bf 4}{(13)}\,\st{\bf 7}{(3)}\,(7)^{2}\,(9)$ \tabularnewline
\hline 
\rule[-2mm]{0mm}{10mm}2 \, 6  & $\st{\bf 3}{(15)}\,(7)^{2}\,(5)^{2}\,(9)$ \tabularnewline
\hline 
\rule[-2mm]{0mm}{10mm}2 \, 7  & $\st{\bf 5}{(16)}\,\st{\bf 7}{(10)^{3}}\,(4)^{2}\,(6)^{2}$ \tabularnewline
\hline 
\rule[-2mm]{0mm}{10mm}3 \, 3  & $-\st{\bf 2}{(14)}\,\st{\bf 7}{(2)}\,(8)^{2}\,(12)^{2}$ \tabularnewline
\hline 
\end{tabular}
~~~~~~~~~~~~~~~%
\begin{tabular}{|c|c|}
\hline 
$a$ \, $b$  & $S_{ab}$ \tabularnewline
\hline 
\hline 
\rule[-2mm]{0mm}{10mm}3 \, 4  & $\st{\bf 1}{(15)}\,(5)^{2}\,(7)^{2}\,(9)$ \tabularnewline
\hline 
\rule[-2mm]{0mm}{10mm}3 \, 5  & $\st{\bf 1}{(16)}\,\st{\bf 6}{(10)^{3}}\,(4)^{2}\,(6)^{2}$ \tabularnewline
\hline 
\rule[-2mm]{0mm}{10mm}3 \, 6  & $-\st{\bf 2}{(16)}\,\st{\bf 5}{(12)^{3}}\,\st{\bf 7}{(8)^{3}}\,(4)^{2}$ \tabularnewline
\hline 
\rule[-2mm]{0mm}{10mm}3 \, 7  & $\st{\bf 3}{(17)}\,\st{\bf 6}{(13)^{3}}\,(3)^{2}\,(7)^{4}\,(9)^{2}$ \tabularnewline
\hline 
\rule[-2mm]{0mm}{10mm}4 \, 4  & $\st{\bf 4}{(12)}\,\st{\bf 5}{(10)^{3}}\,\st{\bf 7}{(4)}\,(2)^{2}$ \tabularnewline
\hline 
\rule[-2mm]{0mm}{10mm}4 \, 5  & $\st{\bf 2}{(15)}\,\st{\bf 4}{(13)^{3}}\,\st{\bf 7}{(7)^{3}}\,(9)$ \tabularnewline
\hline 
\rule[-2mm]{0mm}{10mm}4 \, 6  & $\st{\bf 1}{(17)}\,\st{\bf 6}{(11)^{3}}\,(3)^{2}\,(5)^{2}\,(9)^{2}$ \tabularnewline
\hline 
\rule[-2mm]{0mm}{10mm}4 \, 7  & $\st{\bf 4}{(16)}\,\st{\bf 5}{(14)^{3}}\,(6)^{4}\,(8)^{4}$ \tabularnewline
\hline 
\rule[-2mm]{0mm}{10mm}5 \, 5  & $\st{\bf 5}{(12)^{3}}\,(2)^{2}\,(4)^{2}\,(8)^{4}$ \tabularnewline
\hline 
\rule[-2mm]{0mm}{10mm}5 \, 6  & $\st{\bf 1}{(16)}\,\st{\bf 3}{(14)^{3}}\,(6)^{4}\,(8)^{4}$ \tabularnewline
\hline 
\rule[-2mm]{0mm}{10mm}5 \, 7  & $\st{\bf 2}{(17)}\,\st{\bf 4}{(15)^{3}}\,\st{\bf 7}{(11)^{5}}\,(5)^{4}\,(9)^{3}$ \tabularnewline
\hline 
\rule[-2mm]{0mm}{10mm}6 \, 6  & $-\st{\bf 4}{(14)^{3}}\,\st{\bf 7}{(10)^{5}}\,(12)^{4}\,(16)^{2}$ \tabularnewline
\hline 
\rule[-2mm]{0mm}{10mm}6 \, 7  & $\st{\bf 1}{(17)}\,\st{\bf 3}{(15)^{3}}\,\st{\bf 6}{(13)^{5}}\,(5)^{6}\,(9)^{3}$ \tabularnewline
\hline 
\rule[-2mm]{0mm}{10mm}7 \, 7  & $\st{\bf 2}{(16)^{3}}\,\st{\bf 5}{(14)^{5}}\,\st{\bf 7}{(12)^{7}}\,(8)^{8}$ \tabularnewline
\hline 
\end{tabular}
\egroup
\par\end{centering}
\caption{$S$-matrix amplitudes in the $E_7$ factorised scattering theory}
\label{Smatrixamplitudes} 
\end{table}

\begin{table}[ht]
\begin{center}
\begin{tabular}{|c|c|c|c|c|c|c|}
\hline 
$abc$ & $112$ & $114$ & $123$ & $134$ & $135$ & $146$ \tabularnewline
\hline 
$\left|\Gamma_{abc}\right|$ & $4.83871$ & $1.22581$ & $7.34688$ & $29.0008$ & $19.1002$ & $4.83871$ \tabularnewline
\hline 
\hline 
$abc$ & $156$ & $167$ & $222$ & $224$ & $225$ & $233$\tabularnewline
\hline 
$\left|\Gamma_{abc}\right|$ & $114.477$ & $29.0008$ & $11.1552$ & $19.1002$ & $1.86121$ & $75.3953$ \tabularnewline
\hline 
\hline 
$abc$ & $236$ & $245$ & $257$ & $337$ & $444$ & $447$\tabularnewline
\hline 
$\left|\Gamma_{abc}\right|$ & $29.0008$ & $114.477$ & $19.1002$ & $7.34688$ & $686.12$ & $114.477$\tabularnewline
\hline 
\end{tabular}
\caption{Three-particle couplings $\Gamma_{abc}$ coming from the simple poles of the $S$-matrix amplitudes. They are entirely symmetric in all three indices, so only the independent non-vanishing ones are shown.}
\label{tab:Sgammas}
\end{center}
\end{table}

The on-shell three-particle couplings $\Gamma_{ab}^c = \Gamma_{abc}$, necessary to implement the Form Factor Equations can be determined up to a sign, and are reported in Table~\ref{tab:Sgammas}. In our calculations, we assume that they are positive unless otherwise stated. Their eventual value depends on the phase convention for the multi-particle states. While most of them can be fixed to have a positive real value, the consistency of the form factor equations eventually determines some of them to have a negative sign. For the matrix elements considered here, this is reflected in the cluster property fixing the sign of one-particle matrix elements described at the end of Subsection \ref{sec:TIMFF}. For more details the interested reader is referred to \cite{2022ScPP...12..162C}.

\subsection{Minimal form factors} 

The minimal form factors for the thermal deformation of the tricritical Ising model have the general expression 
\EQ
F^{min}_{ab}(\th)=\left(-i\sinh\frac{\th}{2}\right)^{\delta_{ab}}
\prod_{\alpha\in{\cal A}_{ab}}\left(g_{\alpha}(\th)
\right)^{p_\alpha}\,,
\lab{fmin}
\EN
where
\EQ
g_{\alpha}(\th)=\exp\left\{2\int_0^\infty\frac{dt}{t}\frac{\cosh\left(
\frac{\alpha}{18} - \frac{1}{2}\right)t}{\cosh\frac{t}{2}\sinh
t}\sin^2\frac{(i\pi-\th)t}{2\pi}\right\}\,.
\lab{block}
\EN
For large values of the rapidity ($|\th|\goto\infty$), by saddle point evaluation it is easy to see that this function has the following asymptotic behaviour 
\EQ
g_{\alpha} (\th) \,\sim\, \mathcal{N}_{\alpha}\exp\left\{\frac{|\th|}{2}-\frac{i \pi}{2}\right\} \, , 
\label{dec}
\EN
where 
\beq
\mathcal{N}_{\alpha}(\th)=\exp\left\{\int_0^\infty\frac{dt}{t}\left[\frac{\cosh\left(
\frac{\alpha}{18} - \frac{1}{2}\right)t}{\cosh\frac{t}{2}\sinh
t}-\frac{1}{t^2}\right]\right\}\,.
\eeq
The function $g_\alpha(\theta)$ has an infinite number of poles outside the physical strip, these are shown explicitly in the infinite-product representation:
\beq
g_\alpha(\theta)=\prod_{k=0}^\infty\left[\frac{\left[1+\left(\frac{\hat{\theta}/2\pi}{k+1-\frac{\alpha}{36}}\right)^2\right]\left[1+\left(\frac{\hat{\theta}/2\pi}{k+\frac{1}{2}+\frac{\alpha}{36}}\right)^2\right]}{\left[1+\left(\frac{\hat{\theta}/2\pi}{k+1+\frac{\alpha}{36}}\right)^2\right]\left[1+\left(\frac{\hat{\theta}/2\pi}{k+1-\frac{3}{2}-\frac{\alpha}{36}}\right)^2\right]}\right]^{k+1},\nonumber
\eeq
where $\hat{\theta}=i\pi-\theta$.
The mixed representation:
\begin{eqnarray}
g_\alpha(\theta)&=&\prod_{k=0}^{N-1}\left[\frac{\left[1+\left(\frac{\hat{\theta}/2\pi}{k+1-\frac{\alpha}{36}}\right)^2\right]\left[1+\left(\frac{\hat{\theta}/2\pi}{k+\frac{1}{2}+\frac{\alpha}{36}}\right)^2\right]}{\left[1+\left(\frac{\hat{\theta}/2\pi}{k+1+\frac{\alpha}{36}}\right)^2\right]\left[1+\left(\frac{\hat{\theta}/2\pi}{k+1-\frac{3}{2}-\frac{\alpha}{36}}\right)^2\right]}\right]^{k+1}\nonumber\\
&&\times\exp\left[2\int_0^\infty\frac{dt}{t}\frac{\cosh\left[t\left(\frac{1}{2} -\frac{\alpha}{18}\right)\right]}{\cosh\frac{t}{2}\sinh t}(N+1-Ne^{-2t})e^{-2Nt}\sin^2\frac{\hat{\theta}t}{2\pi}\right],\nonumber
\end{eqnarray}
is particularly useful for numerical computations.

\subsection{Pole structure of the $2$-particle FF} 

The pole terms entering the parameterisation (\ref{param}) can be expressed as
\EQ
D_{ab}(\th) =\prod_{\alpha\in {\cal
A}_{ab}} \left({\cal P}_\alpha(\th)\right)^{i_\alpha}
\left({\cal P}_{1-\alpha}(\th)\right)^{j_\alpha} \,,
\lab{dab}
\EN
where
\EQ
\begin{array}{lll}
i_{\alpha} = n+1\, , & j_{\alpha} = n \, , & 
{\rm if} \hspace{.5cm} p_\alpha=2n+1\,; \\
i_{\alpha} = n \, , & j_{\alpha} = n \, , & 
{\rm if} \hspace{.5cm} p_\alpha=2n\, , 
\end{array} 
\EN
and we have introduced the notation
\EQ
{\cal P}_{\alpha}(\th)\equiv
\frac{\cos\pi\frac{\alpha}{18} - \cosh\th}{2\cos^2\frac{\pi\alpha}{36}}\,.
\lab{polo}
\EN
Notice that both $F^{min}_{ab}(\th)$ and $D_{ab}(\th)$ are normalised to 1 in $\th=i\pi$.

\end{document}